\documentclass[traditabstract]{aa}

\usepackage{natbib}
  \bibpunct{(}{)}{;}{a}{}{,}
\usepackage{amssymb}
\usepackage{graphicx}
\usepackage{epstopdf}
\usepackage{multirow}
\usepackage{placeins}
\usepackage{stfloats}
\usepackage{tabu}
\usepackage{xspace}

\newcommand{\her}{\textit{Herschel}\xspace}
\newcommand{\spz}{\textit{Spitzer}\xspace}
\newcommand{\hii}{\ion{H}{ii}\xspace}
\newcommand{\mum}{\,\ensuremath{\mu}m\xspace}

\newcommand{\degx}{\degr\xspace}
\newcommand{\amin}{\arcmin\xspace}
\newcommand{\asec}{\arcsec\xspace}
\newcommand{\as}{\arcsec$\!$}
\newcommand{\msun}{\ensuremath{_\odot}\xspace}
\newcommand{\getso}{\textit{getsources}\xspace}
\newcommand{\getsover}{1.140127}
\newcommand{\hipe}{11.1.0}
\newcommand{\x}{\,\ensuremath{\times}\,}
\newcommand{\expo}[1]{\ensuremath{\,\times\,10^{#1}\xspace}}
\newcommand{\pcm}{\,cm$^{-2}$\xspace}
\newcommand{\hms}[3]{\mbox{#1$^\mathrm{h}$#2$^\mathrm{m}$#3$^\mathrm{s}$}}
\newcommand{\dms}[3]{\mbox{#1\degr#2\arcmin#3\arcsec}\xspace}

\newcommand{\mspp}{\,M\msun\,pc$^{-1}$\xspace}

\newcommand{\bbeam}{5.6}    % blue beam   - ie pacs 70 - assumes scan speed 20"/s
\newcommand{\rbeam}{11.4}   % red beam    - ie pacs 160
\newcommand{\sbeam}{18.1}   % short beam  - ie spire 250
\newcommand{\mbeam}{24.9}   % medium beam - ie spire 350
\newcommand{\lbeam}{36.4}   % long beam   - ie spire 500

\newcommand{\cfull}{555\xspace} % Actual total number
\newcommand{\cgood}{177\xspace} % Number on catalogue
\newcommand{\nprot}{28\xspace} % Number of protostars
\newcommand{\ngbnd}{118\xspace} % Number of bound cores
\newcommand{\nfila}{109\xspace} % Number on filaments (by getso)
\newcommand{\ncent}{29\xspace}  % Number in hub mask
\newcommand{\ncpro}{11\xspace} % Number of protostars in hub
\newcommand{\ncbnd}{18\xspace} % Number of bound cores in hub
\newcommand{\nfilc}{23\xspace} % Number in filaments and hub
\begin{document}
  \title{Far-infrared observations of a massive cluster forming in the Monoceros R2 filament hub\thanks{Full version of Tables \ref{tab:deriv} and
\ref{tab:base}--\ref{tab:fwise} are only available in electronic form at the CDS
via anonymous ftp to cdsarc.u-strasbg.fr (130.79.128.5) or via 
http://cdsweb.u-strasbg.fr/cgi-bin/qcat?J/A+A/}}
  \titlerunning{Far-infrared observations of a massive cluster forming in Mon~R2}
  \author{T.~S.~M.~Rayner\inst{1}\thanks{E-mail: T.Rayner@astro.cf.ac.uk}
    \and M.~J.~Griffin\inst{1}    
    \and N.~Schneider\inst{2,\,3}   
    \and F.~Motte\inst{4,\,5}         
    \and V.~K\"onyves\inst{5}
    \and P.~Andr\'e\inst{5}
    \and J.~Di~Francesco\inst{6}
    \and P.~Didelon\inst{5}
    \and K.~Pattle\inst{7}
    \and D.~Ward-Thompson\inst{7}
    \and L.~D.~Anderson\inst{8}
    \and M.~Benedettini\inst{9}
    \and J.-P.~Bernard\inst{10}
    \and S.~Bontemps\inst{3}
    \and D.~Elia\inst{9}
    \and A.~Fuente\inst{11}
    \and M.~Hennemann\inst{5}
    \and T.~Hill\inst{5,\,12}
    \and J.~Kirk\inst{7}
    \and K.~Marsh\inst{1}
    \and A.~Men'shchikov\inst{5}
    \and Q.~Nguyen~Luong\inst{13,\,14}
    \and N.~Peretto\inst{1}
    \and S.~Pezzuto\inst{9}
    \and A.~Rivera-Ingraham\inst{15}
    \and A.~Roy\inst{5}
    \and K.~Rygl\inst{16}
    \and \'A.~S\'anchez-Monge\inst{2}
    \and L.~Spinoglio\inst{9}
    \and J.~Tig\'e\inst{17}
    \and S.~P.~Trevi\~no-Morales\inst{18}
    \and G.~J.~White\inst{19,\,20}
  }
  \institute{Cardiff School of Physics and Astronomy, Cardiff University,
      Queen's Buildings, The Parade, Cardiff, Wales, CF24 3AA, UK % 1 - Cardiff
    \and I. Physik. Institut, University of Cologne, 50937 Cologne,
      Germany                                                     % 2 - Cologne
    \and Laboratoire d'Astrophysique de Bordeaux, Univ. Bordeaux, CNRS,
			B18N, all\'ee G. Saint-Hilaire, 33615 Pessac, France				% 3 - Bordeaux
		\and Universit\'e Grenoble Alpes, CNRS, Institut de Planetologie et 
			d'Astrophysique de Grenoble, 38000 Grenoble, France					% 4 - Grenoble
    \and Laboratoire AIM, CEA/IRFU -- CNRS/INSU -- Universit\'e Paris Diderot,
      CEA-Saclay, 91191 Gif-sur-Yvette Cedex, France              % 5 - Saclay
    \and NRC, Herzberg Institute of Astrophysics, Victoria, Canada% 6 - Herzberg
    \and Jeremiah Horrocks Institute, University of Central Lancashire,
      Preston PR1 2HE, UK                                         % 7 - Preston
    \and Department of Physics and Astronomy, West Virginia University,
      Morgantown, WV 26506, USA                                   % 8 - W. Virginia
    \and INAF -- Istituto di Astrofisica e Planetologia Spaziali,
      via Fosso del Cavaliere 100, I-00133 Roma, Italy            % 9 - Rome
    \and Universit\'e de Toulouse, UPS-OMP, IRAP, Toulouse, France% 10- Toulouse
    \and Observatorio Astron\'omico Nacional (OAN), Apdo 112,
      E-28803 Alcal\'a de Henares, Madrid, Spain                  % 11- Madrid
    \and Joint ALMA Observatory, 3107 Alonso de Cordova, Vitacura, Santiago,
      Chile                                                       % 12- ALMA
  	\and Korea Astronomy and Space Science Institute, 776 Daedeokdae-ro,
			Yuseong-gu, Daejeon, 305-348, Republic of Korea							% 13 - South Korea
    \and National Astronomical Observatory of Japan, Chile Observatory,
      2-21-1 Osawa, Mitaka, Tokyo 181-8588, Japan                 % 14- Tokyo
    \and ESA/ESAC, 28691 Villanueva de la Ca\~nada, Madrid, Spain % 15- ESA Spain
    \and INAF -- Istituto di Radioastronomia, Via Gobetti 101, I-40129 Bologna,
      Italy                                                       % 16- Bologna
    \and Aix-Marseille Universit\'e, CNRS, LAM (Laboratoire d'Astrophysique
      de Marseille) UMR 7326, 13388 Marseille, France             % 17- Marseille
    \and Instituto de Ciencia de Materiales de Madrid (ICMM-CSIC), Sor Juana In\'es
      de la Cruz 3, E-28049 Cantoblanco, Madrid, Spain            % 18- IRAM Spain
    \and The Rutherford Appleton Laboratory, Chilton, Didcot OX11 0NL,
      UK                                                          % 19- RAL
    \and Department of Physics and Astronomy, The Open University, Milton Keynes,
      UK                                                          % 20- OU
  }
  \date{Accepted: August 22, 2017}

  \abstract{
    We present far-infrared observations of Monoceros~R2
    (a giant molecular cloud at approximately 830\,pc distance,
    containing several sites of active star formation),
    as observed at 70\mum, 160\mum, 250\mum, 350\mum, and 500\mum
    by the Photodetector Array Camera and Spectrometer (PACS)
		and Spectral and Photometric Imaging Receiver (SPIRE)
	  instruments on the \her Space Observatory as part of the 
    \her imaging survey of OB young stellar objects (HOBYS) Key programme.
    The \her data are complemented by SCUBA-2 data in the submillimetre range,
    and WISE and \spz data in the mid-infrared.
    In addition, C$^{18}$O data from the IRAM 30-m Telescope are presented,
    and used for kinematic information.
    Sources were extracted from the maps with \getso, and from the
    fluxes measured, spectral energy distributions were constructed,
    allowing measurements of source mass and dust temperature.
    Of \cgood \her sources robustly detected in the region
		(a detection with high signal-to-noise and low axis ratio at multiple wavelengths),
		including protostars and starless cores,
    \ncent are found in a filamentary hub at the centre of the region
		(a little over 1\% of the observed area).
    These objects are on average smaller, more massive,
    and more luminous than those in the surrounding regions
    (which together suggest that they are at a later stage of evolution),
    a result that cannot be explained entirely by selection effects.
    These results suggest a picture in which the hub may have begun
    star formation at a point significantly earlier than the outer regions,
    possibly forming as a result of feedback from earlier star formation.
    Furthermore, the hub may be sustaining its star formation by
    accreting material from the surrounding filaments.
  }
  

  \maketitle

  \section{Introduction}
    \label{sec:Intro}
    \subsection*{Star formation and \her}
    The formation of higher-mass (over $\sim$10\,M\msun) stars
    is a process that remains poorly understood, even today
    \citep[see, for example,][]{2015ASSL..412...43K}.
    This situation is not helped by the rarity of such stars,
    or their short evolutionary timescales.
    Indeed, most regions of higher-mass star formation
    are over a kiloparsec from the Sun, meaning that observations require
    very high angular resolution to view the scales relevant to
    star formation (10--1000\,AU).
    Theoretically, there are two main suggested families of models,
    which differ by whether the protostar gains the majority of its
    mass from its prestellar core \citep[core accretion, or monolithic
    collapse;][]{2002Natur.416...59M}
    or from its surroundings, either from protostellar collisions,
    or from a shared potential well \citep[known as competitive
    accretion;][]{2006MNRAS.370..488B}.

    The two models both require large amounts of matter
    to be concentrated in a single part of the cloud,
    and thus require a method by which such high densities can form.
    One method of achieving such densities would be by the flow
    of material along filaments into dense ``hubs'',
    which exist at the points where filaments merge
    \citep{2009ApJ...700.1609M,2012A&A...540L..11S,2013ApJ...766..115K,
    2013A&A...555A.112P}.
    These filaments and hubs generally have dust
    temperatures of $\sim$10--25\,K,
    and thus they have a spectral energy distribution (SED)
    that peaks in the far-infrared (FIR; $\sim$100--500\mum).

    The \her Space Observatory \citep{2010A&A...518L...1P}
    \footnote{\her is an ESA space observatory with science instruments
    provided by European-led Principal
    Investigator consortia and with important participation from NASA}
    was designed to observe such wavelengths,
    using two photometric instruments: Spectral and Photometric Imaging Receiver
	  \citep[SPIRE;][250--500\mum]{2010A&A...518L...3G} and Photodetector Array
	  Camera and Spectrometer \citep[PACS;][70--160\mum]{2010A&A...518L...2P}
    that, together, covered the desired wavelength range.
    
    Among the \her guaranteed time Key programs, the
    ``\her imaging survey of OB young stellar objects''
    \citep[HOBYS, PIs: F. Motte, A. Zavagno, S. Bontemps;][]{2010A&A...518L..77M}
    was proposed to image regions of high- and intermediate-mass
    star formation, with a view to studying the initial
    conditions of medium- and high-mass star formation, and potentially
    allowing for a better understanding of the process.
    The regions observed by the HOBYS program include
    high-density hubs and ridges (high-density filaments; H$_2$ column density
    above $\sim10^{23}$\,cm$^{-2}$) in which conditions are favourable for the
    formation of medium- to high-mass (over $\sim$10\,M\msun) OB stars, such as
    Vela~C \citep{2011A&A...533A..94H,2012A&A...539A.156G},
    Cygnus~X \citep{2012A&A...543L...3H},
    W48 \citep{2011A&A...535A..76N},
    W43 \citep{2013ApJ...775...88N},
    NGC~6334 \citep{2017A&A...602A..77T},
		NGC~7538 \citep{2013ApJ...773..102F},
    and Monoceros R2 \citep{2015A&A...584A...4D}.
    The first surveys were presented in
    \citet{2010A&A...518L..77M} and \citet{2011A&A...535A..76N},
    and the first catalogue is given in \citet{2017A&A...602A..77T}.
    
    Another \her survey, the ``\her Gould Belt survey'' (HGBS, \citealt{2010A&A...518L.102A}),
    was devoted to observing regions within the Gould Belt (or Gould's
    Belt; a ring of stars and regions of star formation approximately
    $700\,\times\,1000$\,pc across).
    The regions observed are mainly regions of low-mass star formation,
    including the Aquila rift \citep{2010A&A...518L.106K,2015A&A...584A..91K};
    Lupus 1, 3, and 4 \citep{2015MNRAS.453.2036B};
    Orion \citep{2013ApJ...766L..17S};
    Pipe Nebula \citep{2012A&A...541A..63P}; and
    Taurus \citep{2013MNRAS.432.1424K,2014MNRAS.439.3683M,2016MNRAS.459..342M}.
    
    The HOBYS and HGBS surveys use specific definitions for various
    observed objects, which are given here.
    A ``dense core'' is a small, gravitationally bound clump of dust and
    gas, which will collapse (or has begun collapsing)
    to form a protostar or protobinary.
    A ``prestellar core'' is a starless dense core, while
    a ``protostellar core'' is a dense core with an
    embedded young stellar object or protostar.
    ``Massive dense cores'' (MDCs) are 0.1\,pc cloud structures
		which are massive enough to have the potential to form
    high-mass OB stars (see \citealt{2017arXiv170600118M}).
    In this paper, ``starless'' refers to any object that is gravitationally
    bound but contains no visible protostar (including both prestellar cores
    and massive dense cores).
    ``Unbound clumps'' are apparent objects visible on maps, but not bound
    by their own self-gravity.
    ``Source'' refers to any apparent object detected by the source-finding
    routine, whether it is a real astrophysical object or a false positive.
    
    \subsection*{Monoceros R2}
    Monoceros R2 (or Mon~R2) was initially described
    by \citet{1966AJ.....71..990V}
    in a study of associations of reflection nebulae.
    The association lies at an estimated 830\,pc from the Sun 
    (see Section~\ref{sec:dist}), and
    is visible as a 2.4\degr (35\,pc)-long string of reflection nebulae
    running roughly east--west \citep[shown in the visible
    part of the spectrum in][]{1970IAUS...38..219R}.
    The molecular cloud is intermediate between the
    Gould Belt regions and the typical HOBYS regions, in both scale and distance.
    Indeed, it is located almost directly between the Orion~A and B molecular clouds
    (in the Gould Belt, $\sim400\,$pc from the Sun)
    and Canis Major OB1 (not a HOBYS region, but as a region of high-mass
    star formation $\sim1200\,$pc from the Sun, it is certainly similar),
    and potentially connects the two \citep{1986ApJ...303..375M}.
    It is this that makes it particularly interesting as a region with properties
    intermediate between observations of the Gould Belt and those of the more
    distant HOBYS regions, such as NGC~6334 \citep{2017A&A...602A..77T}.
    
    Other regions of the Mon~R2 GMC (giant molecular cloud)
    had previously been detected through extinction
    \citep[LDN 1643--6;][]{1962ApJS....7....1L}.
    The ultracompact \hii (UC\hii) region PKS 0605$-$06
    \citep{1966AuJPh..19..649S,1966AuJPh..19..837S,
    1969AuJPA...8....3S}
    lies roughly at the centre of the nebula NGC~2170,
    the brightest and most westerly part of the association.
    The region immediately surrounding the UC\hii region
    is often referred to as the Mon~R2 Central Core
    (or even simply Mon~R2);
    in this work it shall be referred to as the ``central hub''
    to avoid confusion with starless and protostellar cores.
    Three further more-extended \hii regions surround the
    central hub to the north, east and west.

    The molecular cloud has been observed in
    CO and other molecules \citep[such as CS, H$_2$CO, SO and
    HCN;][]{1974ApJ...194L.103L,1975ApJ...199...79K,
    1979MNRAS.186..107W}.
    More recently, the Heterodyne Instrument for the
    Far-Infrared \citep[HIFI;][]{2010A&A...518L...6D} on \her has
    also observed Mon~R2, in CO, CS,
    HCO$^+\!$, NH, CH, and NH$_3$ \citep{2010A&A...521L..23F,
    2012A&A...544A.110P},
    showing a chemical distinction between the \hii region and
    its high-density surroundings.
    CO outflows have been detected stretching several parsecs
    in the NW--SE direction \citep{1983ApJ...265..824B,
    1990AJ....100.1892W},
    roughly in line with the region's magnetic fields
    \citep{1987A&A...172..304H}.

    \citet{1976ApJ...208..390B} studied the Mon~R2 central hub 
    at mid-infrared wavelengths (1.65\mum--20\mum) and discovered
    that it is composed of several embedded young stellar and protostellar objects,
    one of which (Mon~R2 IRS~1, likely a young B0)
    seems to be responsible for ionising the \hii region.
    The region around the central hub also contains a dense cluster
    of around 500 stars and protostars,
    visible in the near-infrared \citep{1997AJ....114..198C},
		with almost 200 stars in the central square arcminute \citep{2006AJ....132.2296A}.
    The cluster extends over $\sim\,1.1\,\times\,2.1\,$pc,
    and is centred around Mon~R2 IRS~1.
    This object (IRS~1) is coincident with the position of the most massive star
    (approximately 10\,M\msun) within the cluster.
    Although a second infrared source, Mon~R2 IRS~3,
    is associated with a bright (550\,Jy) water maser,
    a recent search of other similar cores in the region failed to
    detect any other water masers with intensities in excess of
    0.5\,Jy \citep[][in prep.]{2015white........1W},
    which are commonly associated with high-mass protostars.
    
		Even more recently, \citet{2015ApJ...803...89D} used
    the SMA interferometer to make observations with higher
    angular resolution than before (0.5\asec--3\asec)
    towards the central hub, at millimetre wavelengths.
    Their results fit well with those of \citet{1976ApJ...208..390B},
    providing very-high-resolution measurements of a few
    objects detected in the hub, especially Mon~R2 IRS~5,
    which is classified as an intermediate-to-high-mass young star,
    with prominent CO outflows at scales below 10\arcsec$\!$, or 0.04\,pc.
    Moreover, multiple outflows were also
    observed, associated with other nearby objects.
		Meanwhile, \her dust column density probability distribution functions (N-PDFs)
		of the region show an unusual overdensity around the central hub
		\citep{2015MNRAS.453L..41S,2016MNRAS.461...22P}, which has been suggested as
		being due to feedback from such objects.
  
    In this paper, we present far-infrared (70\mum--500\mum)
    observations towards the Monoceros~R2 GMC.
    The observations were performed using the \her
    PACS and SPIRE instruments.
    We complement these observations using SCUBA-2 data
    (in the submillimetre regime), \spz and WISE data
    (in the mid infrared range) and IRAM 30-m Telescope
    observations (in the millimeter domain).
    The observations and data reduction are presented in Section~\ref{sec:Obs}.
    In Section~\ref{sec:Reg}, we give a qualitative description
    of the overall region.
    Section~\ref{sec:TCD} presents dust temperature and
    column density maps of Mon~R2.
    Section~\ref{sec:Dis_iram} presents the IRAM 30-m kinematic data,
    and discusses its significance for the region.
    Section~\ref{sec:Sou} details the
    methods used to detect starless and protostellar
    cores, fit their fluxes to SEDs and classify them.
    In Section~\ref{sec:Dis}, we investigate star formation in the region,
		comparing and contrasting the properties of the sources inside and outside the central region.
    Section~\ref{sec:Con} summarises the main conclusions.
    Appendix~\ref{app:Pics} gives all maps directly used in
    the paper (all \her maps, together with the 24\mum MIPS map
    and the 850\mum SCUBA-2 map).
    Appendix~\ref{app:getso} describes the \getso source-finding routine,
    and Appendix~\ref{app:comp} tests the completeness of the \getso
    source identifications in Mon~R2.
    Finally, the catalogues (for the most massive nine objects) are presented
    in Appendix~\ref{app:Table}.
    (The full catalogue is available in the online version.)

  \section{Observations and data reduction}
    \label{sec:Obs}
  \subsection{\her observations}

    \begin{figure*}[htb]
      %\centering
      \mbox{\resizebox{\hsize}{!}{
        \includegraphics{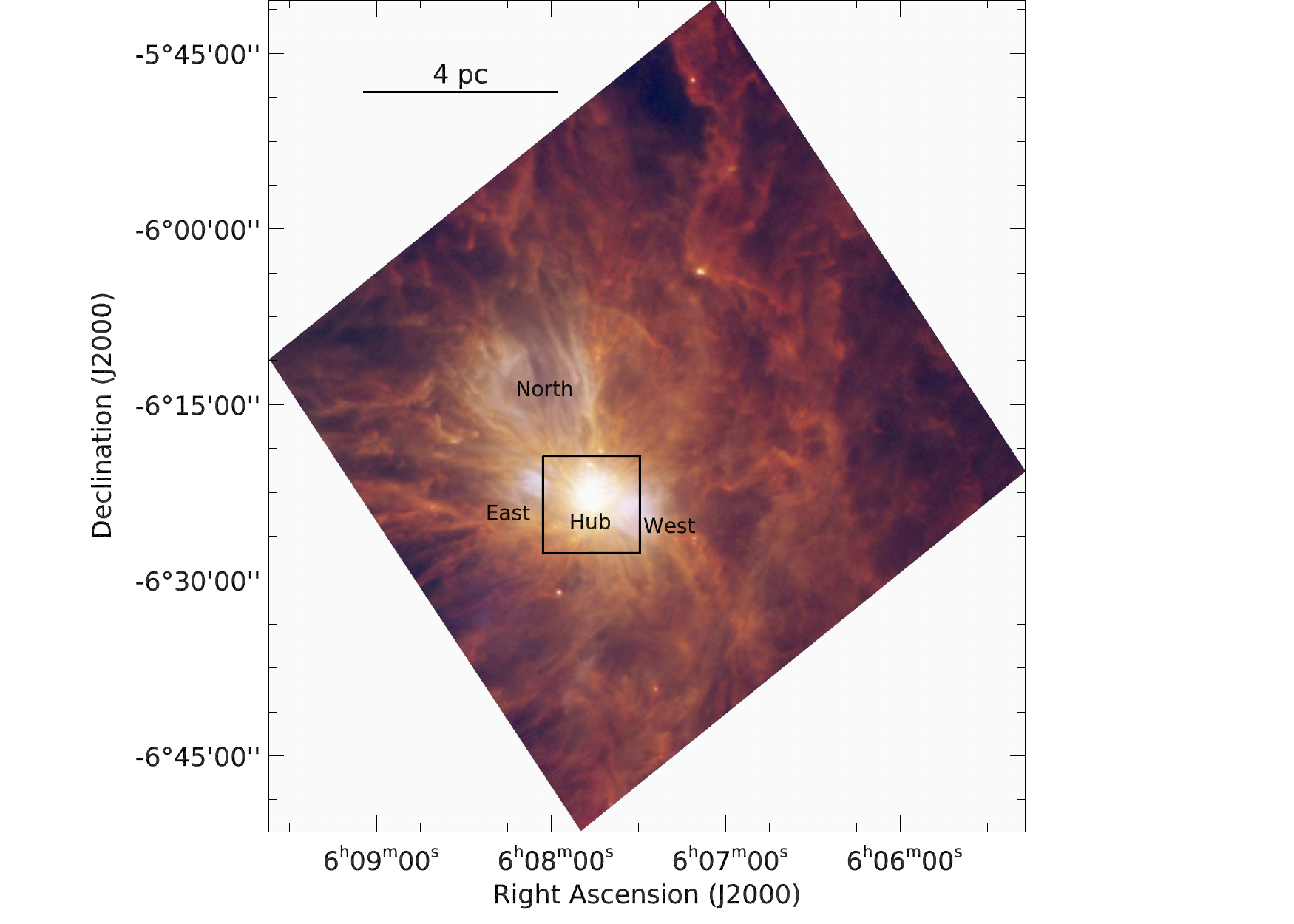}
      }}
      \caption{Monoceros R2 as viewed by \her:
      PACS 70\mum in blue; PACS 160\mum in green; and SPIRE 250\mum in red.
      The image is cropped to the approximate overlap between the
      PACS and SPIRE maps.
      The black box indicates the central hub, which
      is shown in more detail in Figure~\ref{fig:core}
			(due to the large range in brightness, the detailed features of
			the central hub do not appear clearly in this three-colour image).
      The positions of the hub itself, and the three surrounding \hii
      regions are also given.
			The three maps are given separately (with intensity scales) in Appendix~\ref{app:Pics}.
			}
      \label{fig:MonR2}
    \end{figure*}

    Mon~R2 was observed on 4 September 2010, with \her, using the SPIRE
    and PACS instruments.
    The observations (ObsIDs: 1342204052 \& 1342204053)
    were made as part of the HOBYS Key Project.
    The maps were made in Parallel Mode, using both PACS and SPIRE simultaneously,
    with two near-orthogonal (84.8\degr) scans over the region,
    scanning at 20\arcsec$\!$/s,
    and covered five wavebands (70\mum, 160\mum, 250\mum, 350\mum, and 500\mum).
    The bands' mean beam sizes at this scan speed are
    \bbeam\asec, \rbeam\asec, \sbeam\asec, \mbeam\asec, and \lbeam\asec
    (values taken from the \citeauthor{2013PACS.....v2.5.1}, v2.5.1 and
    \citeauthor{2014SPIRE......v2.5}, v2.5) and
    their mean noise levels per beam are
    7.9 (12.0)\,mJy, 7.4 (26.8)\,mJy, 17.6 (7.2)\,mJy,
    10.9 (5.9)\,mJy, and 12.6 (8.5)\,mJy respectively
    (values for Mon~R2, with the mean values from the
    \citeauthor{2014parallel.v2.1} in parentheses).
    The area of overlap between the two instruments' scans was
    about 0.85\degr\x0.85\degr (or 12\,pc\x12\,pc), centred on the J2000 coordinates
    \hms{06}{07}{30}\,$-$\dms{06}{15}{00}.
    A \her composite map (70\mum, 160\mum, 250\mum) is shown in
    Figure~\ref{fig:MonR2}.

    The data were reduced using the
    \her Interactive Processing Environment
    \citep[HIPE, version~\hipe;][]{2010ASPC..434..139O}.
    SPIRE data were reduced using the SPIA \citep[SPIRE Interactive
    Analysis;][]{2011ASPC..442..691S} pipeline.
    This corrects for relative bolometer gains and then
    maps the data with the HIPE Destriper, which adjusts the
    bolometer timelines iteratively until they converge.
    The SPIRE data used the HIPE point source calibration
    \citep{2013MNRAS.433.3062B,2013MNRAS.434..992G}.
    This produces two sets of maps,
    three calibrated for ``point source'' emission (in Jy/beam),
    and three for ``extended emission'' (in MJy/sr),
    which have also been given the zero-point emission offset
    using \textit{Planck} data.
    For consistency with PACS, which only calibrates for
    point sources, in all further analysis only the
    ``point source'' maps have been used.

    The PACS data were reduced using 
    a HIPE-compatible variant of the \textit{Scanamorphos}
    routine \citep{2013PASP..125.1126R}.
    \textit{Scanamorphos} uses the spatial redundancy offered
    by multiple bolometer scans to estimate the overall emission,
    but without making any assumptions about the noise model.
    
    The current PACS processing is unable to correct
    accurately for the diffuse Galactic background levels.
    The absolute levels can however be approximated by
    applying offsets taken from \textit{Planck} data
    to provide the \her maps with background emission offsets,
    which are assumed for each waveband
    to remain constant over the mapped region
    \citep[described in more detail in][]{2010A&A...518L..88B}.
    The SPIRE offsets agree very well with those determined
    using the zero-point correction task in HIPE.

  \subsection{SCUBA-2 observations}
    In addition to the \her observations,
    Mon~R2 was observed with the
    SCUBA-2 instrument \citep{2013MNRAS.430.2513H} on the JCMT
    (data provided by D. Nutter).
		The data provided were used alongside the \her data,
    to provide coverage at longer wavelengths.
    The region was observed eighteen times between November 2011 and January 2012,
    as part of the guaranteed time project M11BGT01.
    Continuum observations at 850\mum and 450\mum were made
    using fully sampled 1\degx diameter
    circular regions \citep[PONG3600 mapping
    mode;][]{2014SPIE.9153E..03B} at resolutions
    of 14.1\asec and 9.6\asec, respectively.
    We present the 850\mum observations here.

    The data were reduced using the iterative map-making technique
    \textit{makemap} in \textsc{smurf} \citep{2013MNRAS.430.2545C},
    and gridded onto 6\asec pixels at 850\mum.
    Areas of low emission were masked out based on their
    signal-to-noise ratio, and a mosaic was formed using this mask
    to define areas of emission.
    Detection of emission structure and calibration
    accuracy are robust within the masked regions,
    and are uncertain outside of the masked regions.

    A spatial filter of 10\amin is used in the reduction
		(described in \citealt{2015MNRAS.450.1094P}).
    Flux recovery is robust for objects with a
    Gaussian FWHM less than 2.5\amin.
    Objects between 2.5\amin and 7.5\amin in size will be detected,
    but both the flux density and the size will be underestimated
    because Fourier components with scales greater
    than 5\amin are removed by the filtering process.
    Detection of objects larger than 7.5\amin is
    dependent on the mask used for reduction.

    The data are calibrated 
    using the peak Flux Conversion Factor (FCF)
    of 537\,Jy/pW at 850\mum, derived from average values
    of JCMT calibrators \citep{2013MNRAS.430.2534D}.
    The PONG scan pattern leads to lower noise levels
    in the map centre, while data reduction and emission
    artefacts can lead to small variations in noise level
    over the whole map.
  
  \subsection{IRAM 30-m observations}
    \label{sec:iram}
    Mon~R2 was also observed with the IRAM 30-m telescope, using the Eight
    MIxer Receiver \citep[EMIR;][]{2012A&A...538A..89C}, between
    12 September and 10 November 2014.
    The observations were carried
    out in the on-the-fly observing mode and covered an area of $\sim$100
    square arcminutes centred on the J2000 coordinates \hms{06}{07}{46.2}
    $-$\dms{06}{23}{08.3}, in two frequency bands
    with coverage of 213.1--220.9\,GHz and 228.8--236.6\,GHz.
    The observations thus cover the molecular emission lines of $^{12}$CO,
    $^{13}$CO, and C$^{18}$O 2$\to$1, DCN 3$\to$2, DCO$^+$ 3$\to$2, and
    two H$_2$CO lines at 218.22\,GHz and 219.16\,GHz.
    In this paper, we only use the 
    C$^{18}$O 2$\to$1 data at 219.56\,GHz.
    The instrument's pointing and focus were
    tested every $\sim$2 hours against nearby bright quasars.
    The emission-free reference position is located at
    \hms{06}{08}{26.2} $-$\dms{06}{33}{08.3}.
    
    The data were reduced with the
    GILDAS\footnote{http://www.iram.fr/IRAMFR/GILDAS/} software. A
    baseline of order 1 was subtracted. In order to convert from antenna
    temperatures (given in this paper) into main beam brightness
    temperatures, a scaling factor of $\sim$0.6 for the main beam
    efficiencies\footnote{http://www.iram.es/IRAMES/mainWiki/\ Iram30mEfficiencies}
    needs to be applied.

  \subsection{Other observations}
    In addition, mid-infrared data were used.
    These included 24\mum observations with the
    Multiband Imaging Photometer for \spz
    \citep[MIPS;][]{2004ApJS..154...25R,2004ApJS..154....1W},
    which were taken from the \spz Heritage Archive.
    Data were also taken from the AllWISE Source Catalog,
    a catalogue of objects discovered as part of the
    Wide-field Infrared Survey Explorer
    \citep[WISE; ][]{2010AJ....140.1868W}
    and NEOWISE \citep{2011ApJ...731...53M} missions.
    In both cases, the data were used only to calculate
    source bolometric luminosities
    \citep[defined here as the integral of flux densities
    from 2\mum to 1\,mm;][]{2017A&A...602A..77T}.
    In addition, 3.6\mum data from \spz's Infrared Array Camera
    \citep[IRAC;][]{2004ApJS..154...10F}
    were used, solely for source visualisation, as seen in
    Appendix~\ref{app:Table}.
    Both \spz and WISE are partially saturated at the central hub
    and cannot be used to determine the luminosities at that location.
    A full list of wavelengths used for this analysis
    is given in Table~\ref{tab:wavs}.

\begin{table}[htb]
\centering
\begin{tabular}{| l | l | c | c |}
\hline
\multirow{2}{*}{Telescope} & \multirow{2}{*}{Instrument} & $\lambda$ & HPBW \\
 & & (\mum) & (arcsec) \\
\hline
\multirow{5}{*}{\her} & \multirow{2}{*}{PACS} &  70 & \bbeam              \\ \cline{3-4}
                      &                       & 160 & \rbeam              \\ \cline{2-4}
                      & \multirow{3}{*}{SPIRE}& 250 & \sbeam              \\ \cline{3-4}
                      &                       & 350 & \mbeam              \\ \cline{3-4}
                      &                       & 500 & \lbeam              \\ \hline
JCMT                  & SCUBA-2               & 850 & 14.1                \\ \hline
\multirow{2}{*}{\spz} & IRAC                  & 3.6 & 1.5                 \\ \cline{2-4}
                      & MIPS                  &  24 & 6.0                 \\ \hline
\multirow{4}{*}{WISE} &                       & 3.4 & 6.1                 \\ \cline{3-4}
                      &                       & 4.6 & 6.4                 \\ \cline{3-4}
                      &                       &  12 & 6.5                 \\ \cline{3-4}
                      &                       &  22 & 12.0                \\ \hline
\end{tabular}
\label{tab:wavs}
\caption{Telescopes and wavelengths used in the analysis.
  We note that the \spz 3.6\mum was only used for source visualisation.
  The half power beam widths (HPBW; generally referred to as beam sizes)
  are taken from the references given in the text.}
\end{table}

  \section{Overview of the Mon~R2 region}
  \label{sec:Reg}
    Figure~\ref{fig:MonR2} shows a three-colour (70\mum, 160\mum, 250\mum) map of the Mon~R2 region, and
    Figures~\ref{fig:app_MonR2_70}--\ref{fig:app_MonR2_500} 
    show individual maps at 70\mum, 160\mum, 250\mum, 350\mum, and 500\mum, respectively.
    In addition, Figures~\ref{fig:app_MonR2_24} and \ref{fig:app_MonR2_850} 
    show the 24\mum MIPS and 850\mum SCUBA-2 data, respectively.

    The most striking feature of the \her maps of Mon~R2 is the central hub,
    which is very bright in all five wavebands.
    This region is located at the junction (hence the term ``hub'') 
    of several filaments, which are also prominent at most wavelengths.
    Most of the large-scale filamentary structure emits most strongly
		at 500\mum, indicating that the filaments consist of cold gas
		(although the poor spatial resolution at this wavelength makes
		identification of features significantly easier at shorter wavelengths).
    However, a significant amount of this filamentary structure
    (most notably the region to the north-east of the hub) is
    even visible at 70\mum, especially around the central region, which indicates
    heating of the filaments, most probably from the associated \hii region.

\begin{figure}[htb]
      \centering
      \mbox{\resizebox{\hsize}{!}{
        \includegraphics[trim=20 0 40 0, clip=true]{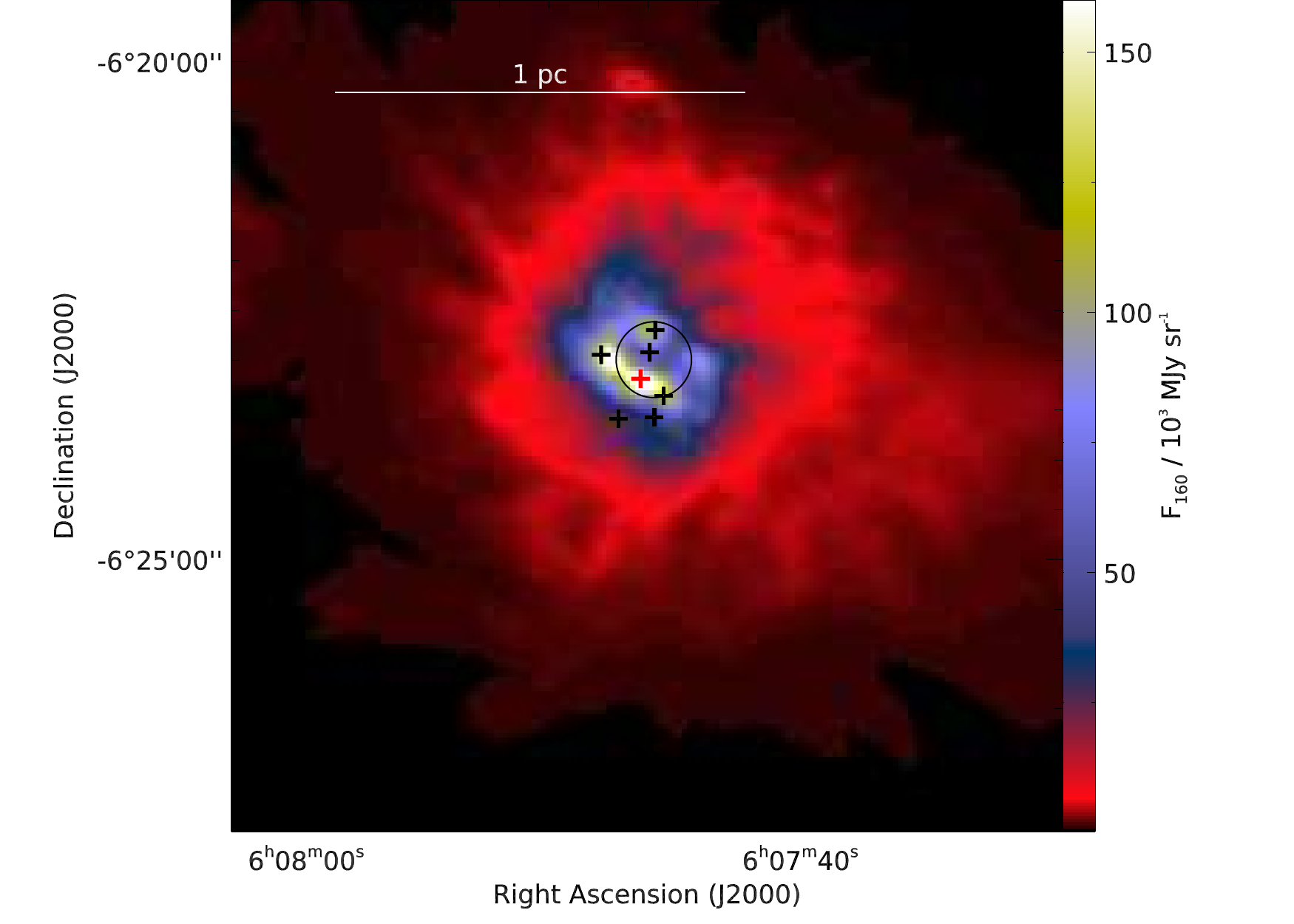}
      }}
      \caption{Monoceros R2 central hub region as seen at 160\mum.
      The region viewed is shown as the black box in Figure~\ref{fig:MonR2}.
      The black circle shows the approximate position and extent of the \hii region
      \citep{2012A&A...544A.110P,2015A&A...584A...4D}.
      The sources are from \citet{1976ApJ...208..390B},
      detected between 1.65\mum and 20\mum
      (the red object is Mon~R2 IRS~1).
      }
      \label{fig:core}
    \end{figure}

    The central hub itself can be seen in detail
    in the PACS and shorter wavelength SPIRE bands,
    as shown in Figure~\ref{fig:core}.
    By inspection, the data appear to contain at least four
    separate infrared sources, arranged approximately to the north, east, south, and west.
    Comparison with \citet{1976ApJ...208..390B} suggests that these objects approximately
    correspond to their sources IRS~1 and 4 (south), IRS~5 (north), IRS~3 (east), and another,
    which was not in the region that they mapped.

    The infrared sources appear to be connected,
    potentially by smaller-scale filamentary
    structures than those that converge on the central hub,
    and form a ring-shape around the \hii region in projection,
    approximately 1\arcmin, or 0.24\,pc, across.
    Molecular line observations suggest that this ring is formed by the
    interactions between the PDR of the \hii region and the denser surrounding material
    \citep{2013A&A...554A..87P,2014A&A...561A..69P,2014A&A...569A..19T,2016arXiv160706265T}.
    These filaments can be seen to radiate from the central hub
    in almost all directions.
    With the help of complementary molecular line data
    (see Section~\ref{sec:Dis_iram})
    we are able to show that most of the (2D projected) filaments
    seen in the \her data correspond to velocity-coherent objects.
    The observed velocity gradients suggest that the filaments are
    gravitationally attracted by the central potential well.
    
  \subsection*{Distance measurements}
    \label{sec:dist}
    The distance to a reflection nebula can be calculated by finding the
    distances to the individual stars associated with that nebula.
    In the case of Mon~R2, magnitude measurements of ten individual
    stars gave a distance of 830$\pm$50\,pc \citep{1968AJ.....73..233R}.
    A similar study of thirty stars confirmed this result
    \citep{1976AJ.....81..840H}.
    A more recent study of over 200 stars in the vicinity found
    a distance of 905$\pm$37\,pc, and the even greater (although less
    precise) value 925$\pm$150\,pc for the hub alone \citep{2011A&A...535A..16L},
    and even more recent parallax measurements give a distance of
    $893^{+44}_{-40}$\,pc \citep{2016arXiv160601757D}.
    The values show good agreement, with a mean value of about 870\,pc.
    For better comparison with prior literature, the
    \citeauthor{1968AJ.....73..233R} value (830$\pm$50\,pc)
    is used throughout this paper.
		Using a value of 900\,pc (in line with the more recent results)
		would systematically increase our source mass estimates by $\sim15\%$.

  \section{Column density and dust temperature maps}
    \label{sec:TCD}
\begin{figure*}[htb]
      \mbox{\resizebox{\hsize}{!}{\includegraphics{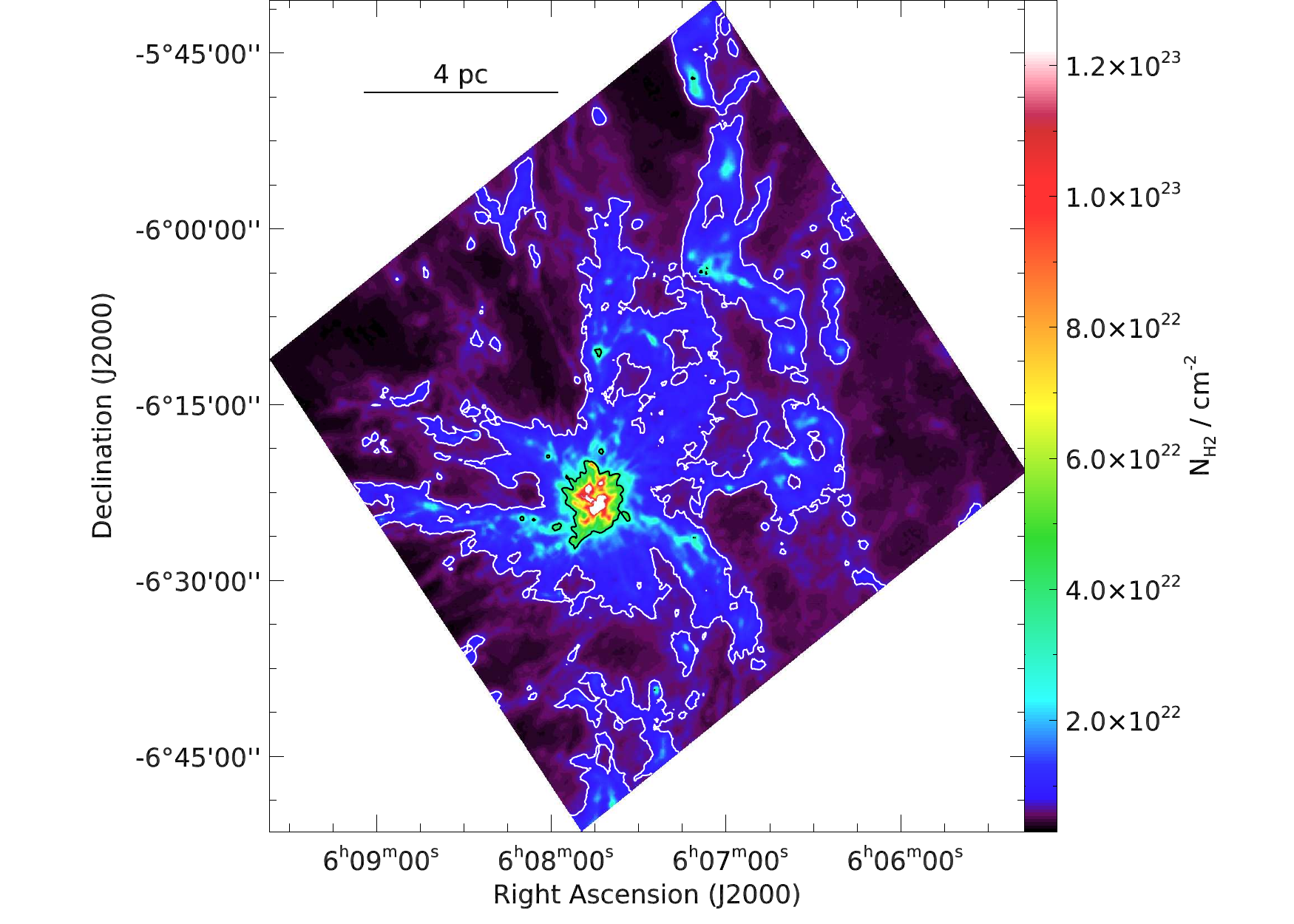}}}
      \caption{Monoceros R2 high-resolution H$_2$ column density map
      ($\tilde{\Sigma}$).
      The resolution is as for the 250\mum map (\sbeam\asec).
      The contours are at the H$_2$ column densities
      7.5\expo{21}\pcm (white) and 3.5\expo{22}\pcm (black).}
      \label{fig:dens}
    \end{figure*}

\begin{figure*}[htb]
      \mbox{\resizebox{\hsize}{!}{\includegraphics{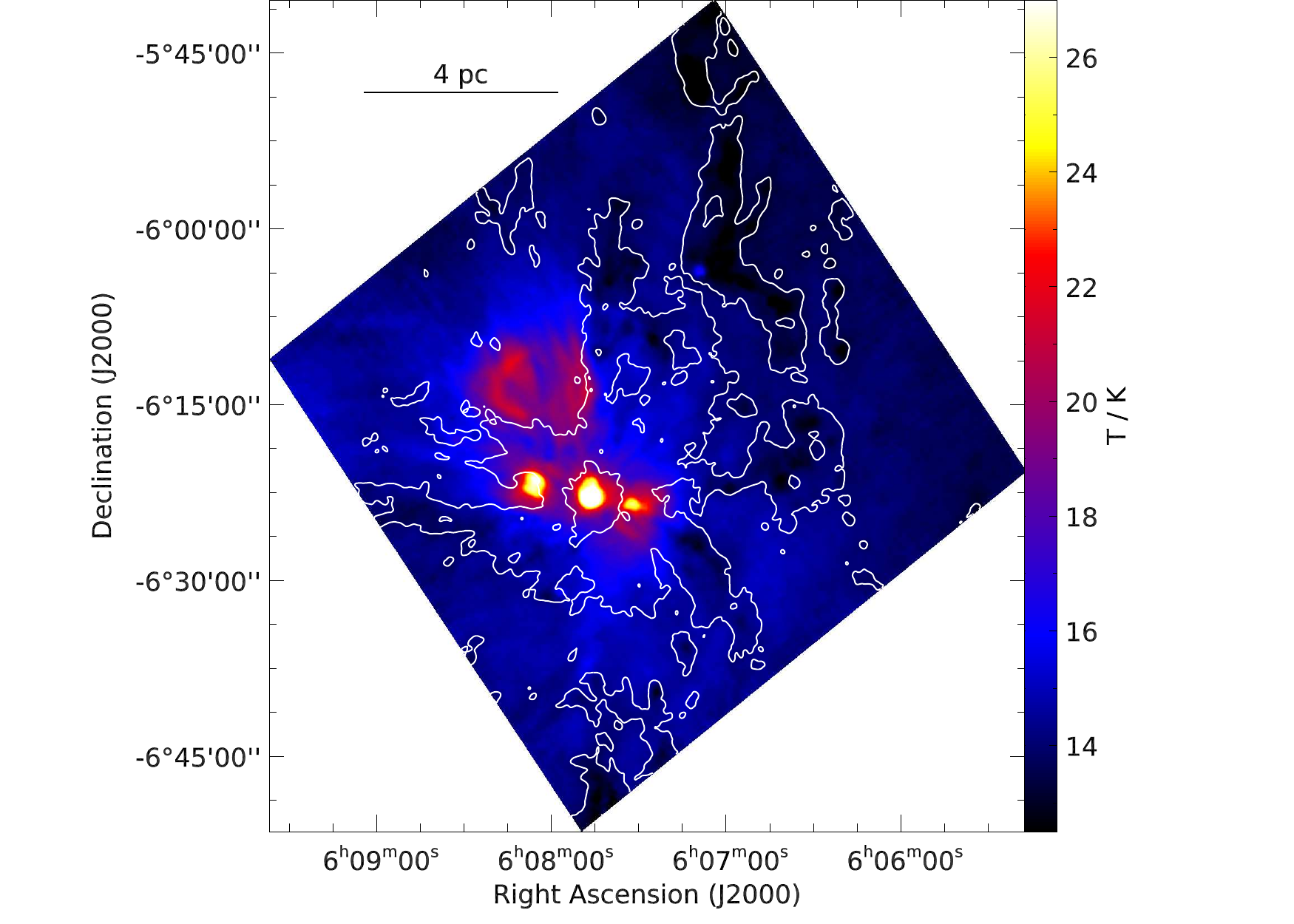}}}
      \caption{Monoceros R2 dust temperature map.
        The resolution is as for the 500\mum map (\lbeam\arcsec).
        The contours (as for Figure~\ref{fig:dens}) are at the
        H$_2$ column densities 7.5\expo{21}\pcm and 3.5\expo{22}\pcm.}
      \label{fig:temp}
    \end{figure*}

    Molecular hydrogen column density
    and dust temperature maps were obtained by a pixel-by-pixel
    modified blackbody (greybody) SED fit to the \her flux maps.
    For more details on the SED fit, see Section~\ref{sec:Sou}.
    We initially produced three column density maps with
    different angular resolutions
    ($\Sigma_{500}$, $\Sigma_{350}$, and $\Sigma_{250}$),
    which were then used to make a ``super-resolution''
    column density map ($\tilde{\Sigma}$),
    using the method given in \citet{2013A&A...550A..38P};
    a similar method is used in \citet{2012A&A...548L...6H}
    (we note that prior to the procedure, all maps were resampled to the
    smallest pixel size, 2.85\asec).
    The first column density map, $\Sigma_{500}$,
    was made by smoothing the 160\mum, 250\mum and 350\mum maps
    to the resolution of the 500\mum map (\lbeam\asec),
    before fitting the SED to all four maps, giving
    a column density map with a \lbeam\asec resolution.
    The second column density map, $\Sigma_{350}$,
    was made by smoothing the 160\mum and 250\mum maps to the
    resolution of the 350\mum map (\mbeam\asec),
    and then fitting the SED to only these three maps.
    The third column density map, $\Sigma_{250}$,
    was made by smoothing the 160\mum map to the resolution of the
    250\mum map (\sbeam\asec), and using the flux ratio
    ($F_{250}/F_{160}$) to create a new temperature map,
    $T_{250}$, using the method given in \citet{2009ApJ...696.2234S}:

    \begin{equation}
      \frac{F_{\nu_1}}{F_{\nu_2}} = \left(\frac{\lambda_1}{\lambda_2}\right)^{3+\beta}
      \frac{\exp(\lambda_T/\lambda_2)-1}{\exp(\lambda_T/\lambda_1)-1},
    \end{equation}

    \noindent where
    $F_{\nu_i}$ are the flux densities at the frequencies
    $\nu_i$ (or wavelengths $\lambda_i$);
    $\beta$, the dust emissivity index, assumed to be 2;
    and $\lambda_T = hc/kT$, where:
    $h$, $c$ and $k$ are the Planck constant, speed of light
    and Boltzmann constant, respectively;
    and $T$ is the dust temperature.
    $T_{250}$ was used to find a column density map,
    $\Sigma_{250}$, using the SED equation (Equation~\ref{equ:sed} in Section~\ref{sec:Sou}),
    and using the 250\mum flux map (and $\lambda = 250$\mum).
    To create the final \sbeam\asec resolution column density
    map, $\tilde{\Sigma}$,
    the three previous column density maps ($\Sigma_{500}$,
    $\Sigma_{350}$, and $\Sigma_{250}$)
    were combined, using the following equation:

    \begin{equation}
      \tilde{\Sigma} = \Sigma_{500}
      + \Sigma_{350} - \Sigma_{350}*G_{500\_350}
      + \Sigma_{250} - \Sigma_{250}*G_{350\_250},
    \end{equation}
    
    \noindent where $G_{500\_350}$ and $G_{350\_250}$
    are circular Gaussians with FWHMs equal to
    26.6\asec ($\sqrt{\lbeam^2-\mbeam^2}$\asec) and
    17.1\asec ($\sqrt{\mbeam^2-\sbeam^2}$\asec).
    This method ensures that the basic shape of $\tilde{\Sigma}$
    comes from the most reliable of the input column density maps
    ($\Sigma_{500}$, which uses four input flux maps),
    while the details come from maps with higher resolution
    but lower reliability ($\Sigma_{350}$ and $\Sigma_{250}$).
    To test the reliability of the final map ($\tilde{\Sigma}$),
    it was smoothed to \lbeam\asec resolution
    and compared with the $\Sigma_{500}$ map,
    yielding a difference of $<$3\% over 90\% of the region.
    Unless otherwise stated, the term ``column density map''
    hereinafter refers to this final map ($\tilde{\Sigma}$).
    The final column density and 500\mum dust temperature maps
    are given in Figures~\ref{fig:dens} and \ref{fig:temp}.

    Looking at the column density map,
    the highly filamentary nature of the region can be clearly seen.
    The filaments correspond well to column densities above
    7.5\expo{21}\pcm (white contour),
    while the central hub corresponds well to column densities above
    3.5\expo{22}\pcm (black contour).
    Meanwhile, in the dust temperature map, it can be seen that the
    filamentary regions often correspond to lower temperatures
    (as expected, since a dense region will be
    better shielded from external radiation),
    while warmer regions (especially the northern and eastern
    \hii regions) can be seen to correspond to lower-density regions
    (both because the heating influence can spread farther in lower densities
    and also because the \hii regions will naturally reduce their own local densities).
    This trend is inverted at the centre of the hub, which has the highest
    dust temperature in the region (this corresponds to the central UC\hii region).
    This is potentially because the UC\hii region has not yet had time
    to disperse its high-density surroundings;
    more examples of this are seen in NGC~6334 \citep{2013A&A...554A..42R,2017A&A...602A..77T},
    W48A \citep{2014MNRAS.440..427R} and part of the DR21 ridge \citep{2012A&A...543L...3H}.
    It should be noted that while the column density is two-dimensional (2D), and thus not
    completely analogous to the true three-dimensional (3D) molecular density,
    for a region with no significant foreground or background emission
    (such as Mon~R2), it can be assumed that high--column-density regions
    are regions of actual high density.

  \section{Velocity structure}
    \label{sec:Dis_iram}
    As described in Section~\ref{sec:iram},
    extended maps of isotopomeric CO 2$\to$1 and 1$\to$0 lines
    have recently been obtained with the IRAM 30-m
    telescope (PIs N. Schneider, A. Fuente, S. Trevi\~no-Morales).
    We present here C$^{18}$O 2$\to$1 data from these projects, which are
    fully shown and discussed in \citet{2016trevino....phdT},
    and in \citet{2015rayner.....phdR}.
    It should be noted that we do not aim to perform explicit
    filament detections and analyses here.
    Rather, we intend to show that what appear as filaments in projected
    column density do indeed correspond to coherent velocity features,
    and can be used to derive first order approximations of their physical properties.
    
\begin{figure*}[htb]
    \begin{centering}  
      \includegraphics[width=14.5cm]{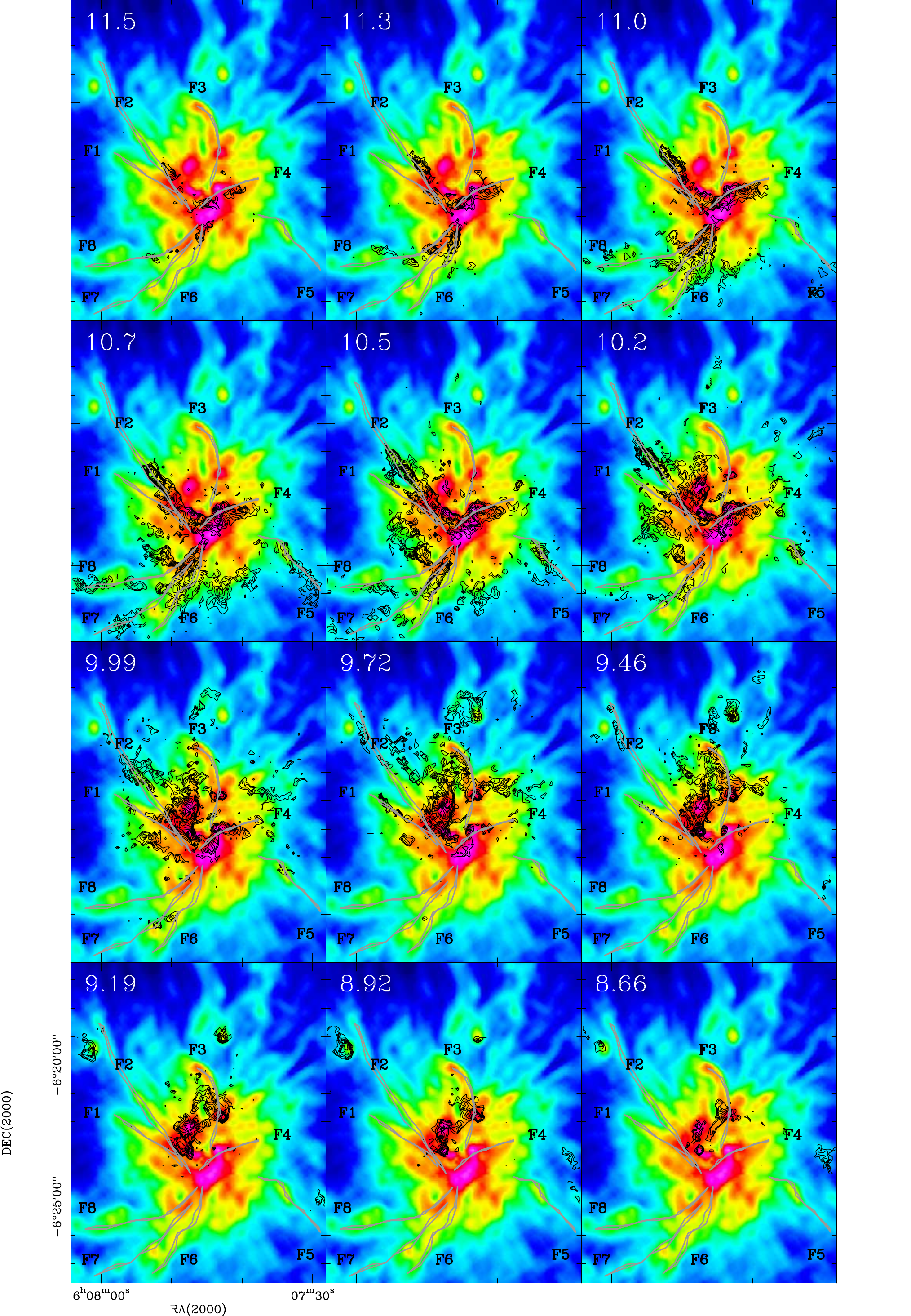}
      \caption{Channel map of C$^{18}$O 2$\to$1 emission.
        Black contours (2.5--7\,K\,km/s by 0.5\,K\,km/s) are overlaid on the
        \her column density map in colour (Figure~\ref{fig:dens}).
        Filaments identified in the C$^{18}$O and in the
        \her map are classified F1--F8 and traced in grey.
        }
      \label{fig:channel-c18o}
    \end{centering}
  \end{figure*}

    Figure~\ref{fig:channel-c18o} shows a channel map of C$^{18}$O
    emission overlaid on the column density map obtained with \her.
    The black contours of C$^{18}$O emission trace very well, in some
    velocity intervals, the filaments seen with \her and reveal
    velocity gradients that are potentially
    caused by inflow along inclined filaments.
    All clear correspondences are labelled (filaments F1--F8;
    identified by eye from the C$^{18}$O structures).
    Fainter filamentary structures that are only seen in
    the \her map were not considered for our census, even though
    they are partly identified in lower-density tracers such as
    $^{12}$CO 1$\to$0.
		
    We note that these filaments do not exactly correspond to the
    filaments detected by the \getso routine (see Appendix~\ref{app:getso}),
    which are detected by measuring large-scale but low width objects
    on the column density maps.
		The main reason for this discrepancy is not the method of identification, but rather
		the fact that the C$^{18}$O filaments have been identified across multiple molecular line maps,
		while the \getso filaments have been identified only on the column density map.
		Consequently, a one-to-one correlation would not be expected, both because
		C$^{18}$O does not directly trace dust column density, and also because
		the by-eye detection focuses on filaments that are coherent structures in velocity space,
		rather than 2-dimensional column density.
		
\begin{figure}[htb]
    \begin{centering}  
      \mbox{\resizebox{\hsize}{!}{\includegraphics[trim=20 0 90 0,
        clip=true]{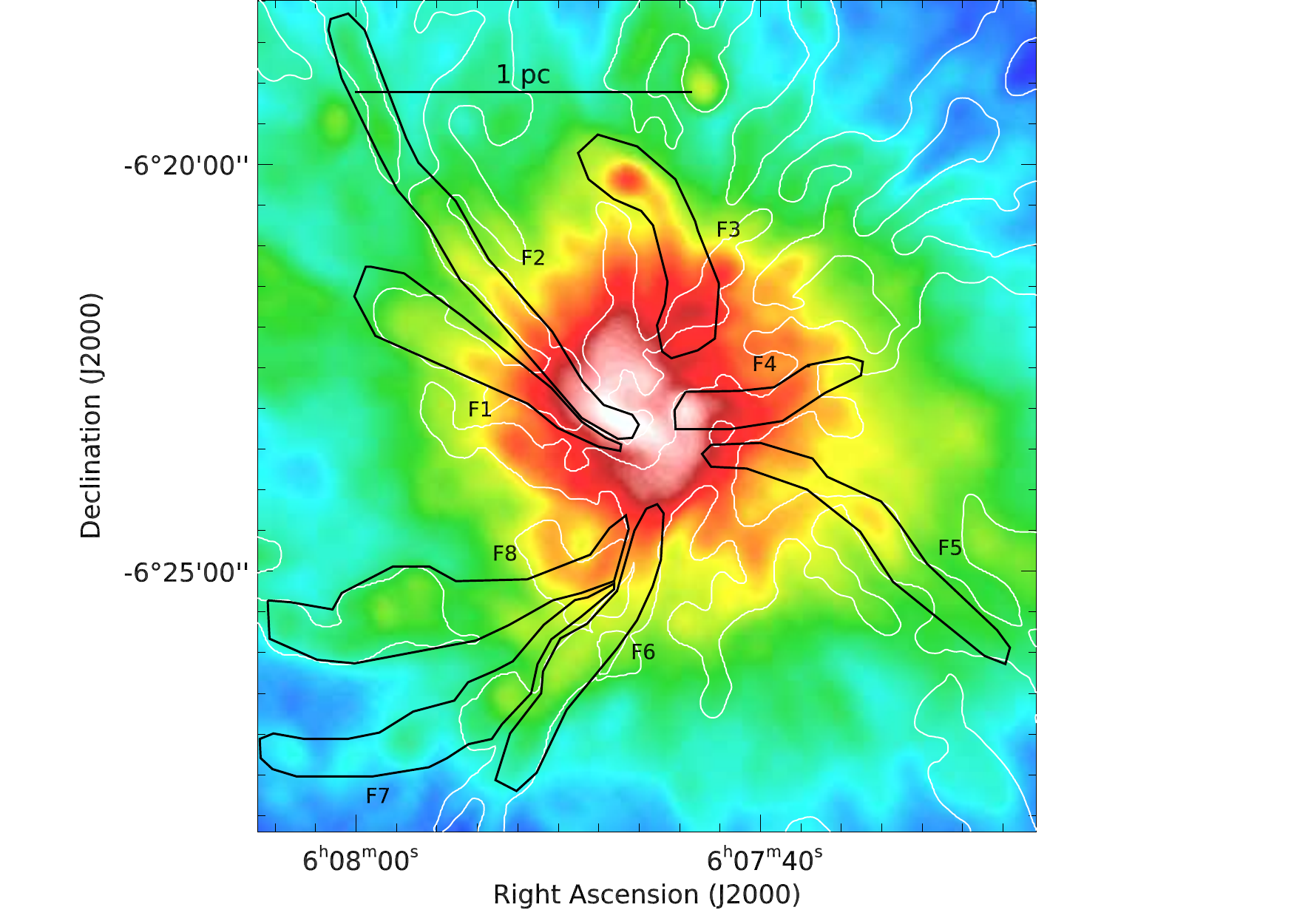}}}
      \caption{Colour map of the \her column density (Figure~\ref{fig:dens}),
        overlaid with polygons (F1--F8) indicating the areas over which the filament
        properties (average column density, density, temperature, and total mass)
        were determined.
        The white contours show the filaments identified by \getso
        in the column density map
        } 
      \label{fig:filaments}
    \end{centering}
  \end{figure}

    The two different sets of filaments are compared in
    Figure~\ref{fig:filaments}, with the C$^{18}$O filaments in black,
    and the \getso filaments in white.
    There is reasonably good overlap in most cases,
    especially filament F2 (to the north-east) and F6 (to the south).
    Only two of the C$^{18}$O filaments are not detected by \getso,
    F3 (to the north) and F7 (to the south-east), and the former
    is certainly visible by eye.
    
    One of the best examples of a velocity gradient is in F2, which shows
    its first emission close to to the cluster centre at 11.5\,km\,s$^{-1}$.
    The C$^{18}$O emission then shifts towards north-east ``along''
    the \her filament until a velocity of 9.5\,km\,s$^{-1}$.
    This emission pattern is consistent with infall along a filament
    that is inclined towards the observer
    (although outflow along a filament inclined away from the observer
    is also a possibility).
    For other filaments, for example, F8, the top of the column lies at
    higher positive velocities, indicating that this part
    of the filament is tilted away from the observer
    (again, assuming infall).
    Overall, the northern filaments (F1--F4) appear visually
    more collimated than the southern ones (F6--F8).
    As outlined below, the two filament groups
    also differ in their physical properties.
    The more widespread C$^{18}$O emission close to the central
    region in the velocity range 10.2--8.9\,km\,s$^{-1}$
    can be interpreted as filaments seen head-on.
    
    For all filaments, we determined the velocity gradient
    relative to the observer, $\Delta v_\mathrm{grad}$,
    from the channel map in order to derive dynamical lifetimes and infall rates.
    We assume a random distribution of orientation
    angles and thus an average angle to the line-of-sight of 57.3\degx
    \citep{2010A&A...520A..49S} meaning
    that the dynamical lifetime $t_l$ of the filament is calculated as
    $t_l=l/(\Delta v_\mathrm{grad} \tan(57.3))$.
    The infall rate $\dot{M}$ is then $\dot{M} = M/t_l$
    with the mass $M$ obtained from the \her maps.
    For the mass estimate, we defined polygons
    around the filament skeletons identified in the C$^{18}$O
    map and approximately following the filaments seen on the column density map,
    with similar widths ($\sim25''\!$, or $\sim0.1$\,pc).
    These are shown in Figure~\ref{fig:filaments}.
		The background
		around the filaments is highly variable, ranging from $\sim10^{21}$\pcm
		over most of the region to over $4~\times~10^{22}$\pcm at the central hub,
		approximately 50\% of the column density values measured over the filaments.
		Since the majority of the mass in the filaments comes from the central hub,
		this suggests that a reasonable lower limit for the masses is
		50\% of the total measured value (which was used as the upper limit).
		As this is an estimate, no further background subtraction was performed.
		From this point, the upper limit value is used.
		
    We also estimated the average column density $N$,
    the average density $n$, the mass per unit length $M_l$, and the
    average dust temperature $T$ from the \her data.
    The critical mass per unit length $M_{l,\mathrm{crit}}$ above which
    filaments become gravitationally unstable \citep{2011A&A...529L...6A}
    was determined following \citet{1964ApJ...140.1056O}:
%critical mass per length 
  \begin{equation}
    M_{l,\mathrm{crit}}~=~\frac{2\,c^2_s}{G}~=~1.67~T~~[M\msun\,pc^{-1}].
    \label{Eqn:critmass}
  \end{equation}

    For the sound speed $c_s$ we assume isothermal gas
    (of temperature $T$) where the ideal gas
    equation of state holds and thus $c_s~=~\sqrt{kT/\mu}$ with
    the Boltzmann constant $k$ and the mean molecular weight per
    free gas particle $\mu$~=~2.33 (accounting for 10\% He and
    trace metals). 
    
    The physical properties of the filaments are reported in Table~\ref{table-fil}.
    The masses of the filaments range between 26 and 114\,M\msun per parsec,
    and are thus a factor of $\sim$10 less massive (though only
    a factor of 2 shorter) than the filaments linked to the DR\,21 ridge
    \citep[100--3700\mspp;][]{2010A&A...520A..49S,2012A&A...543L...3H}
    and a factor of $\sim$3 less massive than the Serpens South filament
    \citep[62--290\mspp;][]{2012A&A...548L...6H,2013ApJ...766..115K},
    but of a similar mass per unit length to many other regions'
    filaments \citep[$\sim$20--100\mspp, in regions as diverse as
    Aquila, Polaris, and IC~5146;][]{2014prpl.conf...27A}.
    
    The filaments fall onto central hub, which, with a mass of
    $\sim$1000 M\msun, is many times more massive
    than even the most massive of the filaments.
    The average temperature within the filaments ranges between 15 and 19.5\,K.
    The southern filaments F6, F7, and F8 are colder (around 15\,K)
    than the northern ones (around 19\,K).
    All upper limit masses for the filaments are supercritical
		($M_l > M_{l,\mathrm{crit}}$) and thus gravitationally unstable
		and only three of the eight filaments have subcritical lower-limit masses.

    Assuming that the full velocity gradient corresponds to infall,
    the dynamic lifetimes of the filaments are short,
    between 1.6 and 5.6~$\times$~10$^5$\,yr, and the mass
    infall rates range between 0.5 and 3.25~$\times$~10$^{-4}$\,M\msun/yr.
    The southern filaments have lower infall rates than the northern ones.
    If a constant average infall rate of 1.4~$\times$~10$^{-3}$\,M\msun/yr over time
    is assumed, it takes $\sim$0.7~$\times$~10$^6$\,yr to build up a central hub of
    1000 M\msun, the current mass of the central region.
		Using the lower-limit mass values, this value would be 
		approximately doubled, to $\sim$1.4~$\times~10^6$\,yr.
    This is of course a rough estimate and the timescale is probably lower because we
    consider only a few prominent filaments, mostly oriented in the plane
    of the sky, and ignore all head-on ones and fainter filaments.
    Nevertheless, a value of about 10$^6$\,yr 
    indicates that the formation of ridges and hubs starts already at a very
    early phase during molecular cloud formation.
    This is consistent with a scenario in which these massive regions are
    formed out of initially atomic flows that quickly transform into filaments, which
    then supply the mass by means of gravitational contraction onto the hub
    \citep{2013ApJ...769..115H,2013ApJ...776...62H}.

  \begin{table*}[htbp] 
%   \centering 
    \begin{tabular}{|l|c|c|c|c|c|c|c|c|c|c|} 
      \hline
      \rule{0pt}{3ex}$\!\!$
      Fila-& Mass   & $\langle N \rangle$  &$\langle n \rangle$&$\langle T \rangle$&$l$ &$M_l$      &$M_{l,\mathrm{crit}}$ &$\Delta v_\mathrm{grad}$&$t_l$&$\dot M$\\ 
      ment &(M\msun)&(10$^{21}$\,cm$^{-2}$)&(10$^4$\,cm$^{-3}$)& (K)      &(pc)&(M\msun/pc)&(M\msun/pc)&(km\,s$^{-1}$)&(10$^5$yr)&(10$^{-4}$\,M\msun/yr) \\ 
      \hline
      \rule{0pt}{3ex}$\!\!$
      F1 &  24--48  & 22                 & 2.7              &  18\,(1.9) & 0.77 &  31--62 & 31 & 2.1 & 2.3 & 2.1 \\
      F2 &  37--73  & 21                 & 1.7              &  19\,(2.9) & 1.6  &  24--47 & 31 & 2.6 & 3.7 & 2.0 \\
      F3 &  33--66  & 29                 & 3.4              &  18\,(1.6) & 0.58 & 55--110 & 30 & 1.4 & 2.6 & 2.5 \\
      F4 &  26--52  & 44                 & 5.1              &  20\,(1.3) & 0.55 &  48--95 & 33 & 2.2 & 1.6 & 3.3 \\
      F5 &  16--32  & 17                 & 2.8              &  19\,(0.6) & 0.78 &  21--41 & 31 & 1.3 & 3.8 & 0.8 \\
      F6 &  24--48  & 23                 & 2.9              &  16\,(1.0) & 0.75 &  32--64 & 26 & 1.5 & 3.1 & 1.6 \\
      F7 &  16--31  & 12                 & 1.3              &  15\,(0.4) & 1.2  &  13--26 & 25 & 1.3 & 5.6 & 0.5 \\
      F8 &  29--58  & 16                 & 1.3              &  16\,(0.7) & 1.1  &  27--53 & 27 & 1.7 & 4.0 & 1.4 \\
      \hline 
    \end{tabular} 
    \caption{Properties of the filaments F1--F8:
      Total mass within the polygon defining the filament from \her H$_2$
        column density map (both lower and upper limit are given);
      $\langle N \rangle$, mean column density from \her;
      $\langle n \rangle$, mean H$_2$ density derived from $N/r$ with the beam deconvolved
        equivalent radius $r=\sqrt{(\mathrm{area})/\pi}$;
      $\langle T \rangle$, mean temperature;
      $l$, projected length;
      $M_l$, mass/length value from $M/l$ given in columns 2 and 6
				(both lower and upper limit are given);
      $M_{l,\mathrm{crit}}$, critical mass/length derived from Equation~\ref{Eqn:critmass};
      $\Delta v_\mathrm{grad}$, velocity gradient of the filament estimated from the C$^{18}$O
        channel map, relative to the observer;
      $t_l$, dynamical lifetime;
      $\dot M$, mass infall rate.}
    \label{table-fil} 
  \end{table*} 

  \section{Source detection and classification}
    \label{sec:Sou}
    The source detection and classification process
    was carried out according to the standard procedures
    of the HOBYS group, as outlined in \citet{2017A&A...602A..77T}.
    To identify the positions of compact objects,
    we used the \getso routine
    \citetext{version~\getsover; \mbox{\citealp{2012A&A...542A..81M}},
    \mbox{\citealp{2013A&A...560A..63M}}}
    which uses maps at all \her wavelengths simultaneously.
    An account of the workings of this routine is given in Appendix~\ref{app:getso}.
    The observed maps input to the routine were the five \her maps
    (70\mum, 160\mum, 250\mum, 350\mum, and 500\mum) and
    two ancillary maps (MIPS 24\mum and SCUBA-2 850\mum).
    Three derived maps are also used as input to \getso:
    the high-resolution column density map ($\tilde{\Sigma}$),
    and two others, versions of the
    160\mum and 250\mum maps corrected for the effects of temperature.
    These are created by using the colour-temperature map, $T_{250}$,
    to remove the effects of temperature from the 160\mum and 250\mum maps
    (these maps are used for detection in place of the
    observed 160\mum and 250\mum maps,
    but measurements are only taken from the original maps).
    We note that neither the 24\mum nor the 850\mum maps were used for detection,
    since the former contains many sources that are not seen at longer wavelengths,
    which could add unwanted MIR sources to the output catalogue,
    and the latter is noisy due to atmospheric effects,
    which could cause errors in position measurement.

    To test the completeness of \getso in Mon~R2, we injected additional
    synthetic sources into the maps, and the routine
    was run again on these source-injected maps.
    The \getso extraction was performed identically
    to the extraction described above, even including the column density and
    temperature-corrected maps.
    This test suggested that $\sim$70\% of sources over 1\,M\msun
    would have been detected.
    Five further extractions were performed on the central hub alone;
    in each extraction only a small number of sources was added
    so as not to increase the crowding in the region too much.
    This test suggested that, within the central hub,
    only $\sim$33\% of sources over 1\,M\msun would have been detected.
    More details of the completeness test are given in Appendix~\ref{app:comp}.

    In the Mon~R2 region, \getso detected \cfull sources,
    but even though these were all detected to a 
    (detection) signal-to-noise level of 5 $\sigma$,
    only the most robust sources (as described below,
    and in \citealp{2017A&A...602A..77T})
    are counted in the final catalogue.
    First, sources were removed from the \getso catalogue
    if they were at the edge of the map. 
    In addition, each wavelength detection for each source was designated ``reliable''
    if its peak intensity (measurement) signal-to-noise ratio
    and its total flux density (measurement) signal-to-noise ratio
    were both over two, and its axis ratio was under two.
    Sources were removed from the catalogue if they did not have reliable measurements
    at either 160\mum or 250\mum. 
    In order to find a single value for the ``size'' of the sources,
    the geometric mean of the measured major and minor FWHM sizes
    was found for each wavelength.
    These values were then deconvolved with the wavelengths' beam sizes,
    with a minimum value set to half the beam size.
    The reference size ($\Theta$) was defined as the smaller of the deconvolved
    sizes at 160\mum and 250\mum, in the case where both wavelengths had reliable detections,
    or the size at the reliable wavelength (out of 160\mum and 250\mum),
    in the case where only one of the two was reliable.
    
    Sources were treated as ``robust'' if they had reliable detections
    at three wavelengths over 100\mum
    (one of which, as mentioned above, had to be at either 160\mum or 250\mum).
    Finally, sources were removed due to poor SED fits, and
    due to being likely extraction artefacts
    (both described in more detail below), leaving \cgood ``robust'' sources.
    For the remainder of the paper, only these robust sources are considered.
    
    \begin{figure*}[htb]
      \mbox{\resizebox{\hsize}{!}{\includegraphics{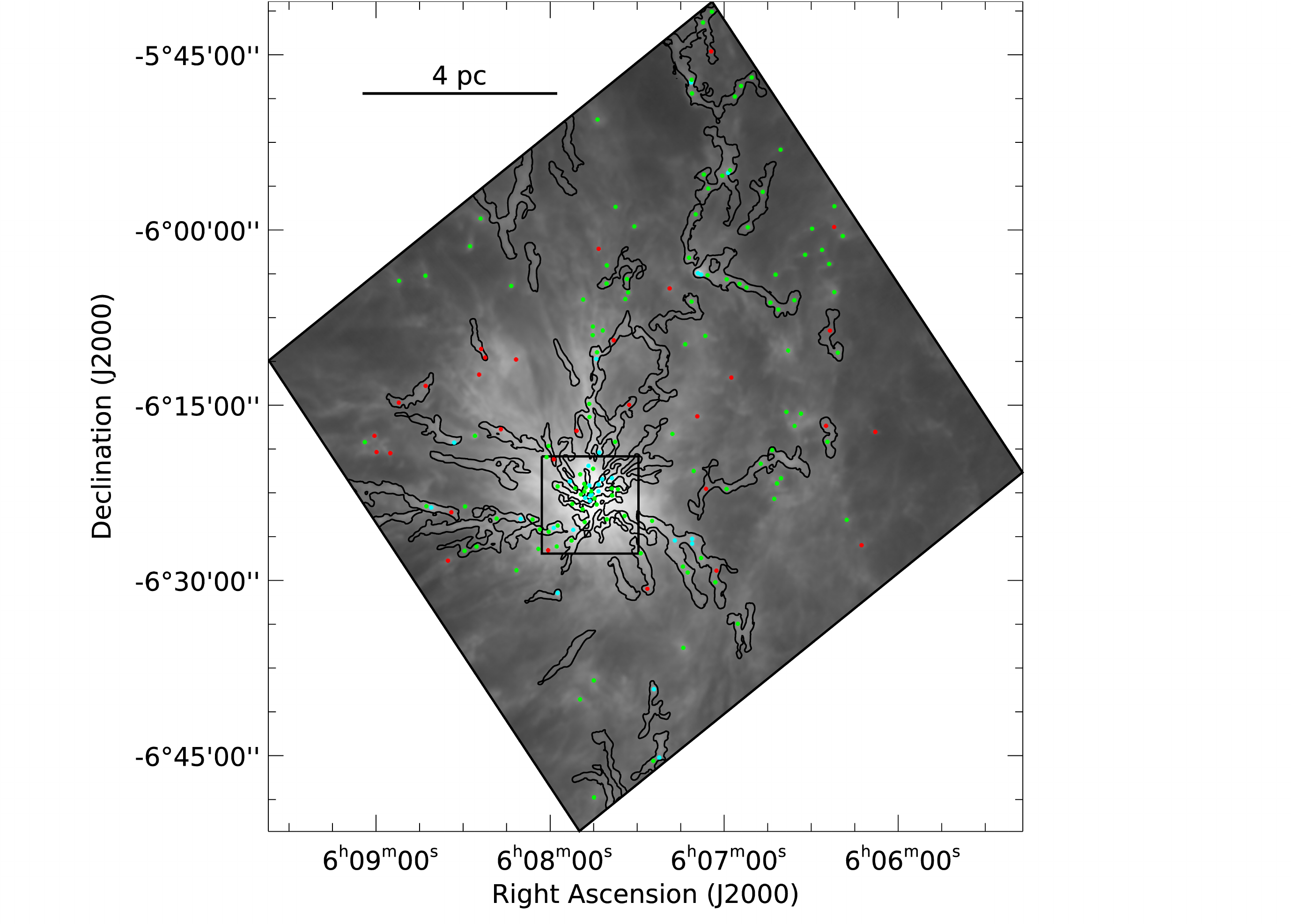}}}
      \caption{Monoceros R2 sources, on the 250\mum flux map (Figure~\ref{fig:app_MonR2_250}).
      The \textit{getfilaments} map is overlaid as black contours
      (strictly, these are the filaments detected on the column density
      map at scales below 72\asec).
      The contours are those from Figure~\ref{fig:dens} (H$_2$ column densities
      3\expo{21}\pcm and 1.5\expo{22}\pcm).
      A zoom-in of the region in the red box (the central hub) is shown in Figure~\ref{fig:hubsources}.
      The points are protostars (blue), bound cores (green), and unbound clumps (red);
      see text for details.
      }
      \label{fig:sources}
    \end{figure*}
    
    From the measured flux densities, the spectral energy distribution
    (SED) of each robust source can be constructed.
    The SED is assumed to fit a greybody with the following form \citep{2008ApJ...681..428C}:

    \begin{equation}
      F_\nu = M_\mathrm{c} \frac{\kappa_0}{R D^2} \left(\frac{\nu}{\nu_0}\right)^\beta B_\nu(T),
      \label{equ:sed}
    \end{equation}

    \noindent where
    $F_\nu$ is the observed flux density (here and elsewhere in this paper)
    at frequency $\nu$;
    $M_\mathrm{c}$ is the source gas mass;
    $\kappa_0$ is the dust absorption coefficient, measured at a reference frequency $\nu_0$;
    $R$ is the assumed mass ratio of gas to dust in the cloud;
    $\beta$ is the dust emissivity index;
    and $B_\nu(T)$ is the Planck function, with temperature $T$.
    Although $\beta$, $R$ and $\kappa_0$ are subject to significant uncertainties, there exist common
    approximations based on measurements made in the literature.
    It has been suggested that the emissivity index, $\beta$, varies between 1 for higher frequencies
    (wavelengths below 250\mum) and 2 for lower frequencies
    \citep[above 250\mum;][]{1983QJRAS..24..267H};
    a value of $\beta = 2$ is generally used in \her
    SAG3\footnote{A SPIRE Consortium Specialist Astronomy Group which implemented
    the HOBYS and \her Gould Belt surveys.} papers
    \citep[including][]{2010A&A...518L..77M,2010A&A...518L..84H,2010A&A...518L.106K,2012A&A...539A.156G}.
    The gas-to-dust ratio is generally taken to be $R = 100$ \citep{1990AJ.....99..924B}, or in
    other words, the dust is assumed to make up about 1\% of the cloud by mass.
    The absorption coefficient, $\kappa_0$, is set to 10\,cm$^2$\,g$^{-1}$ (or 0.1\,cm$^2$\,g$^{-1}$
    when divided by the gas-to-dust ratio) at a reference frequency of 1\,THz
    (or a wavelength of 300\mum);
    this has been found to have an accuracy of better than 50\% over column densities
    between 3\expo{21}\pcm and $10^{23}$\pcm \citep{2014A&A...562A.138R}.
    The distance to Mon~R2, 830\,pc, is discussed in Section~\ref{sec:dist}.
    Based on these assumptions, the source's mass and temperature can be
    found from the best SED fit (in this case, using the MPFIT routine;
    \citealt{2009ASPC..411..251M}).
    
    \begin{figure}[htb]
      \mbox{\resizebox{\hsize}{!}{\includegraphics[trim=0 0 120 0, clip=true]{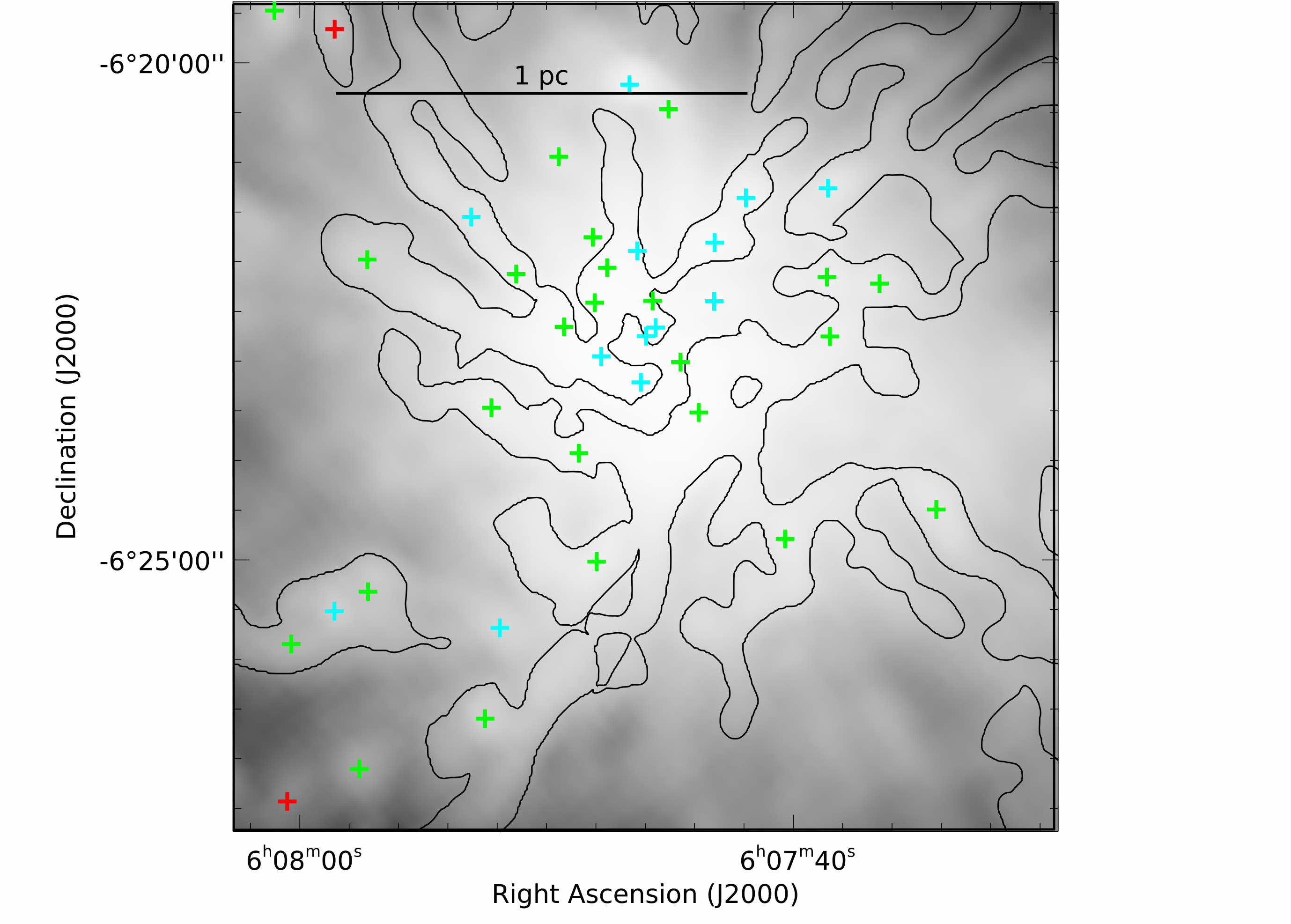}}}
      \caption{Zoom in on the central hub region of Figure~\ref{fig:sources}.
      The contour shows the filaments as detected by \getso on the column density map.
      }
      \label{fig:hubsources}
    \end{figure}

    Before the SED can be fit, there is another correction that needs to be made.
    The total fluxes for wavelengths of 160\mum or greater are scaled for the source size
    (since the fluxes were measured over variable apertures), so that:

\begin{equation}
  F_{\nu\mathrm{sca}} = F_\nu \left(\frac{\Theta}{\Theta_\nu}\right),
\end{equation}

    \noindent where $\Theta_\nu$ is the deconvolved FWHM at frequency $\nu$.
    This flux scaling process is introduced in \citet{2010A&A...518L..77M} and
    explained in more detail in \citet{2011A&A...535A..76N}.
    It should be noted that the flux scaling is merely
    an estimate, and does not account for more complex internal core structure
    (such as subfragmentation or density flattening).
    It is also a practical/empirical approach, and while it does generally give
    better-fitting SEDs, it is likely to increase the uncertainty of the values.
    In order to provide a test for flux scaling, SEDs were also constructed
    using flux measured by aperture photometry on maps convolved to the
    500\mum resolution (\lbeam\asec).
    The RMS of the relative difference between the mass calculated by this method
    and the mass calculated using flux scaling was found to be 1.17
    (strictly, the RMS of $(M-M_\mathrm{conv})/M$, where $M$ is the
    mass calculated by flux scaling, and $M_\mathrm{conv}$ is the mass
    calculated using the convolved fluxes;
    the fluxes used and masses measured are provided in the catalogue files).
    This suggests that an error of $\sim20$\% should be
    applied to the masses, in addition to the mass errors calculated
    purely from the fitting routine (the error due to the fit alone
    is given in the catalogue as $M_\mathrm{err\_fit}$),
		and the 15\% systematic uncertainty from the distance measurements.
    We note that colour corrections were not performed, as any correction applied
    would be smaller than the uncertainties likely introduced by flux scaling.

    For the SED fitting, all reliable detections (as defined earlier)
    above 100\mum are included in the fit.
    The errors on each wavelength are the total flux error, as measured by \getso,
    summed in quadrature to the instrumental and calibration errors (10\% of the flux for all bands).
    Wavelengths that are not reliable due to the source shape
    are included as ``1 $\sigma$ upper-limits''$\!$, meaning that the error is set equal to the flux itself;
    meanwhile, wavelengths that are not reliable due to having flux below 2 $\sigma$ have the
    flux itself (along with the errors) set to the 2 $\sigma$ value.
    The 70\mum data are fit only to those sources that show a temperature
    of over 32\,K when fit without it.
    This is necessary, since such SEDs are poorly described
    by wavelengths of 160\mum and above alone.
    As mentioned above, fits (both including and excluding 70\mum) with reduced $\chi^2$ values above 10 are also excluded
    from the final catalogue, since the fits (and thus output parameters) provided are dubious
    (only about 10\% of the total number removed were cut due to this requirement alone).
    It should be noted that the 24\mum flux is only used to calculate the luminosity,
    and takes no part in the SED determination.

    In addition to the \getso fluxes, WISE all-sky catalogue data
    are used to provide mid-infrared coverage for luminosity calculation.
    All WISE sources within 6\asec of the centre of the detected
    \her source are assumed to contribute.
    The mid-infrared luminosity is derived by calculating the integral of the WISE
    fluxes, the 24\mum flux and the 70\mum flux.
    The far-infrared luminosity is the integral of the SED between 70\mum and 1200\mum.

    Sources are classified as protostars if they have both a reliable detection at 70\mum
    and either a reliable detection at 24\mum, or are present in the WISE catalogue
    (the second requirement was lifted for sources in the central region, as the 24\mum MIPS
    data and two of the four WISE bands are saturated here).
    In addition, the source FWHM at 70\mum is required to be under 11.6\asec (twice the 70\mum beam size).
    A non-protostellar (or starless) source is classified as gravitationally bound if its
    reference size ($\Theta$) is less than twice its Bonnor radius
    (the radius of a critically dense Bonnor-Ebert sphere with the same mass and temperature), which is given by:

\begin{equation}
  R_\mathrm{B} = \frac{G M_\mathrm{c} m_\mathrm{H_2}}{2.4 k_\mathrm{B} T},
  \label{equ:brad}
\end{equation}

    \noindent where $G$ and $k_\mathrm{B}$ are the gravitational and Boltzmann constants
    and $m_\mathrm{H_2}$ is the average molecular mass \citep{1956MNRAS.116..351B}.
    All other robust sources were classified as either gravitationally unbound
    or ``undefined cloud structures'' \citep{2017A&A...602A..77T}, a designation for objects detected by \getso,
    but not associated with either a protostar or a peak in column density.
    These are potentially artefacts of the extraction, especially given that they
    are generally associated with the most crowded parts of the region.
    Consequently, these objects are not included in the catalogue or the analysis.
    
    The positions of these sources are shown in Figure~\ref{fig:sources},
    overlaid on a filament map taken from \getso
    (strictly, this is the map of filaments detected at scales under
    72\as, or 0.3\,pc, on the column density map).
    As can be seen, the majority (60\%, or \nfila out of \cgood)
    of sources are coincident with the \getso filaments.
    In addition, the sources are mainly clustered around the central hub
    (as shown in Figure~\ref{fig:hubsources}), with \ncent robust
    sources detected there (80\%, or \nfilc of them, on \getso filaments).
    
    The basic parameters of the overall source dataset (temperature, mass, bolometric luminosity,
    and reference size) are given in Table~\ref{tab:overall}.
    We note that not all of the bound cores are likely to be true prestellar cores;
    although such objects could be detected separately at this distance
    (the 160\mum resolution is $\sim$0.05\,pc at Mon~R2,
    while prestellar cores have sizes of 0.1--0.2\,pc; \mbox{\citealp{2014A&A...562A.138R}}),
    the larger bound cores are potentially clusters of several prestellar cores.
    These are likely to be on the path to star formation.

\begin{table*}[htb]
  \hskip-25px
  \begin{tabu}{| c | c | c | c | c | c | c | c | c | c | c | c | c | c |}
    \hline
    \rule{0pt}{3ex}$\!\!$
    \multirow{2}{*}{Sources} & \multirow{2}{*}{N} & 
		$\langle\mathrm{T}_\mathrm{d}\rangle$ & \multirow{2}{*}{$\mu$} &
    \multirow{2}{*}{$\sigma$} & Mass & \multirow{2}{*}{$\mu$} & 
		\multirow{2}{*}{$\sigma$} & L$_\mathrm{bol}$ & \multirow{2}{*}{$\mu$} &
		\multirow{2}{*}{$\sigma$} & R$_\mathrm{dec}$ & \multirow{2}{*}{$\mu$} &
		\multirow{2}{*}{$\sigma$} \\
		& & (K) & & & (M\msun) & & & (L\msun) & & & (pc) & & \\
    \hline 
    \rule{0pt}{3ex}$\!\!$
    Robust         & 177 (440) & 9.0--39 & 15 & 5.4 & 0.084--24  & 3.1  & 3.7  & 0.058--5000 & 67   & 430  & 0.023--0.30  & 0.074 & 0.052  \\
    (Hub)          & 29  (145) & 12--39  & 22 & 8.0 & 1.5--24    & 6.9  & 5.9  & 3.5--5000   & 400  & 1000 & 0.023--0.099 & 0.033 & 0.015  \\
    Protostellar   & 28  (50)  & 13--39  & 21 & 7.5 & 0.36--16   & 2.2  & 3.1  & 0.42--5000  & 350  & 1000 & 0.023--0.038 & 0.026 & 0.0051 \\
    (Hub)          & 11  (35)  & 17--39  & 27 & 7.5 & 1.5--16    & 4.3  & 4.3  & 6--5000     & 890  & 1600 & 0.023--0.035 & 0.026 & 0.0049 \\
    Bound          & 118 (200) & 9.0--33 & 13 & 4.0 & 0.46--24   & 3.9  & 3.9  & 0.058--520  & 16   & 66   & 0.023--0.30  & 0.085 & 0.055  \\
    (Hub)          & 18  (80)  & 12--33  & 19 & 6.7 & 1.7--24    & 8.5  & 6.3  & 3.5--520    & 99   & 150  & 0.023--0.099 & 0.037 & 0.018  \\
    Unbound        & 31  (390) & 11--25  & 16 & 3.1 & 0.084--1.4 & 0.56 & 0.33 & 0.13--4.6   & 0.99 & 1.0  & 0.036--0.18  & 0.080 & 0.037  \\
    M$>$10\,M\msun & 11        & 9.8--32 & 15 & 6.3 & 11--24     & 14   & 3.8  & 0.92--5000  & 470  & 1500 & 0.029--0.26  & 0.077 & 0.069  \\
    (Hub)          & 8         & 12--32  & 16 & 6.8 & 11--24     & 15   & 4.0  & 3.5--5000   & 650  & 1800 & 0.029--0.099 & 0.047 & 0.022  \\
    \hline
  \end{tabu}
  \caption{The numbers (N) and parameter ranges for all robust sources.
    The table also gives the numbers for the three individual types of source
		(described in the text),
    both in total, and within the central hub (those rows indicated by ``(Hub)'').
		The number in parentheses is an estimate of the true number of sources,
		based on the completeness calculations (Appendix~\ref{app:comp}).
    The lowest two rows give the ranges for sources with masses over 10\,M\msun;
		due to the small number of such model sources in the completeness calculations,
		no estimate for the true number is given.
    The parameter range mean values ($\mu$) and standard deviations ($\sigma$) are also given.
    No unbound sources were detected in the central hub.}
  \label{tab:overall}
\end{table*}

    As can be seen from Table~\ref{tab:overall}, those sources in the central hub
    (defined here by the N-PDF excess from \citealt{2015MNRAS.453L..41S},
    which is equivalent to column densities above $3.5\expo{22}$\pcm)
    are both hotter and more luminous than those outside.
    There appears to be a difference in the mass ranges, too,
    with the more massive sources found in the hub.
    This effect is partly due to the high completeness limit
    of \getso in the hub region (see Appendix~\ref{app:comp}),
    which could only reliably detect a third of sources above 1\,M\msun
    due to the high crowding and confusion.

    One single source (HOBYS J060740.3 $-$062447) has a mass above 
    20\,M\msun, and is thus potentially a true high-mass star in formation.
    At the edge of the hub region, it is associated with the young star
    2MASS J06074062$-$0624410, and appears to overlap with the western \hii region
    (although this may be a projection effect).
    A further eleven sources have masses over 10\,M\msun, meaning that
    they are potentially intermediate-mass stars (final mass $\gtrsim5$\,M\msun) in formation.
    Four of these sources are larger objects in the outer parts of the region.
    While they are all massive enough to be gravitationally bound,
    none are particularly dense, with densities of 2.2\expo{3}--6.0\expo{4}\,cm$^{-3}$
    (for comparison, the density of HOBYS J060740.3 $-$062447 is 9.0\expo{5}\,cm$^{-3}$).
    The remaining seven sources are (like HOBYS J060740.3 $-$062447)
    within (or close to) the $\gtrsim3.5\expo{22}$\pcm column density hub,
    and include HOBYS J060746.1 $-$062312, which is associated with Mon~R2 IRS~1.
    These objects are mainly bound dense cores, with densities of
    6.2\expo{4}--2.1\expo{6}\,cm$^{-3}$.
    The properties of these sources are given in the lowest rows of Table~\ref{tab:overall},
    and the individual properties are given in the tables in Appendix~\ref{app:Table}
    (measured properties) and Table~\ref{tab:deriv} (derived properties).
    
\begin{table*}[htb]
\centering
\begin{tabu}{| r | c | c | c | c | c | c | c | c |}
\hline
\rule{0pt}{3ex}$\!\!$
\# & $T \pm \sigma_T$ & $M \pm \sigma_M$ & $L_\mathrm{bol}$
  & $R_\mathrm{dec}$ & $\langle n_{\mathrm{H}_2}\rangle$
  & \multirow{2}{*}{Core Type} & \multirow{2}{*}{SIMBAD name} \\ 
 & (K) & (M\msun) & (L\msun) & (pc) & ($\!$\expo{3}\,cm$^{-3}$) & & \\
\hline
\rule{0pt}{3ex}$\!\!$
 1 & 12.0\,$\pm$\,0.8 & 24\,$\pm$\,9  & 6.8  & 0.045 &  7100 & Bound dense core* & 2MASS J$06074062-0624410^\mathrm{a}$    \\ % HOBYS J060740.3 -062447
 2 &   13\,$\pm$\,2   & 17\,$\pm$\,10 & 8.1  & 0.045 &  5200 & Bound dense core  & 2MASS J$06073823-0622411^\mathrm{a}$    \\ % HOBYS J060738.4 -062244
 3 &   32\,$\pm$\,1   & 16\,$\pm$\,4  & 5000 & 0.029 & 18000 & Protostar         & Mon R2 IRS 1$^\mathrm{b}$               \\ % HOBYS J060746.1 -062312
 4 &   14\,$\pm$\,1   & 15\,$\pm$\,6  & 9.8  & 0.036 &  8500 & Bound dense core  & \\ % HOBYS J060751.2 -062206
 5 &   14\,$\pm$\,1   & 14\,$\pm$\,6  & 11   & 0.037 &  8000 & Bound dense core* & \\ % HOBYS J060752.2 -062327
 6 &   20\,$\pm$\,2   & 13\,$\pm$\,5  & 100  & 0.035 &  8600 & Bound dense core* & \\ % HOBYS J060747.5 -062203
 7 &   12\,$\pm$\,1   & 13\,$\pm$\,5  & 3.8  & 0.090 &   480 & Bound dense core  & Mon R2 4$^\mathrm{c}$                   \\ % HOBYS J060743.6 -061028
 8 &   12\,$\pm$\,2   & 12\,$\pm$\,8  & 3.5  & 0.099 &   340 & Bound dense core* & 2MASS J$06075740-0622103^\mathrm{a}$    \\ % HOBYS J060757.2 -062158
 9 & 10.3\,$\pm$\,0.8 & 12\,$\pm$\,4  & 1.3  & 0.265 &    17 & Bound dense core  & \\ % HOBYS J060706.3 -060904
10 &   15\,$\pm$\,1   & 11\,$\pm$\,4  & 12   & 0.047 &  2800 & Bound dense core  & JCMTSF J$060747.9-062502^\mathrm{d}$    \\ % HOBYS J060747.9 -062501
11 &  9.8\,$\pm$\,0.5 & 11\,$\pm$\,4  & 0.92 & 0.115 &   190 & Bound dense core  & \\ % HOBYS J060621.9 -060519
\hline
\end{tabu}
\caption{
  Derived core properties:
  temperature, $T$, and source mass, $M$, both with errors, $\sigma$,
  bolometric luminosity, $L$, reference size ($R_\mathrm{dec}$),
  and mean density, $\langle n_{\mathrm{H}_2}\rangle$.
  Also, the table has the most likely core type for the object, whether protostellar,
  bound starless, or unbound (see Section~\ref{sec:Sou}),
  and a potential identity, from comparisons with other observations of the region.
	We note that the mass errors given here are solely from the SED fit,
	and do not include the 25\% errors from flux-scaling and distance estimates.
  Source references:
  a: \citet{2003yCat.2246....0C};
  b: \citet{1976ApJ...208..390B};
  c: \citet{2009ApJS..184...18G};
  d: \citet{2008ApJS..175..277D}.
  The full catalogue also references
  \citet{1997ApJ...474..329T},
  \citet{2007AJ....134.2020H}, and
  \citet{1998AJ....115.1693C}.
}
\label{tab:deriv}
\end{table*}

    Looking at total masses, we can see that the sources in the central hub
    (total mass $\sim$220\,M\msun) make up about half of the mass of sources in the region
    (total mass $\sim$580\,M\msun), even though they account for less than a quarter by number.
		Taking completeness into account, we find an even greater discrepancy,
		with potentially 620\,M\msun in central hub sources,
		two thirds the completeness-corrected mass of all sources in the region (980\,M\msun).
    From the column density map, we can see that the total masses of material in the
    regions are approximately 2200\,M\msun (central) and 30\,000\,M\msun (total),
    meaning that 10--30\% of the central region is associated
    with star-forming cores, compared to only 2--4\% of the total region.
    Such behaviour has been seen in other high density regions,
    including the W43 ridge \citep{2014A&A...570A..15L}.
    
  \FloatBarrier
  \section{Source properties in and outside the central hub}
    \label{sec:Dis}
\begin{figure}[htb]
      \mbox{\resizebox{\hsize}{!}{\includegraphics[trim=0 0 0 0, clip=true]{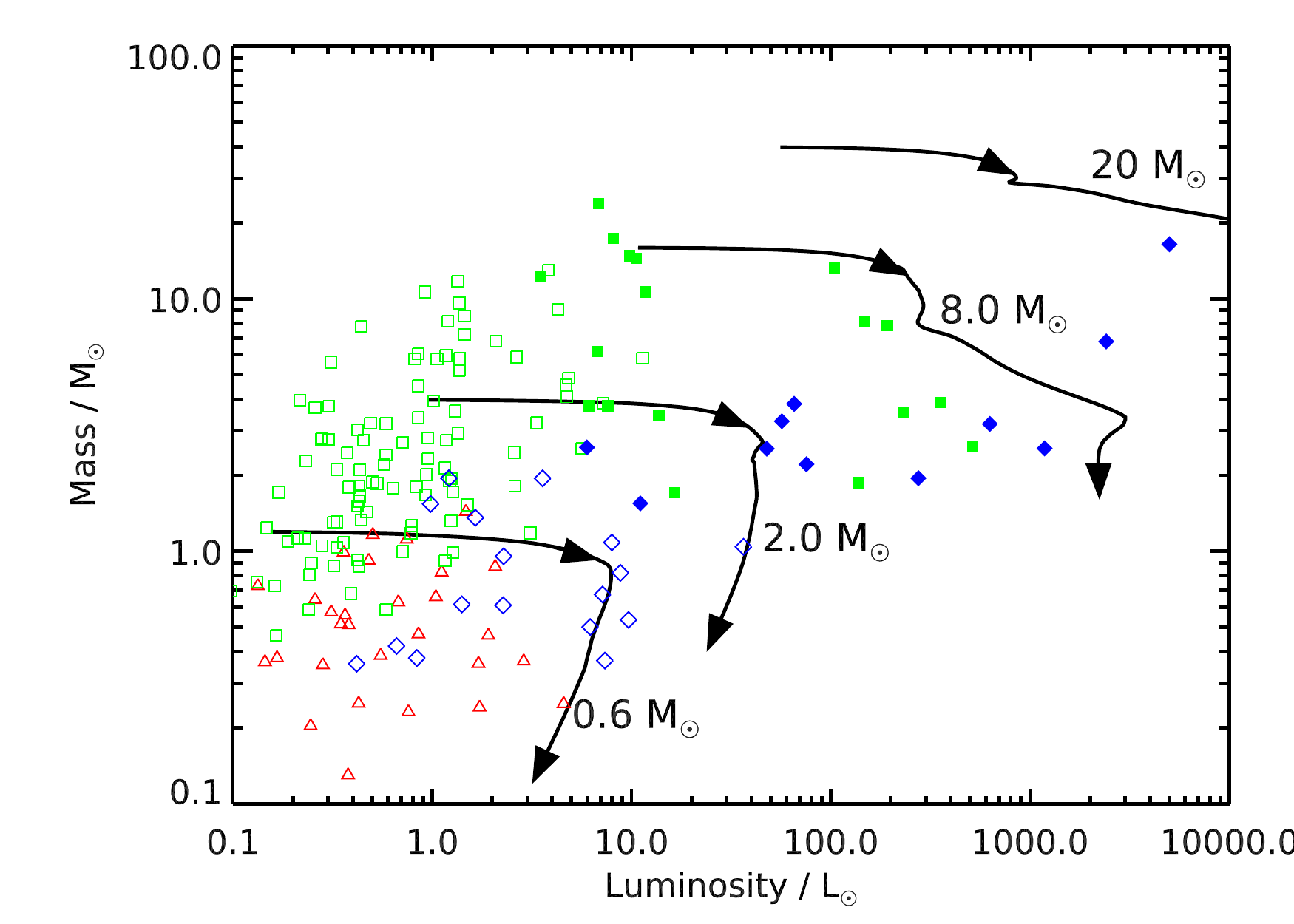}}}
      \caption{Bolometric luminosity-mass plot for robust sources in Mon~R2.
      The coloured symbols represent protostars (blue diamonds), bound cores (green squares),
      and unbound clumps (red triangles). 
      Filled shapes are sources within the central hub.
      The black lines represent evolutionary tracks for
      protostars with final protostellar masses of 0.6\,M\msun, 2\,M\msun,
      8\,M\msun, and 20\,M\msun (described in Section~\ref{sec:Dis}).}
      \label{fig:lm}
    \end{figure}

\begin{figure}[htb]
      \mbox{\resizebox{\hsize}{!}{\includegraphics[trim=10 0 0 0, clip=true]{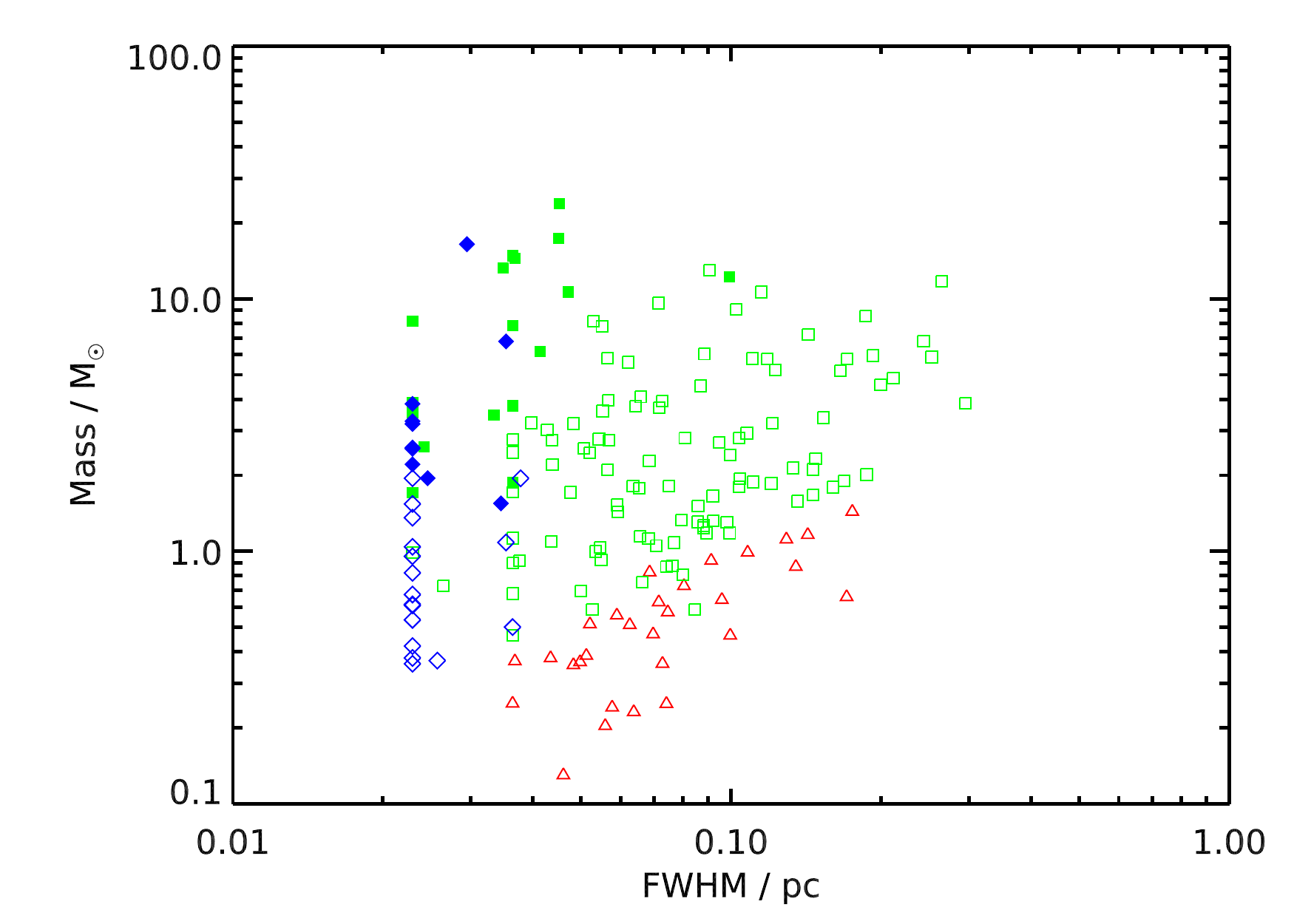}}}
      \caption{Mon~R2 plot of mass against reference size
      (full width at half maximum at either 160\mum or 250\mum;
      defined in Section~\ref{sec:Sou}).
      As for Figure~\ref{fig:lm}.
      The two lines of protostellar (blue) sources are at half the
      160\mum and 250\mum beam sizes (5.7\asec and 9.1\as, respectively,
      corresponding to sizes 0.023\,pc and 0.037\,pc),
      which were taken as the minimum source sizes
      (see Section~\ref{sec:Sou} for more details).
      }
      \label{fig:mr}
    \end{figure}

    Contrasting the parameters of the sources inside the
    central hub with those outside it, a difference can be seen.
    Plots of bolometric luminosity and reference size against mass
    are shown in Figures~\ref{fig:lm} and \ref{fig:mr}, respectively,
    and they show that the central hub sources (filled shapes) are generally smaller,
    more luminous, and potentially even more massive than those outside the hub,
    occupying distinct locations on each plot
    (masses over $\sim$2\,M\msun; luminosities over $\sim$10\,L\msun;
    reference size under $\sim$0.04\,pc).
    The luminosity-mass plot shows evolutionary tracks for four
    protostars of final protostellar mass ($M$) 0.6\,M\msun, 2\,M\msun, 8\,M\msun, and 20\,M\msun
    \citep[from][ Duarte-Cabral, private communication]{2013A&A...558A.125D},
    in which $L_\mathrm{bol} = L_* + G \epsilon M M_* / \tau R_*$,
    where: $L_*$, $M_*$, and $R_*$ are the luminosity, mass, and radius of the protostar itself;
    $G$ is the gravitational constant;
    $\epsilon$ is the efficiency for an individual core's formation
    (the fraction of the core that eventually joins the protostar), set to 50\%;
    and $\tau$ is the characteristic timescale for protostellar evolution, set to $10^5$\,yr
    \citep{2008A&A...490L..27A,2017A&A...602A..77T}.
    
    The objects from the central hub seem to be segregated from those in other parts of the region,
    with the inner objects generally located above the 2\,M\msun evolutionary track,
    and the majority of outer objects located below the track.
    In addition, those outer objects above 2\,M\msun appear to show a segregation based on
    evolutionary level, with the inner objects being exclusively located at the
    start of the tracks.
    This suggests that these objects form a secondary population within
    the region, one that potentially began evolution significantly earlier than those
    outside the hub.
    This is in contrast to what was seen in Cygnus~X \citep{2013A&A...558A.125D},
    in which the massive objects are far less evolved than the lower-mass objects,
    suggesting that for the Cygnus~X region, at least, the massive star formation
    occurring is comparatively more recent than in Mon~R2.
    The different types of objects found outside the central hub
    (protostellar, bound, or unbound) all occupy distinct positions
    in the two plots (albeit with some overlap);
    those inside show no such differentiation.
    Similarly, the size-mass positions show a distinction between the
    hub sources (small and massive) and the outer sources, which can have
    similar masses, but only with much lower densities.
		
    It should be noted that the tracks represent the evolution of objects heated only from within;
    the externally heated bound cores in the central hub are likely to be higher
    up the tracks than they would be in isolation.
		While this could affect the positions of such sources on Figure~\ref{fig:lm},
		it does not explain their small size and high mass, as seen in Figure~\ref{fig:mr}.
		This positioning can be partially explained by the high completeness limit of the central hub
    (see Appendix~\ref{app:comp}), as
    less massive, less luminous, and larger sources within the central hub
    will be missed simply due to the complexity of the region.
    The number of sources missed in the central hub (66\% above 1\,M\msun and 80\% above 0.1\,M\msun;
		likely over a hundred in total)
    could certainly explain the absence of central hub sources
    in the high-size, low-mass and low-luminosity parts of the plots.
    The poor completeness of the central hub does not, however,
    explain the absence of ``outer'' sources (sources from outside the hub
    in the low-size, high-mass and high-luminosity parts of the plots;
		indeed, the completeness tests suggest that 90\% of sources under 0.035\,pc and over 1\,M\msun
		have been detected in the outer regions.
		Only four such sources have been found outside the central hub ($\sim$3\%),
		while eighteen of the central hub sources ($\sim$62\%) fit these parameters.
		If the ratios were equivalent, then over 500 extra sources over 0.1\,M\msun would
		be needed in the central hub, rather than the 150 suggested by the completeness analysis.
    This at the very least suggests that the central hub of Mon~R2
    has an abundance of these objects when compared to the surroundings,
    and that the source populations are indeed different.
    Such crowding of high-mass protostars has also been observed in HOBYS ridges
    \citep{2011A&A...533A..94H,2011A&A...535A..76N,2014A&A...570A..15L}.

    These features suggest a region in which star formation initially begins
    at the meeting-point of a network of filaments (the hub),
    and commences in the outer regions at a later time.
    It is also possible that the star-formation occurring in the central hub is
    fuelled by material from the filaments themselves,
    which would in turn partially deplete star-forming material in the surrounding areas
    making it harder for stars to form in surrounding regions.
    It is possible that the surrounding \hii regions played
    a part in the formation of the central hub by providing external pressure
    to enhance the gravitational collapse, although it is
    unlikely to have played a major part \citep{2015A&A...584A...4D}.
    The ionising stars of these three regions are young B types
    \citep[BD$-$06 1415, B1; BD$-$06 1418, B2.5V; HD 42004, B1.5V;][]{1968AJ.....73..233R,2003AJ....125.2531R}
    that are too evolved to be detected by \her and \getso.
    Their probable masses \citep[$\sim$10\,M\msun for a B1--2 type star;][]{1981A&AS...46..193H}
    would thus mean that at least some intermediate-to-high mass star formation must have occurred
    in the region prior to the formation of the hub,
    but the environment at the time of their forming is impossible to determine.
    The three stars are all at most $\sim$2\,pc from the edge of the hub region
    (assuming minimal projection effects), which, given the age of the region
    ($\sim$5\expo{6}\,yr as determined here, which is of the same order of
    magnitude as that of \citealt{2015A&A...584A...4D}, $\sim10^6$\,yr),
    could allow the stars to have begun forming in the vicinity of the hub,
    before moving to their current positions at velocities of no more than 1\,km\,s$^{-1}$.
		Indeed, there is already a sizable population of young stars ($\sim$1\,Myr) already in existence at the
		Mon~R2 central hub, with at least 300 stars (and likely over 500) detected over the entire hub
		\citep{1997AJ....114..198C} and almost 200 detected in the central square arcminute
		(approximately coincident with the \hii region shown in Figure~\ref{fig:core};
		\citealp{2006AJ....132.2296A}).

    Mon~R2 is an unusual, but not necessarily unique, region;
    as mentioned earlier, both Cep~OB3 and NGC~6334
    show similar N-PDF excesses \citep{2015MNRAS.453L..41S},
    as do W3 \citep{2015arXiv150702055R},
    and NGC~2264 \citep{2015rayner.....phdR},
    the latter of which also showing a similar split
    in its source population, although without any
    obvious filamentary hub.
    While both W3 and NGC~6334 are more than twice as distant
    as Mon~R2, and thus their internal structure (both sources
    and filaments) is significantly harder to resolve,
    both Cep~OB3 and NGC~2264 are at similar distances to
    Mon~R2, and thus follow-up studies on these two regions
    could help add to the observations from Mon~R2.
    Other high-density ridges, such as those in
    W43 \citep{2014A&A...570A..15L} and Cygnus~X \citep{2012A&A...543L...3H}
    could also provide similar excesses,
    even though neither has been examined for the N-PDF excess.
    Cygnus~X could thus provide another region for follow-up studies,
    although as W43 is more than six times as distant as Mon~R2,
    it is unlikely to have sufficient resolution to rival Cep~OB3 or NGC~2264.
    
\section{Conclusions}
  \label{sec:Con}
    In Mon~R2, we detect \cgood robust sources,
    including \nprot protostars and \ngbnd bound cores.
    About a sixth of these by number (\ncent: \ncpro protostars
    and \ncbnd bound cores)
    and a third by mass (200\,M\msun, out of 540\,M\msun)
    are found in a filamentary hub structure at the centre of the region.
    These sources are also smaller and more luminous (on average),
    and thus likely more highly evolved,
    than the sources in the outer regions, and while this is partly
    attributable to poor completeness in the central hub,
    this cannot account entirely for the difference.

    In addition, the central hub was observed in C$^{18}$O,
    giving the kinematics of the system.
    Matter is observed to be moving along the filaments,
    with mass infall rates of $\sim2~\expo{-4}$\,M\msun/yr,
    indicating that the $\sim$1000\,M\msun central hub
    has likely been forming for $\sim5~\expo{6}$\,yr,
    a significant portion of the lifetime of the molecular cloud.

    This all comes together to suggest a model in which the hub forms
    very early in the life of the molecular cloud
    (although it itself may have been formed by young stars forming around it),
    being fed by infall from filaments around it.
    It then began star formation in advance of the rest
    of the region, fuelled by the increased densities there.
    In addition, it is possible that material flowing into the hub from the
    filaments could decrease the amount of star formation in surrounding regions.

  \begin{acknowledgements}
		%Referee
		We are grateful for the comments of an anonymous referee, which have helped us to improve this paper.
    %Herschel
    The \her spacecraft was designed, built, tested, and launched under a contract to ESA
    managed by the \her/\textit{Planck} Project team by an industrial consortium under the overall
    responsibility of the prime contractor Thales Alenia Space (Cannes), and including Astrium
    (Friedrichshafen) responsible for the payload module and for system testing at spacecraft
    level, Thales Alenia Space (Turin) responsible for the service module, and Astrium
    (Toulouse) responsible for the telescope, with in excess of a hundred subcontractors.
    %SPIRE
    SPIRE has been developed by a consortium of institutes led by Cardiff Univ. (UK) and
    including Univ. Lethbridge (Canada); NAOC (China); CEA, LAM (France); IFSI, Univ. Padua (Italy);
    IAC (Spain); Stockholm Observatory (Sweden); Imperial College London, RAL, UCL-MSSL, UKATC,
    Univ. Sussex (UK); Caltech, JPL, NHSC, Univ. Colorado (USA). This development has been supported
    by national funding agencies: CSA (Canada); NAOC (China); CEA, CNES, CNRS (France); ASI (Italy);
    MCINN (Spain); SNSB (Sweden); STFC and UKSA (UK); and NASA (USA).
    %PACS
    PACS has been developed by a consortium of institutes led by MPE (Germany) and
    including UVIE (Austria); KUL, CSL, IMEC (Belgium); CEA, OAMP (France); MPIA (Germany);
    IFSI, OAP/AOT, OAA/CAISMI, LENS, SISSA (Italy); IAC (Spain). This development has been supported
    by the funding agencies BMVIT (Austria), ESA-PRODEX (Belgium), CEA/CNES (France), DLR (Germany),
    ASI (Italy), and CICT/MCT (Spain).
    %HIPE
    HIPE is a joint development by the \her Science Ground Segment Consortium, consisting of ESA,
    the NASA Herschel Science Center, and the HIFI, PACS and SPIRE consortia.
    %Spitzer
    This work is based in part on observations made with the \spz Space Telescope,
    obtained from the NASA/IPAC Infrared Science Archive, both of which are operated
    by the Jet Propulsion Laboratory, California Institute of Technology under a contract with NASA.
    IRSA and the \spz Heritage Archive utilize technology developed for the Virtual Astronomical Observatory (VAO),
    funded by the National Science Foundation and NASA under Cooperative Agreement AST-0834235.
    %JCMT + SCUBA2
    The James Clerk Maxwell Telescope has historically been operated by the Joint Astronomy Centre on behalf
    of the Science and Technology Facilities Council of the United Kingdom,
    the National Research Council of Canada,
    and the Netherlands Organisation for Scientific Research.
    Additional funds for the construction of SCUBA-2 were provided by the Canada Foundation for Innovation.
    %SIMBAD
    This research has made use of the SIMBAD database, operated at CDS, Strasbourg, France.
    %WISE
    This publication makes use of data products from the Wide-field Infrared Survey Explorer,
    which is a joint project of the University of California, Los Angeles, and the Jet Propulsion
    Laboratory/California Institute of Technology, and NEOWISE, which is a project of the Jet Propulsion
    Laboratory/California Institute of Technology.
    WISE and NEOWISE are funded by the National Aeronautics and Space Administration.
    %Collaborator funding:
    N.\,S. acknowledges support by the ANR-11-BS56-010 project ``STARFICH''.
    Part of this work was supported by the French National Agency for Research (ANR) project ``PROBeS'',
    number ANR-08-BLAN-0241.
    N.\,S. acknowledges support by the DFG through project number Os 177/2-1 and 177/2-2
    and central funds of the program 1573 (ISM-SPP).
    GJW gratefully acknowledges support from the Leverhulme Trust.
    SPTM and AF thank the Spanish MINECO for funding support from grants AYA2012-32032, CSD2009-00038, FIS2012-32096, and ERC under
    ERC-2013-SyG, G. A. 610256 NANOCOSMOS.
    This work has received support from the ERC under the European Union's Seventh 
    Framework Programme (ERC Advanced Grant Agreements no. 291294 --  `ORISTARS') 
  \end{acknowledgements}

  \bibliographystyle{aa}
  \bibliography{aabib}

\begin{thebibliography}{117}
\expandafter\ifx\csname natexlab\endcsname\relax\def\natexlab#1{#1}\fi

\bibitem[{{Andersen} {et~al.}(2006){Andersen}, {Meyer}, {Oppenheimer},
  {Dougados}, \& {Carpenter}}]{2006AJ....132.2296A}
{Andersen}, M., {Meyer}, M.~R., {Oppenheimer}, B., {Dougados}, C., \&
  {Carpenter}, J. 2006, \aj, 132, 2296

\bibitem[{{Andr\'e} {et~al.}(2014){Andr\'e}, {Di Francesco}, {Ward-Thompson},
  {Inutsuka}, {Pudritz}, \& {Pineda}}]{2014prpl.conf...27A}
{Andr\'e}, P., {Di Francesco}, J., {Ward-Thompson}, D., {et~al.} 2014,
  Protostars and Planets VI, 27

\bibitem[{{Andr\'e} {et~al.}(2010){Andr\'e}, {Men'shchikov}, {Bontemps},
  {K\"onyves}, {Motte}, {Schneider}, {Didelon}, {Minier}, {Saraceno},
  {Ward-Thompson}, {Di Francesco}, {White}, {Molinari}, {Testi}, {Abergel},
  {Griffin}, {Henning}, {Royer}, {Mer\'in}, {Vavrek}, {Attard}, {Arzoumanian},
  {Wilson}, {Ade}, {Aussel}, {Baluteau}, {Benedettini}, {Bernard}, {Blommaert},
  {Cambr\'esy}, {Cox}, {Di Giorgio}, {Hargrave}, {Hennemann}, {Huang}, {Kirk},
  {Krause}, {Launhardt}, {Leeks}, {Le Pennec}, {Li}, {Martin}, {Maury},
  {Olofsson}, {Omont}, {Peretto}, {Pezzuto}, {Prusti}, {Roussel}, {Russeil},
  {Sauvage}, {Sibthorpe}, {Sicilia-Aguilar}, {Spinoglio}, {Waelkens},
  {Woodcraft}, \& A.}]{2010A&A...518L.102A}
{Andr\'e}, P., {Men'shchikov}, A., {Bontemps}, S., {et~al.} 2010, \aap, 518,
  L102

\bibitem[{{Andr\'e} {et~al.}(2008){Andr\'e}, {Minier}, {Gallais}, {Rev\'eret},
  {Le Pennec}, {Rodriguez}, {Boulade}, {Doumayrou}, {Dubreuil}, {Lortholary},
  {Martignac}, {Talvard}, {De Breuck}, {Hamon}, {Schneider}, {Bontemps},
  {Lagage}, {Pantin}, {Roussel}, {Miller}, {Purcell}, {Hill}, \&
  {Stutzki}}]{2008A&A...490L..27A}
{Andr\'e}, P., {Minier}, V., {Gallais}, P., {et~al.} 2008, \aap, 490, L27

\bibitem[{{Arzoumanian} {et~al.}(2011){Arzoumanian}, {Andr\'e}, {Didelon},
  {K\"onyves}, {Schneider}, {Men'shchikov}, {Sousbie}, {Zavagno}, {Bontemps},
  {Di Francesco}, {Griffin}, {Hennemann}, {Hill}, {Kirk}, {Martin}, {Minier},
  {Molinari}, {Motte}, {Peretto}, {Pezzuto}, {Spinoglio}, {Ward-Thompson},
  {White}, \& {Wilson}}]{2011A&A...529L...6A}
{Arzoumanian}, D., {Andr\'e}, P., {Didelon}, P., {et~al.} 2011, \aap, 529, L6

\bibitem[{{Bally} \& {Lada}(1983)}]{1983ApJ...265..824B}
{Bally}, J. \& {Lada}, C.~J. 1983, \apj, 265, 824

\bibitem[{{Beckwith} {et~al.}(1976){Beckwith}, {Evans}~II, {Becklin}, \&
  {Neugebauer}}]{1976ApJ...208..390B}
{Beckwith}, S., {Evans}~II, N.~J., {Becklin}, E.~E., \& {Neugebauer}, G. 1976,
  \apj, 208, 390

\bibitem[{{Beckwith} {et~al.}(1990){Beckwith}, {Sargent}, {Chini}, \&
  {G\"usten}}]{1990AJ.....99..924B}
{Beckwith}, S.~V.~W., {Sargent}, A.~I., {Chini}, R.~S., \& {G\"usten}, R. 1990,
  \aj, 99, 924

\bibitem[{{Bendo} {et~al.}(2013){Bendo}, {Griffin}, {Bock}, {Conversi},
  {Dowell}, {Lim}, {Lu}, {North}, {Papageorgiou}, {Pearson}, {Pohlen},
  {Polehampton}, {Schulz}, {Shupe}, {Sibthorpe}, {Spencer}, {Swinyard},
  {Valtchanov}, \& {Xu}}]{2013MNRAS.433.3062B}
{Bendo}, G.~J., {Griffin}, M.~J., {Bock}, J.~J., {et~al.} 2013, \mnras, 433,
  3062

\bibitem[{{Benedettini} {et~al.}(2015){Benedettini}, {Schisano}, {Pezzuto},
  {Elia}, {Andr{\'e}}, {K{\"o}nyves}, {Schneider}, {Tremblin}, {Arzoumanian},
  {Di Giorgio}, {Di Francesco}, {Hill}, {Molinari}, {Motte}, {Nguyen Luong},
  {Palmeirim}, {Rivera-Ingraham}, {Roy}, {Rygl}, {Spinoglio}, {Ward-Thompson},
  \& {White}}]{2015MNRAS.453.2036B}
{Benedettini}, M., {Schisano}, E., {Pezzuto}, S., {et~al.} 2015, \mnras, 453,
  2036

\bibitem[{{Bernard} {et~al.}(2010){Bernard}, {Paradis}, {Marshall}, {Montier},
  {Lagache}, {Paladini}, {Veneziani}, {Brunt}, {Mottram}, {Martin},
  {Ristorcelli}, {Noriega-Crespo}, {Compi\`egne}, {Flagey}, {Anderson},
  {Popescu}, {Tuffs}, {Reach}, {White}, {Benedettini}, {Calzoletti}, {Di
  Giorgio}, {Faustini}, {Juvela}, {Joblin}, {Joncas}, {Mivilles-Deschenes},
  {Olmi}, {Traficante}, {Piacentini}, {Zavagno}, \&
  {Molinari}}]{2010A&A...518L..88B}
{Bernard}, J.-P., {Paradis}, D., {Marshall}, D.~J., {et~al.} 2010, \aap, 518,
  L88

\bibitem[{{Bertin} {et~al.}(2002){Bertin}, {Mellier}, {Radovich}, {Missonnier},
  {Didelon}, \& {Morin}}]{2002ASPC..281..228B}
{Bertin}, A., {Mellier}, Y., {Radovich}, M., {et~al.} 2002, in Astronomical
  Society of the Pacific Conference Series, Vol. 281, Astronomical Data
  Analysis Software and Systems XI, 228--237

\bibitem[{{Bintley} {et~al.}(2014){Bintley}, {Holland}, {MacIntosh}, {Friberg},
  {Bell}, {Berke}, {Berry}, {Berthold}, {Cookson}, {Coulson}, {Currie},
  {Dempsey}, {Gibb}, {Gorges}, {Graves}, {Jenness}, {Johnstone}, {Parsons},
  {Thomas}, {Walther}, \& {Wouterloot}}]{2014SPIE.9153E..03B}
{Bintley}, D., {Holland}, W.~S., {MacIntosh}, M.~J., {et~al.} 2014, in Society
  of Photo-Optical Instrumentation Engineers (SPIE) Conference Series, Vol.
  9153, Society of Photo-Optical Instrumentation Engineers (SPIE) Conference
  Series, 3

\bibitem[{{Bonnell} \& {Bate}(2006)}]{2006MNRAS.370..488B}
{Bonnell}, I.~A. \& {Bate}, M.~R. 2006, \mnras, 370, 488

\bibitem[{{Bonnor}(1956)}]{1956MNRAS.116..351B}
{Bonnor}, W.~B. 1956, \mnras, 116, 351

\bibitem[{{Carpenter} {et~al.}(1997){Carpenter}, {Meyer}, {Dougados}, {Strom},
  \& {Hillenbrand}}]{1997AJ....114..198C}
{Carpenter}, J.~M., {Meyer}, M.~R., {Dougados}, C., {Strom}, S.~E., \&
  {Hillenbrand}, L.~A. 1997, \aj, 114, 198, {Erratum:} 1997, \aj, 114, 1275

\bibitem[{{Carter} {et~al.}(2012){Carter}, {Lazareff}, {Maier}, {Chenu},
  {Fontana}, {Bortolotti}, {Boucher}, {Navarrini}, {Blanchet}, {Greve}, {John},
  {Kramer}, {Morel}, {Navarro}, {Pe{\~n}alver}, {Schuster}, \&
  {Thum}}]{2012A&A...538A..89C}
{Carter}, M., {Lazareff}, B., {Maier}, D., {et~al.} 2012, \aap, 538, A89

\bibitem[{{Chapin} {et~al.}(2008){Chapin}, {Ade}, {Bock}, {Brunt}, {Devlin},
  {Dicker}, {Griffin}, {Gundersen}, {Halpern}, {Hargrave}, {Hughes}, {Klein},
  {Marsden}, {Martin}, {Mauskopf}, {Netterfield}, {Olmi}, {Pascale},
  {Patanchon}, {Rex}, {Scott}, {Semisch}, {Truch}, {Tucker}, {Viero}, \&
  {Weibe}}]{2008ApJ...681..428C}
{Chapin}, E.~L., {Ade}, P.~A.~R., {Bock}, J.~J., {et~al.} 2008, \apj, 681, 428

\bibitem[{{Chapin} {et~al.}(2013){Chapin}, {Berry}, {Gibb}, {Jenness}, {Scott},
  {Tilanus}, {Economou}, \& {Holland}}]{2013MNRAS.430.2545C}
{Chapin}, E.~L., {Berry}, D.~S., {Gibb}, A.~G., {et~al.} 2013, \mnras, 430,
  2545

\bibitem[{{Condon} {et~al.}(1998){Condon}, {Cotton}, {Greisen}, {Yin},
  {Perley}, {Taylor}, \& {Broderick}}]{1998AJ....115.1693C}
{Condon}, J.~J., {Cotton}, W.~D., {Greisen}, E.~W., {et~al.} 1998, \aj, 115,
  1693

\bibitem[{{Cutri} {et~al.}(2003){Cutri}, {Skrutskie}, {van Dyk}, {Beichman},
  {Carpenter}, {Chester}, {Cambresy}, {Evans}, {Fowler}, {Gizis}, {Howard},
  {Huchra}, {Jarrett}, {Kopan}, {Kirkpatrick}, {Light}, {Marsh}, {McCallon},
  {Schneider}, {Stiening}, {Sykes}, {Weinberg}, {Wheaton}, {Wheelock}, \&
  {Zacarias}}]{2003yCat.2246....0C}
{Cutri}, R.~M., {Skrutskie}, M.~F., {van Dyk}, S., {et~al.} 2003, VizieR Online
  Data Catalog, 2246, 0

\bibitem[{{De Graauw} {et~al.}(2010){De Graauw}, {Helmich}, {Phillips},
  {Stutzki}, {Caux}, {Whyborn}, {Dieleman}, {Roelfsema}, {Aarts}, {Assendorp},
  {Bachiller}, {Baechtold}, {Barcia}, {Beintema}, {Belitsky}, {Benz}, {Bieber},
  {Boogert}, {Borys}, {Bumble}, {Ca\"is}, {Caris}, {Cerulli-Irelli},
  {Chattopadhyay}, {Cherednichenko}, {Ciechanowicz}, {Coeur-Joly}, {Comito},
  {Cros}, {de Jonge}, {de Lange}, {Delforges}, {Delorme}, {den Boggende},
  {Desbat}, {Diez-Gonz\'alez}, {Di Giorgio}, {Dubbeldam}, {Edwards}, {Eggens},
  {Erickson}, {Evers}, {Fich}, {Finn}, {Franke}, {Gaier}, {Gal}, {Gao},
  {Gallego}, {Gauffre}, {Gill}, {Glenz}, {Golstein}, {Goulooze}, {Gunsing},
  {G\"usten}, {Hartogh}, {Hatch}, {Higgins}, {Honingh}, {Huisman}, {Jackson},
  {Jacobs}, {Jacobs}, {Jarchow}, {Javadi}, {Jellema}, {Justen}, {Karpov},
  {Kasemann}, {Kawamura}, {Keizer}, {Kester}, {Klapwijk}, {Klein}, {Kollberg},
  {Kooi}, {Kooiman}, {Kopf}, {Krause}, {Krieg}, {Kramer}, {Kruizenga}, {Kuhn},
  {Laauwen}, {Lai}, {Larsson}, {Leduc}, {Leinz}, {Lin}, {Liseau}, {Liu},
  {Loose}, {L\'opez-Fernandez}, {Lord}, {Luinge}, {Marston},
  {Mart\'in-Pintado}, {Maestrini}, {Maiwald}, {McCoey}, {Mehdi}, {Megej},
  {Melchior}, {Meinsma}, {Merkel}, {Michalska}, {Monstein}, {Moratschke},
  {Morris}, {Muller}, {Murphy}, {Naber}, {Natale}, {Nowosielski}, {Nuzzolo},
  {Olberg}, {Olbrich}, {Orfei}, {Orleanski}, {Ossenkopf}, {Peacock}, {Pearson},
  {Peron}, {Phillip-May}, {Piazzo}, {Planesas}, {Rataj}, {Ravera}, {Risacher},
  {Salez}, {Samoska}, {Saraceno}, {Schieder}, {Schlecht}, {Schl\"oder},
  {Schm\"ulling}, {Schultz}, {Schuster}, {Siebertz}, {Smit}, {Szczerba},
  {Shipman}, {Steinmetz}, {Stern}, {Stokroos}, {Teipen}, {Teyssier}, {Tils},
  {Trappe}, {van Baaren}, {van Leeuwen}, {van de Stadt}, {Visser}, {Wildeman},
  {Wafelbakker}, {Ward}, {Wesselius}, {Wild}, {Wulff}, {Wunsch}, {Tielens},
  {Zaal}, {Zirath}, {Zmuidzinas}, \& {Zwart}}]{2010A&A...518L...6D}
{De Graauw}, T., {Helmich}, F.~P., {Phillips}, T.~G., {et~al.} 2010, \aap, 518,
  L6

\bibitem[{{Dempsey} {et~al.}(2013){Dempsey}, {Friberg}, {Jenness}, {Tilanus},
  {Thomas}, {Holland}, {Bintley}, {Berry}, {Chapin}, {Chrysostomou}, {Davis},
  {Gibb}, {Parsons}, \& {Robson}}]{2013MNRAS.430.2534D}
{Dempsey}, J.~T., {Friberg}, P., {Jenness}, T., {et~al.} 2013, \mnras, 430,
  2534

\bibitem[{{Di~Francesco} {et~al.}(2008){Di~Francesco}, {Johnstone}, {Kirk},
  {MacKenzie}, \& {Ledwosinska}}]{2008ApJS..175..277D}
{Di~Francesco}, J., {Johnstone}, D., {Kirk}, H., {MacKenzie}, T., \&
  {Ledwosinska}, E. 2008, \apjs, 175, 277

\bibitem[{{Didelon} {et~al.}(2015){Didelon}, {Motte}, {Tremblin}, {Hill},
  {Hony}, {Hennemann}, {Hennebelle}, {Anderson}, {Galliano}, {Schneider},
  {Rayner}, {Rygl}, {Louvet}, {Zavagno}, {K\"onyves}, {Sauvage}, {Andr\'e},
  {Bontemps}, {Peretto}, {Griffin}, {Gonz\'alez}, {Lebouteiller},
  {Arzoumanian}, {Bernard}, {Benedettini}, {Di Francesco}, {Men'shchikov},
  {Minier}, {Nguyen Luong}, {Bernard}, {Palmeirim}, {Pezzuto},
  {Rivera-Ingraham}, {Russeil}, {Ward-Thompson}, \&
  {White}}]{2015A&A...584A...4D}
{Didelon}, P., {Motte}, F., {Tremblin}, P., {et~al.} 2015, \aap, 584, A4

\bibitem[{{Dierickx} {et~al.}(2015){Dierickx}, {Jim\'enez-Serra}, {Rivilla}, \&
  {Zhang}}]{2015ApJ...803...89D}
{Dierickx}, M., {Jim\'enez-Serra}, I., {Rivilla}, V.~M., \& {Zhang}, Q. 2015,
  \apj, 803, 89

\bibitem[{{Duarte-Cabral} {et~al.}(2013){Duarte-Cabral}, {Bontemps}, {Motte},
  {Hennemann}, {Schneider}, \& {Andr{\'e}}}]{2013A&A...558A.125D}
{Duarte-Cabral}, A., {Bontemps}, S., {Motte}, F., {et~al.} 2013, \aap, 558,
  A125

\bibitem[{{Dzib} {et~al.}(2016){Dzib}, {Ortiz-Le{\'o}n}, {Loinard},
  {Mioduszewski}, {Rodr{\'{\i}}guez}, {Torres}, \&
  {Deller}}]{2016arXiv160601757D}
{Dzib}, S.~A., {Ortiz-Le{\'o}n}, G.~N., {Loinard}, L., {et~al.} 2016, \apj,
  826, 201, {arXiv:1606.01757}

\bibitem[{{Fallscheer} {et~al.}(2013){Fallscheer}, {Reid}, {Di~Francesco},
  {Martin}, {Hill}, {Hennemann}, {Nguyen Luong}, {Motte}, {Men'shchikov},
  {Andr\'e}, {Ward-Thompson}, {Griffin}, {Kirk}, {K\"onyves}, {Rygl},
  {Sadavoy}, {Sauvage}, {Schneider}, {Anderson}, {Benedettini}, {Bernard},
  {Bontemps}, {Ginsburg}, {Molinari}, {Polichroni}, {Rivera-Ingraham},
  {Roussel}, {Testi}, {White}, {Williams}, {Wilson}, {Wong}, \&
  {Zavagno}}]{2013ApJ...773..102F}
{Fallscheer}, C., {Reid}, M.~A., {Di~Francesco}, J., {et~al.} 2013, \apj, 773,
  102

\bibitem[{{Fazio} {et~al.}(2004){Fazio}, {Hora}, {Allen}, {Ashby}, {Barmby},
  {Deutsch}, {Huang}, {Kleiner}, {Marengo}, {Megeath}, {Melnick}, {Pahre},
  {Patten}, {Polizotti}, {Smith}, {Taylor}, {Wang}, {Willner}, {Hoffmann},
  {Pipher}, {Forrest}, {McMurty}, {McCreight}, {McKelvey}, {McMurray}, {Koch},
  {Moseley}, {Arendt}, {Mentzell}, {Marx}, {Losch}, {Mayman}, {Eichhorn},
  {Krebs}, {Jhabvala}, {Gezari}, {Fixsen}, {Flores}, {Shakoorzadeh}, {Jungo},
  {Hakun}, {Workman}, {Karpati}, {Kichak}, {Whitley}, {Mann}, {Tollestrup},
  {Eisenhardt}, {Stern}, {Gorjian}, {Bhattacharya}, {Carey}, {Nelson},
  {Glaccum}, {Lacy}, {Lowrance}, {Laine}, {Reach}, {Stauffer}, {Surace},
  {Wilson}, {Wright}, {Hoffman}, {Domingo}, \& {Cohen}}]{2004ApJS..154...10F}
{Fazio}, G.~G., {Hora}, J.~L., {Allen}, L.~E., {et~al.} 2004, \apjs, 154, 10

\bibitem[{{Fuente} {et~al.}(2010){Fuente}, {Bern\'e}, {Cernicharo}, {Rizzo},
  {Gonz\'alez-Garc\'ia}, {Goicoechea}, {Pilleri}, {Ossenkopf}, {Gerin},
  {G\"usten}, {Akyilmaz}, {Benz}, {Boulanger}, {Bruderer}, {Dedes}, {France},
  {Garc\'ia-Burillo}, {Harris}, {Joblin}, {Klein}, {Kramer}, {Le Petit},
  {Lord}, {Martin}, {Mart\'in-Pintado}, {Mookerjea}, {Neufeld}, {Okada},
  {Pety}, {Phillips}, {R\"ollig}, {Simon}, {Stutzki}, {van der Tak},
  {Teyssier}, {Usero}, {Yorke}, {Schuster}, {Melchior}, {Lorenzani},
  {Szczerba}, {Fich}, {Mc Coey}, {Pearson}, \&
  {Dieleman}}]{2010A&A...521L..23F}
{Fuente}, A., {Bern\'e}, O., {Cernicharo}, J., {et~al.} 2010, \aap, 521, L23

\bibitem[{{Giannini} {et~al.}(2012){Giannini}, {Elia}, {Lorenzetti},
  {Molinari}, {Motte}, {Schisano}, {Pezzuto}, {Pestalozzi}, {Di Giorgio},
  {Andr\'e}, {Hill}, {Benedettini}, {Bontemps}, {Di Francesco}, {Fallscheer},
  {Hennemann}, {Kirk}, {Minier}, {Nguyen Luong}, {Polychroni}, {Rygl},
  {Saraceno}, {Schneider}, {Spinoglio}, {Testi}, {Ward-Thompson}, \&
  {White}}]{2012A&A...539A.156G}
{Giannini}, T., {Elia}, D., {Lorenzetti}, D., {et~al.} 2012, \aap, 539, A156

\bibitem[{{Griffin} {et~al.}(2010){Griffin}, {Abergel}, {Abreu}, {Ade},
  {Andr\'e}, {Augueres}, {Babbedge}, {Bae}, {Baillie}, {Baluteau}, {Barlow},
  {Bendo}, {Benielli}, {Bock}, {Bonhomme}, {Brisbin}, {Brockley-Blatt},
  {Caldwell}, {Cara}, {Castro-Rodriguez}, {Cerulli}, {Chanial}, {Chen},
  {Clark}, {Clements}, {Clerc}, {Coker}, {Communal}, {Conversi}, {Cox},
  {Crumb}, {Cunningham}, {Daly}, {Davis}, {De Antoni}, {Delderfield}, {Devin},
  {Di Giorgio}, {Didschuns}, {Dohlen}, {Donati}, {Dowell}, {Dowell}, {Duband},
  {Dumaye}, {Emery}, {Ferlet}, {Ferrand}, {Fontignie}, {Fox}, {Franceschini},
  {Frerking}, {Fulton}, {Garcia}, {Gastaud}, {Gear}, {Glenn}, {Goizel},
  {Griffin}, {Grundy}, {Guest}, {Guillemet}, {Hargrave}, {Harwit}, {Hastings},
  {Hatziminaoglou}, {Herman}, {Hinde}, {Hristov}, {Huang}, {Imhof}, {Isaak},
  {Israelsson}, {Ivison}, {Jennings}, {Kiernan}, {King}, {Lange}, {Latter},
  {Laurent}, {Laurent}, {Leeks}, {Lellouch}, {Levenson}, {Li}, {Li},
  {Lilienthal}, {Lim}, {Liu}, {Lu}, {Madden}, {Mainetti}, {Marliani}, {McKay},
  {Mercier}, {Molinari}, {Morris}, {Moseley}, {Mulder}, {Mur}, {Naylor},
  {Nguyen}, {O'Halloran}, {Oliver}, {Olofsson}, {Olofsson}, {Orfei}, {Page},
  {Pain}, {Panuzzo}, {Papageorgiou}, {Parks}, {Parr-Burman}, {Pearce},
  {Pearson}, {P\'erez-Fournon}, {Pinsard}, {Pisano}, {Podosek}, {Pohlen},
  {Polehampton}, {Pouliquen}, {Rigopoulou}, {Rizzo}, {Roseboom}, {Roussel},
  {Rowan-Robinson}, {Rownd}, {Saraceno}, {Sauvage}, {Savage}, {Savini},
  {Sawyer}, {Scharmberg}, {Schmitt}, {Schneider}, {Schulz}, {Schwartz},
  {Shafer}, {Shupe}, {Sibthorpe}, {Sidher}, {Smith}, {Smith}, {Smith},
  {Spencer}, {Stobie}, {Sudiwala}, {Sukhatme}, {Surace}, {Stevens}, {Swinyard},
  {Trichas}, {Tourette}, {Triou}, {Tseng}, {Tucker}, {Turner}, {Vaccari},
  {Valtchanov}, {Vigroux}, {Virique}, {Voellmer}, {Walker}, {Ward}, {Waskett},
  {Weilert}, {Wesson}, {White}, {Whitehouse}, {Wilson}, {Winter}, {Woodcraft},
  {Wright}, {Xu}, {Zavagno}, {Zemcov}, {Zhang}, \&
  {Zonca}}]{2010A&A...518L...3G}
{Griffin}, M.~J., {Abergel}, A., {Abreu}, A., {et~al.} 2010, \aap, 518, L3

\bibitem[{{Griffin} {et~al.}(2013){Griffin}, {North}, {Schulz},
  {Amaral-Rogers}, {Bendo}, {Bock}, {Conversi}, {Conley}, {Dowell}, {Ferlet},
  {Glenn}, {Lim}, {Pearson}, {Pohlen}, {Sibthorpe}, {Spencer}, {Swinyard}, \&
  {Valtchanov}}]{2013MNRAS.434..992G}
{Griffin}, M.~J., {North}, C.~E., {Schulz}, B., {et~al.} 2013, \mnras, 434, 992

\bibitem[{{Gutermuth} {et~al.}(2009){Gutermuth}, {Megeath}, {Myers}, {Allen},
  {Pipher}, \& {Fazio}}]{2009ApJS..184...18G}
{Gutermuth}, R.~A., {Megeath}, S.~T., {Myers}, P.~C., {et~al.} 2009, \apjs,
  184, 18

\bibitem[{{Habets} \& {Heintze}(1981)}]{1981A&AS...46..193H}
{Habets}, G.~M.~H.~J. \& {Heintze}, J.~R.~W. 1981, \aaps, 193

\bibitem[{{Heitsch}(2013{\natexlab{a}})}]{2013ApJ...769..115H}
{Heitsch}, F. 2013{\natexlab{a}}, \apj, 769, 115

\bibitem[{{Heitsch}(2013{\natexlab{b}})}]{2013ApJ...776...62H}
{Heitsch}, F. 2013{\natexlab{b}}, \apj, 776, 62

\bibitem[{{Hennemann} {et~al.}(2010){Hennemann}, {Motte}, {Bontemps},
  {Schneider}, {Csengeri}, {Balog}, {Di Francesco}, {Zavagno}, {Andr\'e},
  {Men'shchikov}, {Abergel}, {Ali}, {Baluteau}, {Bernard}, {Cox}, {Didelon},
  {di Giorgio}, {Griffin}, {Hargrave}, {Hill}, {Horeau}, {Huang}, {Kirk},
  {Leeks}, {Li}, {Marston}, {Martin}, {Molinari}, {Nguyen Luong}, {Olofsson},
  {Persi}, {Pezzuto}, {Russeil}, {Saraceno}, {Sauvage}, {Sibthorpe},
  {Spinoglio}, {Testi}, {Ward-Thompson}, {White}, {Wilson}, \&
  {Woodcraft}}]{2010A&A...518L..84H}
{Hennemann}, M., {Motte}, F., {Bontemps}, S., {et~al.} 2010, \aap, 518, L84

\bibitem[{{Hennemann} {et~al.}(2012){Hennemann}, {Motte}, {Schneider},
  {Didelon}, {Hill}, {Arzoumanian}, {Bontemps}, {Csengeri}, {Andr\'e},
  {K\"onyves}, {Louvet}, {Marston}, {Men'shchikov}, {Minier}, {Nguyen Luong},
  {Palmeirim}, {Peretto}, {Sauvage}, {Zavagno}, {Anderson}, {Bernard}, {Di
  Francesco}, {Elia}, {Li}, {Martin}, {Molinari}, {Pezzuto}, {Russeil}, {Rygl},
  {Schisano}, {Spinoglio}, {Sousbie}, {Ward-Thompson}, \&
  {White}}]{2012A&A...543L...3H}
{Hennemann}, M., {Motte}, F., {Schneider}, N., {et~al.} 2012, \aap, 543, L3

\bibitem[{{Herbst} \& {Racine}(1976)}]{1976AJ.....81..840H}
{Herbst}, W. \& {Racine}, R. 1976, \aj, 81, 840

\bibitem[{{Hildebrand}(1983)}]{1983QJRAS..24..267H}
{Hildebrand}, R.~H. 1983, \qjras, 24, 267

\bibitem[{{Hill} {et~al.}(2012){Hill}, {Andr\'e}, {Arzoumanian}, {Motte},
  {Minier}, {Men'shchikov}, {Didelon}, {Hennemann}, {K\"onyves}, {Nguyen
  Luong}, {Palmeirim}, {Peretto}, {Schneider}, {Bontemps}, {Louvet}, {Elia},
  {Giannini}, {Rev\'eret}, {Le Pennec}, {Rodriguez}, {Boulade}, {Doumayrou},
  {Dubreuil}, {Gallais}, {Lortholary}, {Martignac}, {Talvard}, \& {De
  Breuck}}]{2012A&A...548L...6H}
{Hill}, T., {Andr\'e}, P., {Arzoumanian}, D., {et~al.} 2012, \aap, 548, L6

\bibitem[{{Hill} {et~al.}(2011){Hill}, {Motte}, {Didelon}, {Bontemps},
  {Minier}, {Hennemann}, {Schneider}, {Andr\'e}, {Men'shchikov}, {Anderson},
  {Arzoumanian}, {Bernard}, {Di Francesco}, {Elia}, {Giannini}, {Griffin},
  {K\"onyves}, {Kirk}, {Marston}, {Martin}, {Molinari}, {Nguyen Luong},
  {Peretto}, {Pezzuto}, {Roussel}, {Sauvage}, {Sousbie}, {Testi},
  {Ward-Thompson}, {White}, {Wilson}, \& {Zavagno}}]{2011A&A...533A..94H}
{Hill}, T., {Motte}, F., {Didelon}, P., {et~al.} 2011, \aap, 533, A94

\bibitem[{{Hodapp}(1987)}]{1987A&A...172..304H}
{Hodapp}, K.~W. 1987, \aap, 172, 304

\bibitem[{{Hodapp}(2007)}]{2007AJ....134.2020H}
{Hodapp}, K.~W. 2007, \aj, 134, 2020

\bibitem[{{Holland} {et~al.}(2013){Holland}, {Bintley}, {Chapin},
  {Chrysostomou}, {Davis}, {Dempsey}, {Duncan}, {Fich}, {Friberg}, {Halpern},
  {Irwin}, {Jenness}, {Kelly}, {MacIntosh}, {Robson}, {Scott}, {Ade},
  {Atad-Ettedgui}, {Berry}, {Craig}, {Gao}, {Gibb}, {Hilton}, {Hollister},
  {Kycia}, {Lunney}, {McGregor}, {Montgomery}, {Parkes}, {Tilanus}, {Ullom},
  {Walther}, {Walton}, {Woodcraft}, {Amiri}, {Atkinson}, {Burger}, {Chuter},
  {Coulson}, {Doriese}, {Dunare}, {Economou}, {Niemack}, {Parsons},
  {Reintsema}, {Sibthorpe}, {Smail}, {Sudiwala}, \&
  {Thomas}}]{2013MNRAS.430.2513H}
{Holland}, W.~S., {Bintley}, D., {Chapin}, E.~L., {et~al.} 2013, \mnras, 430,
  2513

\bibitem[{{Kirk} {et~al.}(2013{\natexlab{a}}){Kirk}, {Myers}, {Bourke},
  {Gutermuth}, {Hedden}, \& {Wilson}}]{2013ApJ...766..115K}
{Kirk}, H., {Myers}, P.~C., {Bourke}, T.~L., {et~al.} 2013{\natexlab{a}}, \apj,
  766, 115

\bibitem[{{Kirk} {et~al.}(2013{\natexlab{b}}){Kirk}, {Ward-Thompson}, \&
  {Palmeirim}}]{2013MNRAS.432.1424K}
{Kirk}, J.~M., {Ward-Thompson}, D., \& {Palmeirim}, P. 2013{\natexlab{b}},
  \mnras, 432, 1424

\bibitem[{{K\"onyves} {et~al.}(2015){K\"onyves}, {Andr\'e}, {Men'shchikov},
  {Palmeirim}, {Arzoumanian}, {Schneider}, {Roy}, {Didelon}, {Maury},
  {Shimajiri}, {Di Francesco}, {Bontemps}, {Peretto}, {Benedettini}, {Bernard},
  {Elia}, {Griffin}, {Hill}, {Kirk}, {Ladjelate}, {Marsh}, {Martin}, {Motte},
  {Nguyen Luong}, {Pezzuto}, {Roussel}, {Rygl}, {Sadavoy}, {Schisano},
  {Spinoglio}, {Ward-Thompson}, \& {White}}]{2015A&A...584A..91K}
{K\"onyves}, V., {Andr\'e}, P., {Men'shchikov}, A., {et~al.} 2015, \aap, 584,
  A91

\bibitem[{{K\"onyves} {et~al.}(2010){K\"onyves}, {Andr\'e}, {Men'shchikov},
  {Schneider}, {Arzoumanian}, {Bontemps}, {Attard}, {Motte}, {Didelon},
  {Maury}, {Abergel}, {Ali}, {Baluteau}, {Bernard}, {Cambr\'esy}, {Cox}, {Di
  Francesco}, {Di Giorgio}, {Griffin}, {Hargrave}, {Huang}, {Kirk}, {Li},
  {Martin}, {Minier}, {Molinari}, {Olofsson}, {Pezzuto}, {Russeil}, {Roussel},
  {Saraceno}, {Sauvage}, {Sibthorpe}, {Spinoglio}, {Testi}, {Ward-Thompson},
  {White}, {Wilson}, {Woodcraft}, \& {Zavagno}}]{2010A&A...518L.106K}
{K\"onyves}, V., {Andr\'e}, P., {Men'shchikov}, A., {et~al.} 2010, \aap, 518,
  L106

\bibitem[{{Krumholz}(2015)}]{2015ASSL..412...43K}
{Krumholz}, M.~R. 2015, in Astrophysics and Space Science Library, Vol. 412,
  Very Massive Stars in the Local Universe, ed. J.~S. {Vink}, 43--75

\bibitem[{{Kutner} \& {Tucker}(1975)}]{1975ApJ...199...79K}
{Kutner}, M.~L. \& {Tucker}, K.~D. 1975, \apj, 199, 79

\bibitem[{{Lombardi} {et~al.}(2011){Lombardi}, {Alves}, \&
  {Lada}}]{2011A&A...535A..16L}
{Lombardi}, M., {Alves}, J., \& {Lada}, C.~J. 2011, \aap, 535, A16

\bibitem[{{Loren} {et~al.}(1974){Loren}, {Peters}, \& {Vanden
  Bout}}]{1974ApJ...194L.103L}
{Loren}, R.~B., {Peters}, W.~L., \& {Vanden Bout}, P.~A. 1974, \apj, 194, L103

\bibitem[{{Louvet} {et~al.}(2014){Louvet}, {Motte}, {Hennebelle}, {Maury},
  {Bonnell}, {Bontemps}, {Gusdorf}, {Hill}, {Gueth}, {Peretto},
  {Duarte-Cabral}, {Stephan}, {Schilke}, {Csengeri}, {Nguyen Luong}, \&
  {Lis}}]{2014A&A...570A..15L}
{Louvet}, F., {Motte}, F., {Hennebelle}, P., {et~al.} 2014, \aap, 570, A15

\bibitem[{{Lynds}(1962)}]{1962ApJS....7....1L}
{Lynds}, B.~T. 1962, \apjs, 7, 1

\bibitem[{{Maddalena} {et~al.}(1986){Maddalena}, {Morris}, J., \&
  {Thaddeus}}]{1986ApJ...303..375M}
{Maddalena}, R.~J., {Morris}, M., J., M., \& {Thaddeus}, P. 1986, \apj, 303,
  375

\bibitem[{{Mainzer} {et~al.}(2011){Mainzer}, {Bauer}, {Grav}, {Masiero},
  {Cutri}, {Dailey}, {Eisenhardt}, {McMillan}, {Wright}, {Walker}, {Jedicke},
  {Spahr}, {Tholen}, {Alles}, {Beck}, {Brandenburg}, {Conrow}, {Evans},
  {Fowler}, {Jarrett}, {Marsh}, {Masci}, {McCallon}, {Wheelock}, {Wittman},
  {Wyatt}, {DeBaun}, {Elliott}, {Elsbury}, {Gautier}, {Gomillion}, {Leisawitz},
  {Maleszewski}, {Micheli}, \& {Wilkins}}]{2011ApJ...731...53M}
{Mainzer}, A., {Bauer}, J., {Grav}, T., {et~al.} 2011, \apj, 731, 53

\bibitem[{{Markwardt}(2009)}]{2009ASPC..411..251M}
{Markwardt}, C.~B. 2009, in Astronomical Society of the Pacific Conference
  Series, Vol. 411, {Astronomical Data Analysis Software and Systems XVIII},
  251--254

\bibitem[{{Marsh} {et~al.}(2014){Marsh}, {Griffin}, {Palmeirim}, {Andr\'e},
  {Kirk}, {Stamatellos}, {Ward-Thompson}, {Roy}, {Bontemps}, {Di Francesco},
  {Elia}, {Hill}, {K\"onyves}, {Motte}, {Nguyen Luong}, {Peretto}, {Pezzuto},
  {Rivera-Ingraham}, {Schneider}, {Spinoglio}, \&
  {White}}]{2014MNRAS.439.3683M}
{Marsh}, K.~A., {Griffin}, M.~J., {Palmeirim}, P., {et~al.} 2014, \mnras, 439,
  3683

\bibitem[{{Marsh} {et~al.}(2016){Marsh}, {Kirk}, {Andr{\'e}}, {Griffin},
  {K{\"o}nyves}, {Palmeirim}, {Men'shchikov}, {Ward-Thompson}, {Benedettini},
  {Bresnahan}, {Francesco}, {Elia}, {Motte}, {Peretto}, {Pezzuto}, {Roy},
  {Sadavoy}, {Schneider}, {Spinoglio}, \& {White}}]{2016MNRAS.459..342M}
{Marsh}, K.~A., {Kirk}, J.~M., {Andr{\'e}}, P., {et~al.} 2016, \mnras, 459, 342

\bibitem[{{McKee} \& {Tan}(2002)}]{2002Natur.416...59M}
{McKee}, C.~F. \& {Tan}, J.~C. 2002, \nat, 416, 59

\bibitem[{{Men'shchikov}(2013)}]{2013A&A...560A..63M}
{Men'shchikov}, A. 2013, \aap, 560, A63

\bibitem[{{Men'shchikov} {et~al.}(2012){Men'shchikov}, {Andr\'e}, {Didelon},
  {Motte}, {Hennemann}, \& {Schneider}}]{2012A&A...542A..81M}
{Men'shchikov}, A., {Andr\'e}, P., {Didelon}, P., {et~al.} 2012, \aap, 542, A81

\bibitem[{{Mink}(2002)}]{2002ASPC..281..169M}
{Mink}, D.~J. 2002, in Astronomical Society of the Pacific Conference Series,
  Vol. 281, Astronomical Data Analysis Software and Systems XI, 169--172

\bibitem[{{Motte} {et~al.}(2017){Motte}, {Bontemps}, \&
  {Louvet}}]{2017arXiv170600118M}
{Motte}, F., {Bontemps}, S., \& {Louvet}, F. 2017, ArXiv e-prints

\bibitem[{{Motte} {et~al.}(2010){Motte}, {Zavagno}, {Bontemps}, {Schneider},
  {Hennemann}, {Di Francesco}, {Andr\'e}, {Saraceno}, {Griffin}, {Marston},
  {Ward-Thompson}, {White}, {Minier}, {Men'shchikov}, {Hill}, {Anderson},
  {Aussel}, {Balog}, {Baluteau}, {Bernard}, {Cox}, {Csengeri}, {Deharveng},
  {Didelon}, {di Giorgio}, {Hargrave}, {Huang}, {Kirk}, {Leeks}, {Li},
  {Martin}, {Molinari}, {Nguyen Luong}, {Olofsson}, {Persi}, {Peretto},
  {Pezzuto}, {Roussel}, {Russeil}, {Sadavoy}, {Sauvage}, {Sibthorpe},
  {Spinoglio}, {Testi}, {Teyssier}, {Vavrek}, {Wilson}, \&
  {Woodcraft}}]{2010A&A...518L..77M}
{Motte}, F., {Zavagno}, A., {Bontemps}, S., {et~al.} 2010, \aap, 518, L77

\bibitem[{{Myers}(2009)}]{2009ApJ...700.1609M}
{Myers}, P.~C. 2009, \apj, 700, 1609

\bibitem[{{Nguyen Luong} {et~al.}(2013){Nguyen Luong}, {Motte}, {Carlhoff},
  {Louvet}, {Lesaffre}, {Schilke}, {Hill}, {Hennemann}, {Gusdorf}, {Didelon},
  {Schneider}, {Bontemps}, {Duarte-Cabral}, {Menten}, {Martin}, {Wyrowski},
  {Bendo}, {Roussel}, {Bernard}, {Bronfman}, {Henning}, {Kramer}, \&
  {Heitsch}}]{2013ApJ...775...88N}
{Nguyen Luong}, Q., {Motte}, F., {Carlhoff}, P., {et~al.} 2013, \apj, 775, 88

\bibitem[{{Nguyen Luong} {et~al.}(2011){Nguyen Luong}, {Motte}, {Hennemann},
  {Hill}, {Rygl}, {Schneider}, {Bontemps}, {Men'shchikov}, {Andr\'e},
  {Peretto}, {Anderson}, {Arzoumanian}, {Deharveng}, {Didelon}, {Di Francesco},
  {Griffin}, {Kirk}, {K\"onyves}, {Martin}, {Maury}, {Minier}, {Molinari},
  {Pestalozzi}, {Pezzuto}, {Reid}, {Roussel}, {Sauvage}, {Schuller}, {Testi},
  {Ward-Thompson}, {White}, \& {Zavagno}}]{2011A&A...535A..76N}
{Nguyen Luong}, Q., {Motte}, F., {Hennemann}, M., {et~al.} 2011, \aap, 535, A76

\bibitem[{{Ostriker}(1964)}]{1964ApJ...140.1056O}
{Ostriker}, J. 1964, \apj, 140, 1056

\bibitem[{{Ott}(2010)}]{2010ASPC..434..139O}
{Ott}, S. 2010, in Astronomical Society of the Pacific Conference Series, Vol.
  434, Astronomical Data Analysis Software and Systems XIX, 139--142

\bibitem[{{PACS Observer's Manual}(2013)}]{2013PACS.....v2.5.1}
{PACS Observer's Manual}. 2013, {Version 2.5.1,}, {PACS Consortium}

\bibitem[{{Palmeirim} {et~al.}(2013){Palmeirim}, {Andr\'e}, {Kirk},
  {Ward-Thompson}, {Arzoumanian}, {K\"onyves}, {Didelon}, {Schneider},
  {Benedettini}, {Bontemps}, {Di Francesco}, {Elia}, {Griffin}, {Hennemann},
  {Hill}, {Martin}, {Men'shchikov}, {Molinari}, {Motte}, {Nguyen Luong},
  {Nutter}, {Peretto}, {Pezzuto}, {Roy}, {Rygl}, {Spinoglio}, \&
  {White}}]{2013A&A...550A..38P}
{Palmeirim}, P., {Andr\'e}, P., {Kirk}, J., {et~al.} 2013, \aap, 550, A38

\bibitem[{{Pattle} {et~al.}(2015){Pattle}, {Ward-Thompson}, {Kirk}, {White},
  {Drabek-Maunder}, {Buckle}, {Beaulieu}, {Berry}, {Broekhoven-Fiene},
  {Currie}, {Fich}, {Hatchell}, {Kirk}, {Jenness}, {Johnstone}, {Mottram},
  {Nutter}, {Pineda}, {Quinn}, {Salji}, {Tisi}, {Walker-Smith}, {Francesco},
  {Hogerheijde}, {Andr{\'e}}, {Bastien}, {Bresnahan}, {Butner}, {Chen},
  {Chrysostomou}, {Coude}, {Davis}, {Duarte-Cabral}, {Fiege}, {Friberg},
  {Friesen}, {Fuller}, {Graves}, {Greaves}, {Gregson}, {Griffin}, {Holland},
  {Joncas}, {Knee}, {K{\"o}nyves}, {Mairs}, {Marsh}, {Matthews},
  {Moriarty-Schieven}, {Rawlings}, {Richer}, {Robertson}, {Rosolowsky},
  {Rumble}, {Sadavoy}, {Spinoglio}, {Thomas}, {Tothill}, {Viti}, {Wouterloot},
  {Yates}, \& {Zhu}}]{2015MNRAS.450.1094P}
{Pattle}, K., {Ward-Thompson}, D., {Kirk}, J.~M., {et~al.} 2015, \mnras, 450,
  1094

\bibitem[{{Pence}(1999)}]{1999ASPC..172..487P}
{Pence}, W. 1999, in Astronomical Society of the Pacific Conference Series,
  Vol. 172, Astronomical Data Analysis Software and Systems VIII, 487--489

\bibitem[{{Peretto} {et~al.}(2012){Peretto}, {Andr\'e}, {K\"onyves},
  {Schneider}, {Arzoumanian}, {Palmeirim}, {Didelon}, {Attard}, {Bernard}, {Di
  Francesco}, {Elia}, {Hennemann}, {Hill}, {Kirk}, {Men'shchikov}, {Motte},
  {Nguyen Luong}, {Roussel}, {Sousbie}, {Testi}, {Ward-Thompson}, {White}, \&
  {Zavagno}}]{2012A&A...541A..63P}
{Peretto}, N., {Andr\'e}, P., {K\"onyves}, V., {et~al.} 2012, \aap, 541, A63

\bibitem[{{Peretto} {et~al.}(2013){Peretto}, {Fuller}, {Duarte-Cabral},
  {Avison}, {Hennebelle}, {Andr\'e}, {Bontemps}, {Motte}, {Schneider}, \&
  {Molinari}}]{2013A&A...555A.112P}
{Peretto}, N., {Fuller}, G.~A., {Duarte-Cabral}, A., {et~al.} 2013, \aap, 555,
  A112

\bibitem[{{Pilbratt} {et~al.}(2010){Pilbratt}, {Riedinger}, {Passvogel},
  {Crone}, {Doyle}, {Gageur}, {Heras}, {Jewell}, {Metcalfe}, {Ott}, \&
  {Schmidt}}]{2010A&A...518L...1P}
{Pilbratt}, G.~L., {Riedinger}, J.~R., {Passvogel}, T., {et~al.} 2010, \aap,
  518, L1

\bibitem[{{Pilleri} {et~al.}(2012){Pilleri}, {Fuente}, {Cernicharo},
  {Ossenkopf}, {Bern\'e}, {Gerin}, {Pety}, {Goicoechea}, {Rizzo}, {Montillaud},
  {Gonz\'alez-Garc\'ia}, {Joblin}, {Le Bourlot}, {Le Petit}, \&
  {Kramer}}]{2012A&A...544A.110P}
{Pilleri}, P., {Fuente}, A., {Cernicharo}, J., {et~al.} 2012, \aap, 544, A110

\bibitem[{{Pilleri} {et~al.}(2014){Pilleri}, {Fuente}, {Gerin}, {Cernicharo},
  {Goicochea}, {Ossenkopf}, {Joblin}, {Gonz\'alez-Garc\'ia},
  {Trevi\~no-Morales}, {S\'anchez-Monge}, {Pety}, {Bern\'e}, \&
  {Kramer}}]{2014A&A...561A..69P}
{Pilleri}, P., {Fuente}, A., {Gerin}, M., {et~al.} 2014, \aap, 561, A69

\bibitem[{{Pilleri} {et~al.}(2013){Pilleri}, {Trevi\~no-Morales}, {Fuente},
  {Joblin}, {Cernicharo}, {Gerin}, {Viti}, {Bern\'e}, {Goicochea}, {Pety},
  {Gonz\'alez-Garc\'ia}, {Montillaud}, {Ossenkopf}, {Kramer},
  {Garc\'ia-Burillo}, {Le Petit}, \& {Le Bourlot}}]{2013A&A...554A..87P}
{Pilleri}, P., {Trevi\~no-Morales}, S., {Fuente}, A., {et~al.} 2013, \aap, 554,
  A87

\bibitem[{{Poglitsch} {et~al.}(2010){Poglitsch}, {Waelkens}, {Geis},
  {Feuchtgruber}, {Vandenbussche}, {Rodriguez}, {Krause}, {Renotte}, {van
  Hoof}, {Saraceno}, {Cepa}, {Kerschbaum}, {Agn\`ese}, {Ali}, {Altieri},
  {Andreani}, {Augueres}, {Balog}, {Barl}, {Bauer}, {Belbachir}, {Benedettini},
  {Billot}, {Boulade}, {Bischof}, {Blommaert}, {Callut}, {Cara}, {Cerulli},
  {Cesarsky}, {Contursi}, {Creten}, {De Meester}, {Doublier}, {Doumayrou},
  {Duband}, {Exter}, {Genzel}, {Gillis}, {Gr\"ozinger}, {Henning}, {Herreros},
  {Huygen}, {Inguscio}, {Jakob}, {Jamar}, {Jean}, {de Jong}, {Katterloher},
  {Kiss}, {Klaas}, {Lemke}, {Lutz}, {Madden}, {Marquet}, {Martignac}, {Mazy},
  {Merken}, {Montfort}, {Morbidelli}, {M\"uller}, {Nielbock}, {Okumura},
  {Orfei}, {Ottensamer}, {Pezzuto}, {Popesso}, {Putzeys}, {Regibo}, {Reveret},
  {Royer}, {Sauvage}, {Schreiber}, {Stegmaier}, {Schmitt}, {Schubert}, {Sturm},
  {Thiel}, {Tofani}, {Vavrek}, {Wetzstein}, {Wieprecht}, \&
  {Wiezorrek}}]{2010A&A...518L...2P}
{Poglitsch}, A., {Waelkens}, C., {Geis}, N., {et~al.} 2010, \aap, 518, L2

\bibitem[{{Pokhrel} {et~al.}(2016){Pokhrel}, {Gutermuth}, {Ali}, {Megeath},
  {Pipher}, {Myers}, {Fischer}, {Henning}, {Wolk}, {Allen}, \&
  {Tobin}}]{2016MNRAS.461...22P}
{Pokhrel}, R., {Gutermuth}, R., {Ali}, B., {et~al.} 2016, \mnras, 461, 22

\bibitem[{{Racine}(1968)}]{1968AJ.....73..233R}
{Racine}, R. 1968, \aj, 73, 233

\bibitem[{{Racine} \& {van den Bergh}(1970)}]{1970IAUS...38..219R}
{Racine}, R. \& {van den Bergh}, S. 1970, in The spiral structure of our
  galaxy, Proceedings of the International Astronomical Union, 219--221

\bibitem[{{Rayner}(2015)}]{2015rayner.....phdR}
{Rayner}, T.~S.~M. 2015, PhD thesis, {Cardiff University}

\bibitem[{{Reed}(2003)}]{2003AJ....125.2531R}
{Reed}, B.~C. 2003, \aj, 125, 2531

\bibitem[{{Rieke} {et~al.}(2004){Rieke}, {Young}, {Engelbracht}, {Kelly},
  {Low}, {Haller}, {Beeman}, {Gordon}, {Stansberry}, {Misselt}, {Cadien},
  {Morrison}, {Rivlis}, {Latter}, {Noriega-Crespo}, {Padgett}, {Stapelfeldt},
  {Hines}, {Egami}, {Muzerolle}, {Alonso-Herrero}, {Blaylock}, {Dole}, {Hinz},
  {Le Floc'h}, {Papovich}, {P\'erez-Gonz\'alez}, {Smith}, {Su}, {Bennett},
  {Frayer}, {Henderson}, {Lu}, {Masci}, {Pesenson}, {Rebull}, {Rho}, {Keene},
  {Stolovy}, {Wachter}, {Wheaton}, {Werner}, \&
  {Richards}}]{2004ApJS..154...25R}
{Rieke}, G.~H., {Young}, E.~T., {Engelbracht}, C.~W., {et~al.} 2004, \apjs,
  154, 25

\bibitem[{{Rivera-Ingraham} {et~al.}(2015){Rivera-Ingraham}, {Martin},
  {Polychroni}, {Schneider}, {Motte}, {Bontemps}, {Hennemann}, {Men'shchikov},
  {Nguyen Luong}, {Zavagno}, {Andre}, {Bernard}, {Di Francesco}, {Fallscheer},
  {Hill}, {K\"onyves}, {Marston}, {Pezzuto}, {Rygl}, {Spinoglio}, \&
  {White}}]{2015arXiv150702055R}
{Rivera-Ingraham}, A., {Martin}, P.~G., {Polychroni}, D., {et~al.} 2015, arXiv
  e-print

\bibitem[{{Roussel}(2013)}]{2013PASP..125.1126R}
{Roussel}, H. 2013, \pasp, 125, 1126

\bibitem[{{Roy} {et~al.}(2014){Roy}, {Andr\'e}, {Palmeirim}, {Attard},
  {K\"onyves}, {Schneider}, {Peretto}, {Men'shchikov}, {Ward-Thompson}, {Kirk},
  {Griffin}, {Marsh}, {Abergel}, {Arzoumanian}, {Benedettini}, {Hill}, {Motte},
  {Nguyen Luong}, {Pezzuto}, {Rivera-Ingraham}, {Roussel}, {Rygl}, {Spinoglio},
  {Stamatellos}, \& {White}}]{2014A&A...562A.138R}
{Roy}, A., {Andr\'e}, P., {Palmeirim}, P., {et~al.} 2014, aap, 562, A138

\bibitem[{{Russeil} {et~al.}(2013){Russeil}, {Schneider}, {Anderson},
  {Zavagno}, {Molinari}, {Persi}, {Bontemps}, {Motte}, {Ossenkopf}, {Andr\'e},
  {Arzoumanian}, {Bernard}, {Deharveng}, {Didelon}, {Di Francesco}, {Elia},
  {Hennemann}, {Hill}, {K\"onyves}, {Li}, {Martin}, {Nguyen Luong}, {Peretto},
  {Pezzuto}, {Polychroni}, {Roussel}, {Rygl}, {Spinoglio}, {Testi}, {Tig\'e},
  {Vavrek}, {Ward-Thompson}, \& {White}}]{2013A&A...554A..42R}
{Russeil}, D., {Schneider}, N., {Anderson}, L.~D., {et~al.} 2013, \aap, 554,
  A42

\bibitem[{{Rygl} {et~al.}(2014){Rygl}, {Goedhart}, {Polychroni}, {Wyrowski},
  {Motte}, {Elia}, {Nguyen Luong}, {Didelon}, {Pestalozzi}, {Benedettini},
  {Molinari}, {Andr\'e}, {Fallscheer}, {Gibb}, {Giorgio}, {Hill}, {K\"onyves},
  {Marston}, {Pezzuto}, {Rivera-Ingraham}, {Schisano}, {Schneider},
  {Spinoglio}, {Ward-Thompson}, \& {White}}]{2014MNRAS.440..427R}
{Rygl}, K.~L.~J., {Goedhart}, S., {Polychroni}, D., {et~al.} 2014, \mnras, 440,
  427

\bibitem[{{Schneider} {et~al.}(2013){Schneider}, {Andr\'e}, {K\"onyves},
  {Bontemps}, {Motte}, {Federrath}, {Ward-Thompson}, {Arzoumanian},
  {Benedettini}, {Bressert}, {Didelon}, {Di Francesco}, {Griffin}, {Hennemann},
  {Hill}, {Palmeirim}, {Pezzuto}, {Peretto}, {Roy}, {Rygl}, {Spinoglio}, \&
  {White}}]{2013ApJ...766L..17S}
{Schneider}, N., {Andr\'e}, P., {K\"onyves}, V., {et~al.} 2013, \apj, 766, L17

\bibitem[{{Schneider} {et~al.}(2015){Schneider}, {Bontemps}, {Girichidis},
  {Rayner}, {Motte}, {Andr{\'e}}, {Russeil}, {Abergel}, {Anderson},
  {Arzoumanian}, {Benedettini}, {Csengeri}, {Didelon}, {Di Francesco},
  {Griffin}, {Hill}, {Klessen}, {Ossenkopf}, {Pezzuto}, {Rivera-Ingraham},
  {Spinoglio}, {Tremblin}, \& {Zavagno}}]{2015MNRAS.453L..41S}
{Schneider}, N., {Bontemps}, S., {Girichidis}, P., {et~al.} 2015, \mnras, 453,
  41

\bibitem[{{Schneider} {et~al.}(2010){Schneider}, {Csengeri}, {Bontemps},
  {Motte}, {Simon}, {Hennebelle}, {Federrath}, \&
  {Klessen}}]{2010A&A...520A..49S}
{Schneider}, N., {Csengeri}, T., {Bontemps}, S., {et~al.} 2010, \aap, 520, A49

\bibitem[{{Schneider} {et~al.}(2012){Schneider}, {Csengeri}, {Hennemann},
  {Motte}, {Didelon}, {Federrath}, {Bontemps}, {Di Francesco}, {Arzoumanian},
  {Minier}, {Andr\'e}, {Hill}, {Zavagno}, {Nguyen Luong}, {Attard}, {Bernard},
  {Elia}, {Fallscheer}, {Griffin}, {Kirk}, {Klessen}, {K\"onyves}, {Martin},
  {Men'shchikov}, {Palmeirim}, {Peretto}, {Pestalozzi}, {Russeil}, {Sadavoy},
  {Sousbie}, {Testi}, {Tremblin}, {Ward-Thompson}, \&
  {White}}]{2012A&A...540L..11S}
{Schneider}, N., {Csengeri}, T., {Hennemann}, M., {et~al.} 2012, \aap, 540,
  L11, {Erratum:} 2013, \aap, 551, C1

\bibitem[{{Schulz}(2011)}]{2011ASPC..442..691S}
{Schulz}, B. 2011, in Astronomical Society of the Pacific Conference Series,
  Vol. 442, Astronomical Data Analysis Software and Systems XX, 691--694

\bibitem[{{Shetty} {et~al.}(2009){Shetty}, {Kauffmann}, {Schnee}, {Goodman}, \&
  {Ercolano}}]{2009ApJ...696.2234S}
{Shetty}, R., {Kauffmann}, J., {Schnee}, S., {Goodman}, A.~A., \& {Ercolano},
  B. 2009, \apj, 696, 2234

\bibitem[{{Shimmins} {et~al.}(1966{\natexlab{a}}){Shimmins}, {Clarke}, \&
  {Ekers}}]{1966AuJPh..19..649S}
{Shimmins}, A.~J., {Clarke}, M.~E., \& {Ekers}, R.~D. 1966{\natexlab{a}}, Aust.
  J. Phys., 19, 649

\bibitem[{{Shimmins} {et~al.}(1966{\natexlab{b}}){Shimmins}, {Day}, {Ekers}, \&
  {Cole}}]{1966AuJPh..19..837S}
{Shimmins}, A.~J., {Day}, G.~A., {Ekers}, R.~D., \& {Cole}, D.~J.
  1966{\natexlab{b}}, Aust. J. Phys., 19, 837

\bibitem[{{Shimmins} {et~al.}(1969){Shimmins}, {Manchester}, \&
  {Harris}}]{1969AuJPA...8....3S}
{Shimmins}, A.~J., {Manchester}, R.~N., \& {Harris}, B.~J. 1969, Aust. J. Phys.
  Astrophys. Suppl., 8, 3

\bibitem[{{SPIRE Handbook}(2014)}]{2014SPIRE......v2.5}
{SPIRE Handbook}. 2014, {Version 2.5,}, {SPIRE Consortium}

\bibitem[{{SPIRE PACS Parallel Mode Observers'
  Manual}(2011)}]{2014parallel.v2.1}
{SPIRE PACS Parallel Mode Observers' Manual}. 2011, {Version 2.1,}

\bibitem[{{Tafalla} {et~al.}(1997){Tafalla}, {Bachiller}, {Wright}, \&
  {Welch}}]{1997ApJ...474..329T}
{Tafalla}, M., {Bachiller}, R., {Wright}, M.~C.~H., \& {Welch}, W.~J. 1997,
  \apj, 474, 329

\bibitem[{{Tig{\'e}} {et~al.}(2017){Tig{\'e}}, {Motte}, {Russeil}, {Zavagno},
  {Hennemann}, {Schneider}, {Hill}, {Nguyen Luong}, {Di Francesco}, {Bontemps},
  {Louvet}, {Didelon}, {K{\"o}nyves}, {Andr{\'e}}, {Leuleu}, {Bardagi},
  {Anderson}, {Arzoumanian}, {Benedettini}, {Bernard}, {Elia}, {Figueira},
  {Kirk}, {Martin}, {Minier}, {Molinari}, {Nony}, {Persi}, {Pezzuto},
  {Polychroni}, {Rayner}, {Rivera-Ingraham}, {Roussel}, {Rygl}, {Spinoglio}, \&
  {White}}]{2017A&A...602A..77T}
{Tig{\'e}}, J., {Motte}, F., {Russeil}, D., {et~al.} 2017, \aap, 602, A77

\bibitem[{{Trevi\~no-Morales}(2016)}]{2016trevino....phdT}
{Trevi\~no-Morales}, S.~P. 2016, PhD thesis, {University of Granada}

\bibitem[{{Trevi\~no-Morales} {et~al.}(2016){Trevi\~no-Morales}, {Fuente},
  {S\'anchez-Monge}, {Pilleri}, {Goicoechea}, {Ossenkopf-Okada}, {Roueff},
  {Rizzo}, {Gerin}, {Bern\'e}, {Cernicharo}, {G\'onzalez-Garc\'ia}, {Kramer},
  {Garc\'ia-Burillo}, \& {Pety}}]{2016arXiv160706265T}
{Trevi\~no-Morales}, S.~P., {Fuente}, A., {S\'anchez-Monge}, A., {et~al.} 2016,
  \aap, 593, L12

\bibitem[{{Trevi\~no-Morales} {et~al.}(2014){Trevi\~no-Morales}, {Pilleri},
  {Fuente}, {Kramer}, {Roueff}, {Gonz\'alez-Garc\'ia}, {Cernicharo}, {Gerin},
  {Goicoechea}, {Pety}, {Bern\'e}, {Ossenkopf}, {Ginard}, {Garc\'ia-Burillo},
  {Rizzo}, \& {Viti}}]{2014A&A...569A..19T}
{Trevi\~no-Morales}, S.~P., {Pilleri}, P., {Fuente}, A., {et~al.} 2014, \aap,
  569, A19

\bibitem[{{Van den Bergh}(1966)}]{1966AJ.....71..990V}
{Van den Bergh}, S. 1966, \aj, 71, 990

\bibitem[{{Werner} {et~al.}(2004){Werner}, {Roellig}, {Low}, {Rieke}, {Rieke},
  {Hoffmann}, {Young}, {Houck}, {Brandl}, {Fazio}, {Hora}, {Gehrz}, {Helou},
  {Soifer}, {Stauffer}, {Keene}, {Eisenhardt}, {Gallagher}, {Gautier}, {Irace},
  {Lawrence}, {Simmons}, {Van Cleve}, {Jura}, {Wright}, \&
  {Cruikshank}}]{2004ApJS..154....1W}
{Werner}, M.~W., {Roellig}, T.~L., {Low}, F.~J., {et~al.} 2004, \apjs, 154, 1

\bibitem[{{White} {et~al.}(1979){White}, {Watt}, {Cronin}, \& {van
  Vliet}}]{1979MNRAS.186..107W}
{White}, G.~J., {Watt}, G.~D., {Cronin}, N.~J., \& {van Vliet}, A.~H.~F. 1979,
  \mnras, 186, 107

\bibitem[{{White}(2016)}]{2015white........1W}
{White}, G.~J. e.~a. 2016, in prep.

\bibitem[{{Wolf} {et~al.}(1990){Wolf}, {Lada}, \&
  {Bally}}]{1990AJ....100.1892W}
{Wolf}, G.~A., {Lada}, C.~J., \& {Bally}, J. 1990, \aj, 100, 1892

\bibitem[{{Wright} {et~al.}(2010){Wright}, {Eisenhardt}, {Mainzer}, {Ressler},
  {Cutri}, {Jarrett}, {Kirkpatrick}, {Padgett}, {McMillan}, {Skrutskie},
  {Stanford}, {Cohen}, {Walker}, {Mather}, {Leisawitz}, {Gautier}, {McLean},
  {Benford}, {Lonsdale}, {Blain}, {Mendez}, {Irace}, {Duval}, {Liu}, {Royer},
  {Heinrichsen}, {Howard}, {Shannon}, {Kendall}, {Walsh}, {Larsen}, {Cardon},
  {Schick}, {Schwalm}, {Abid}, {Fabinsky}, {Naes}, \&
  {Tsai}}]{2010AJ....140.1868W}
{Wright}, E.~L., {Eisenhardt}, P.~R.~M., {Mainzer}, A.~K., {et~al.} 2010, \aj,
  140, 1868

\end{thebibliography}

  \appendix

  \section{Individual wavelength maps}
    \label{app:Pics}

\begin{figure}[htb]
      \mbox{\resizebox{\hsize}{!}{\includegraphics[trim=30 0 55 0, clip=true]{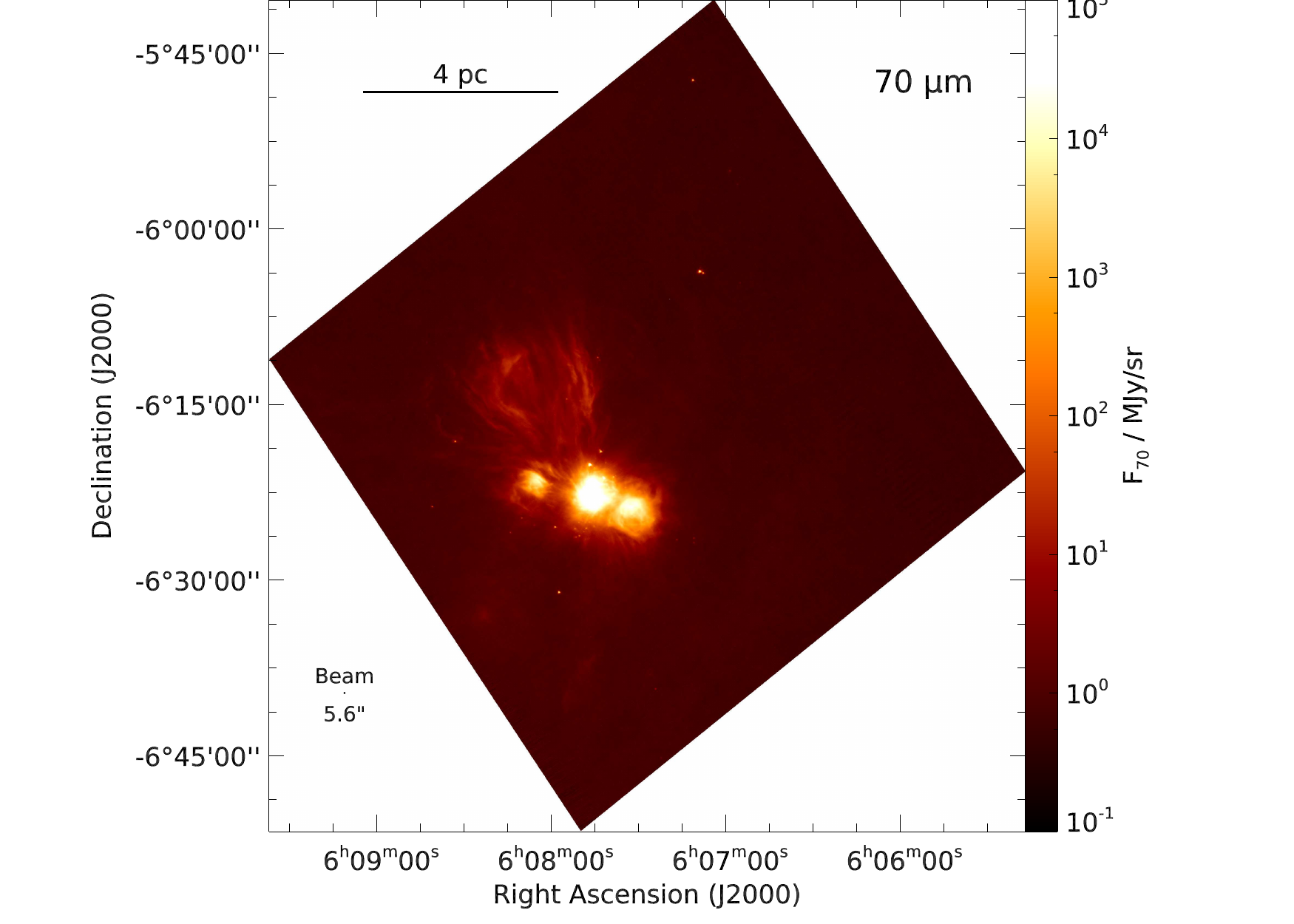}}}
      \caption{Monoceros R2 as viewed by \her PACS 70\mum.
        The image is 1.15\degr$\times$1.3\degr (16.7$\times$19.1\,pc) in size.
        The diffraction-limited instrumental beam (\bbeam\asec) is shown to the lower left of the image.
        }
      \label{fig:app_MonR2_70}
    \end{figure}

\begin{figure}[htb]
      \mbox{\resizebox{\hsize}{!}{\includegraphics[trim=30 0 55 0, clip=true]{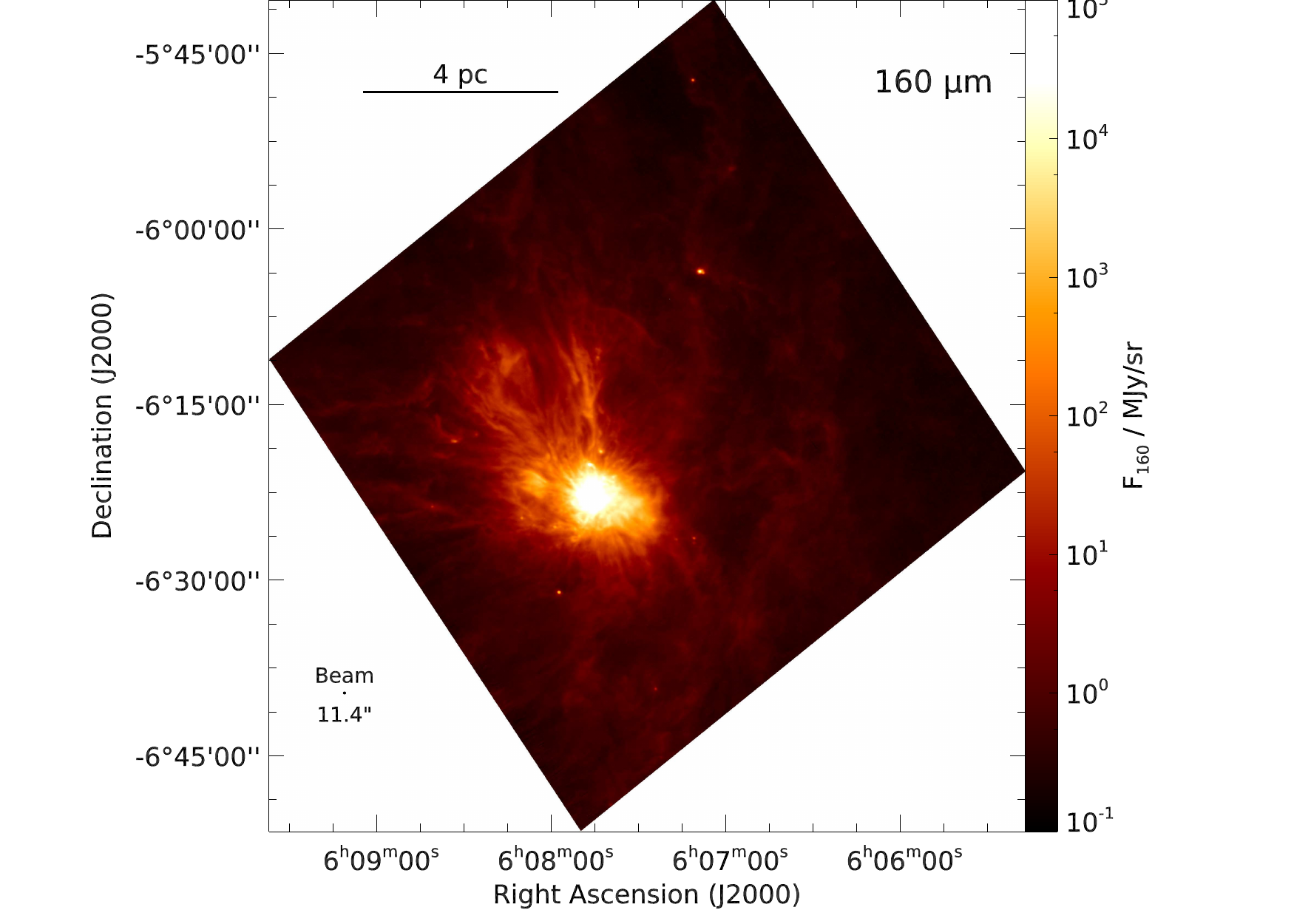}}}
      \caption{Monoceros R2 as viewed by \her PACS 160\mum.
        The image is the same area as covered by Figure~\ref{fig:app_MonR2_70}.
        The instrumental beam (\rbeam\asec) is shown to the lower left of the image.
        }
      \label{fig:app_MonR2_160}
    \end{figure}

\begin{figure}[htb]
      \mbox{\resizebox{\hsize}{!}{\includegraphics[trim=30 0 30 0, clip=true]{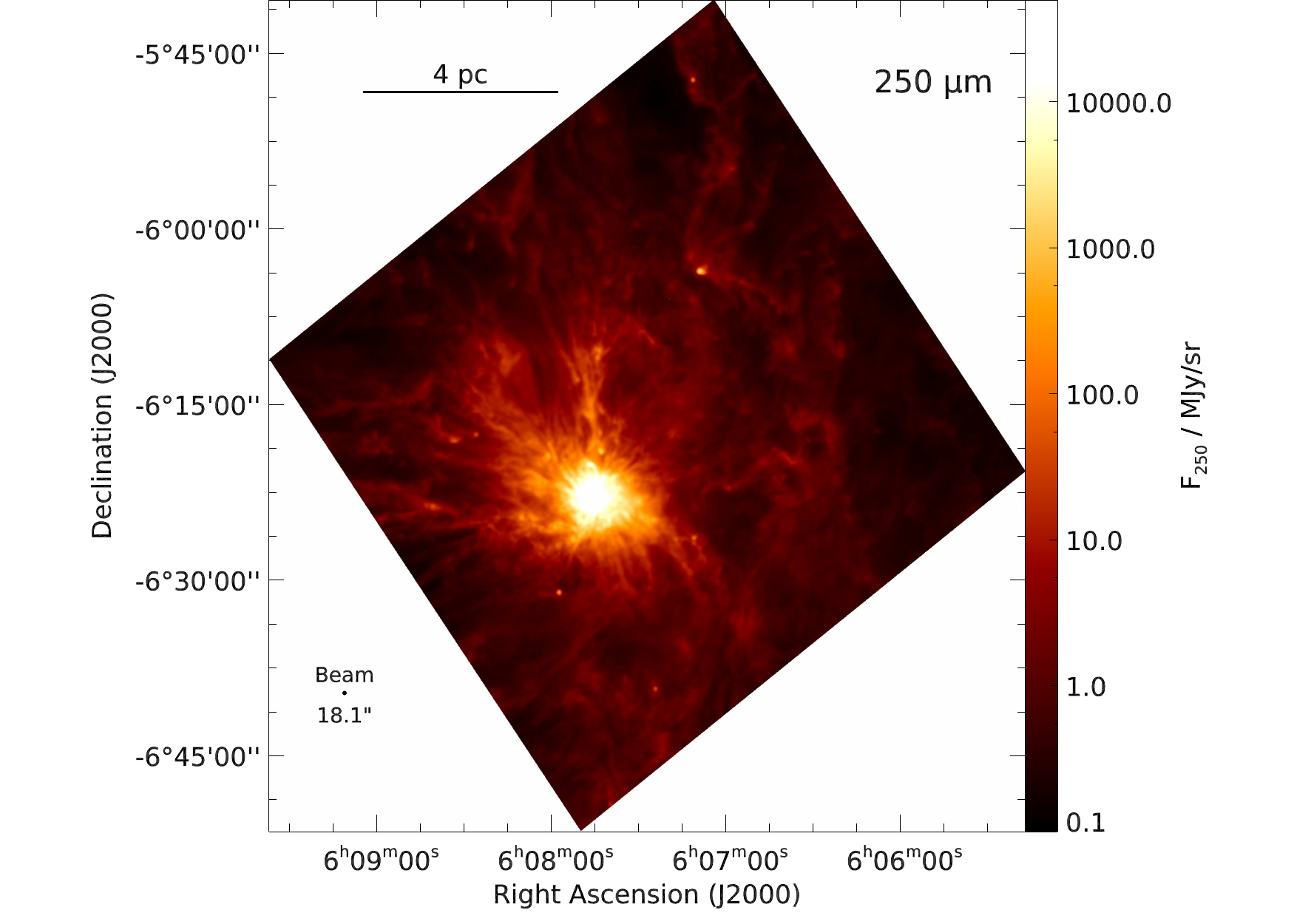}}}
      \caption{Monoceros R2 as viewed by \her SPIRE 250\mum.
        The image is the same area as covered by Figure~\ref{fig:app_MonR2_70}
        (there is a $\sim 20'$ offset between the SPIRE and PACS images).
        The instrumental beam (\sbeam\asec) is shown to the lower left of the image.
        }
      \label{fig:app_MonR2_250}
    \end{figure}

\begin{figure}[htb]
      \mbox{\resizebox{\hsize}{!}{\includegraphics[trim=30 0 30 0, clip=true]{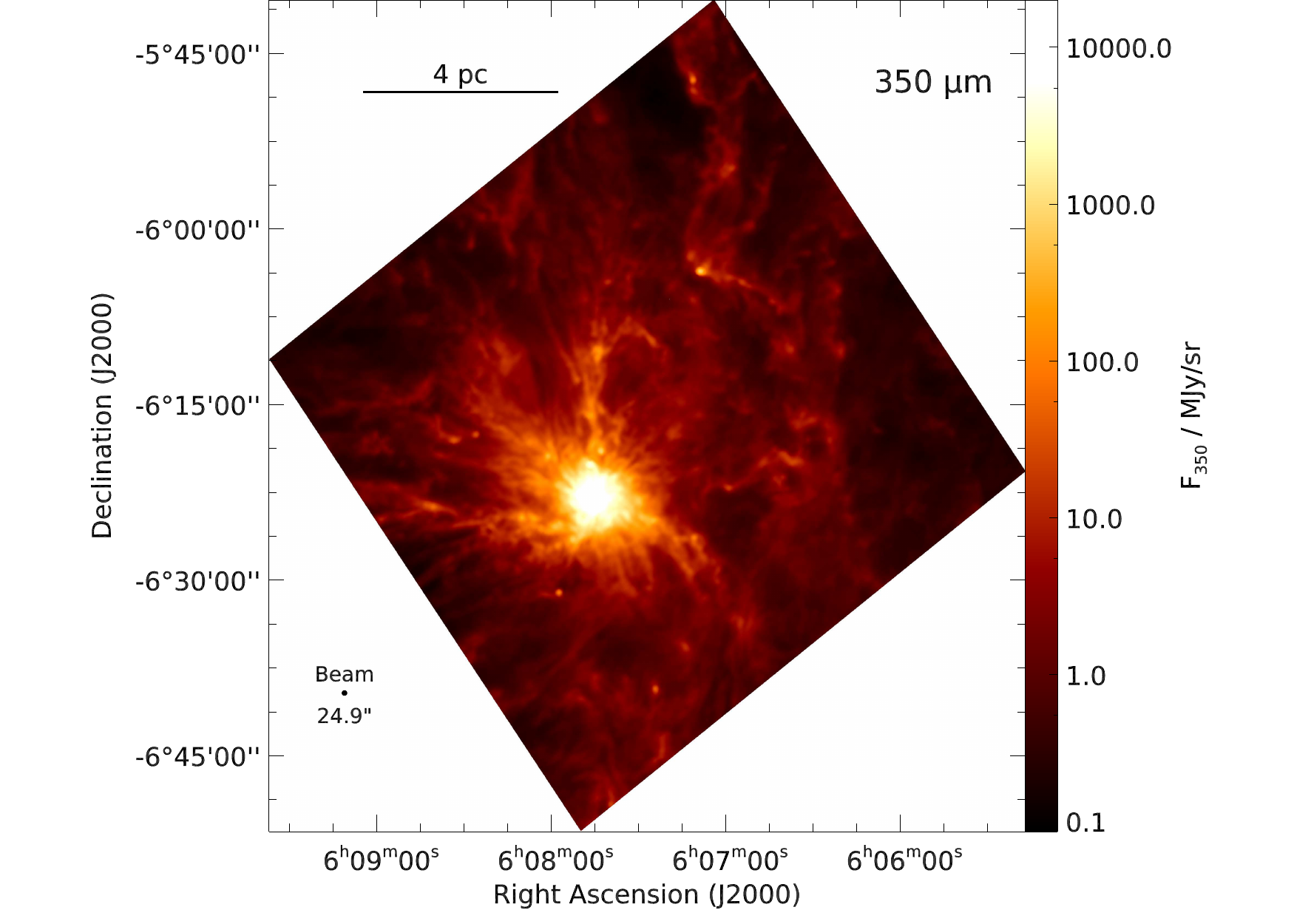}}}
      \caption{Monoceros R2 as viewed by \her SPIRE 350\mum.
        Otherwise, the same as for Figure~\ref{fig:app_MonR2_70}.
        The instrumental beam (\mbeam\asec) is shown to the lower left of the image.
        }
      \label{fig:app_MonR2_350}
    \end{figure}

\begin{figure}[htb]
      \mbox{\resizebox{\hsize}{!}{\includegraphics[trim=30 0 30 0, clip=true]{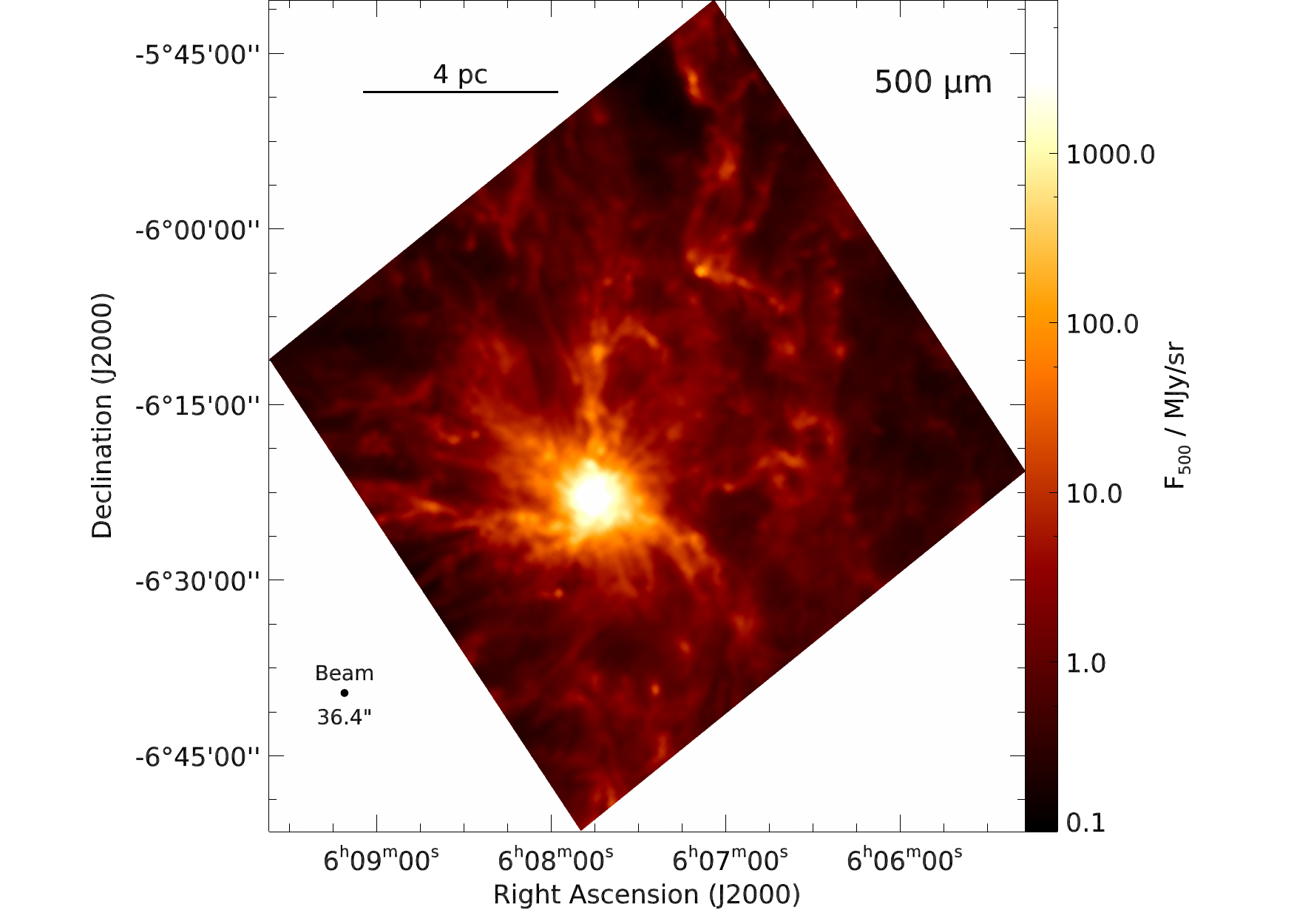}}}
      \caption{Monoceros R2 as viewed by \her SPIRE 500\mum.
        Otherwise, the same as for Figure~\ref{fig:app_MonR2_70}.
        The instrumental beam (\lbeam\asec) is shown to the lower left of the image.
        }
      \label{fig:app_MonR2_500}
    \end{figure}

\begin{figure}[htb]
      \mbox{\resizebox{\hsize}{!}{\includegraphics[trim=30 0 30 0, clip=true]{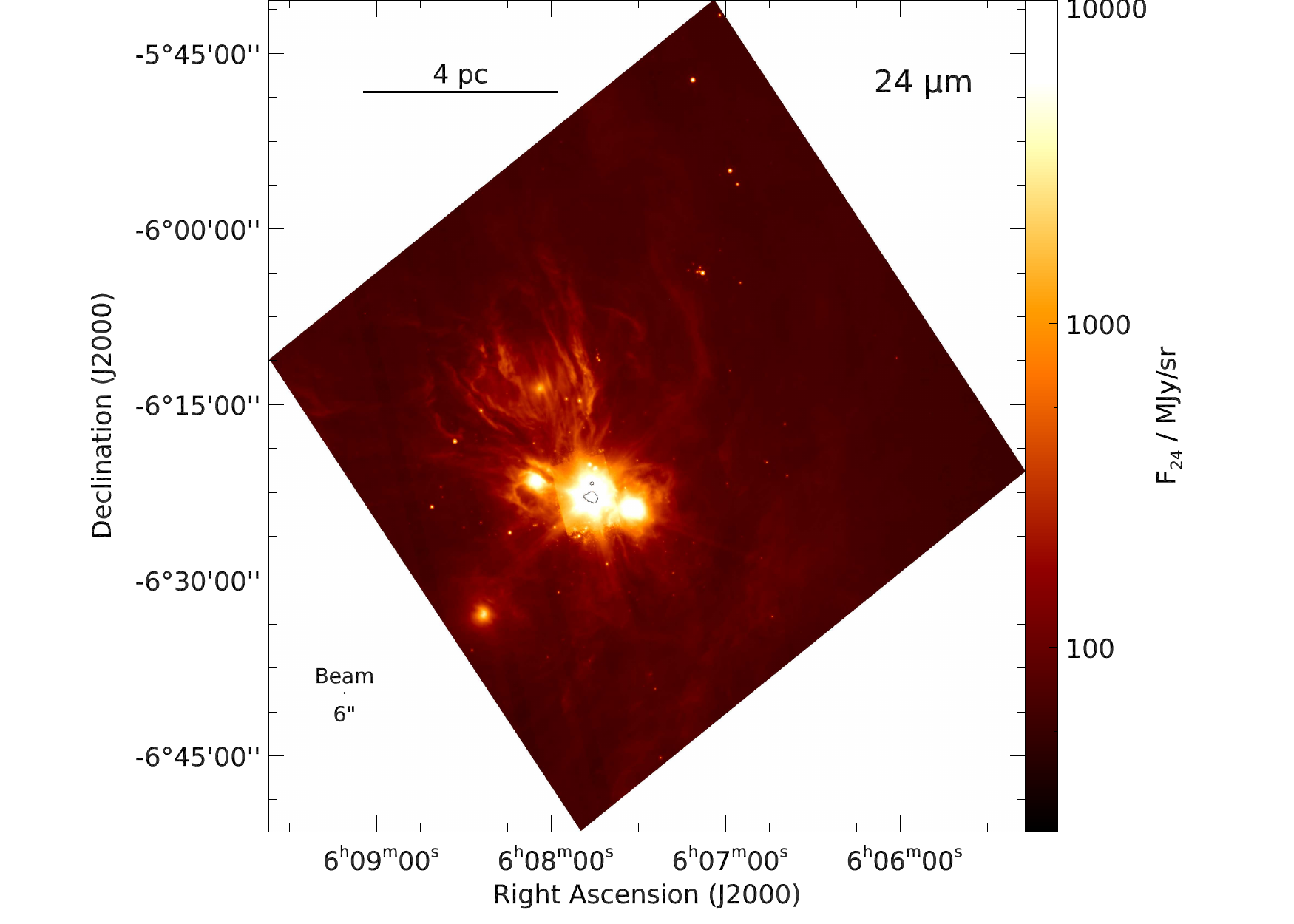}}}
      \caption{Monoceros R2 as viewed by MIPS, 24\mum.
        Otherwise, the same as for Figure~\ref{fig:app_MonR2_70}.
        The instrumental beam (6\asec) is shown to the lower left of the image.
        }
      \label{fig:app_MonR2_24}
    \end{figure}

\begin{figure}[htb]
      \mbox{\resizebox{\hsize}{!}{\includegraphics[trim=30 0 30 0, clip=true]{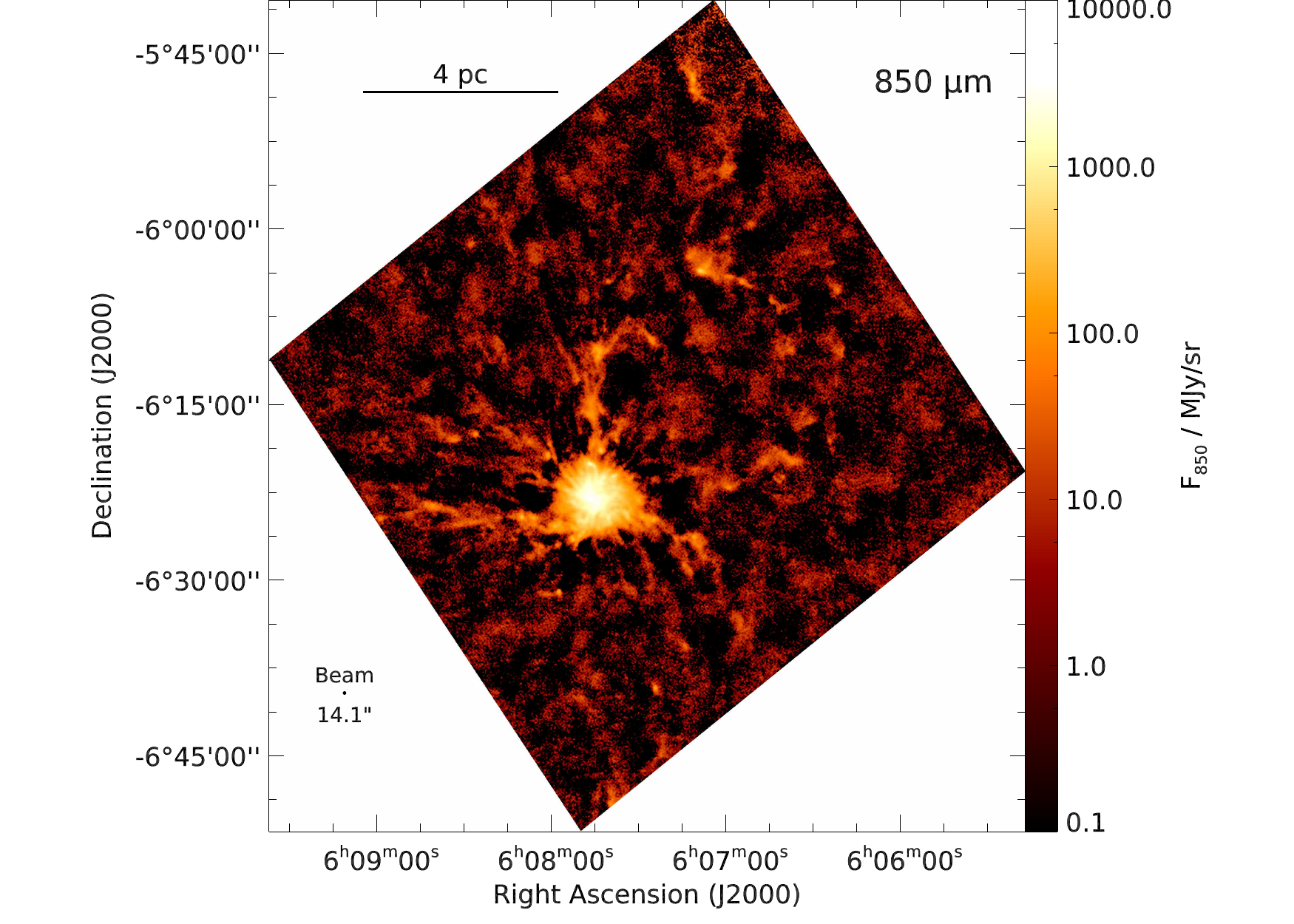}}}
      \caption{Monoceros R2 as viewed by SCUBA-2, 850\mum.
        Otherwise, the same as for Figure~\ref{fig:app_MonR2_70}.
        The instrumental beam (14.1\asec) is shown to the lower left of the image.
        }
      \label{fig:app_MonR2_850}
    \end{figure}

    The maps for all wavelengths used in the \getso extraction are given here:
    70\mum (Figure~\ref{fig:app_MonR2_70});
    160\mum (Figure~\ref{fig:app_MonR2_160});
    250\mum (Figure~\ref{fig:app_MonR2_250});
    350\mum (Figure~\ref{fig:app_MonR2_350});
    500\mum (Figure~\ref{fig:app_MonR2_500});
    MIPS 24\mum (Figure~\ref{fig:app_MonR2_24});
    and SCUBA-2 850\mum (Figure~\ref{fig:app_MonR2_850}).

\section{The \getso routine}
  \label{app:getso}

The \getso routine was designed by the SAG-3 consortium
to provide the best source extractions possible.
It works on multiple wavelengths simultaneously,
disentangles sources based on their spatial scales,
and  comes with a built-in filament-extraction routine (\textit{getfilaments}).
Full details of the routine are given in
\citet{2012A&A...542A..81M}, which gives an overview of the routine,
and \citet{2013A&A...560A..63M}, which looks in detail at the filament extraction.
The routine requires the use of three external libraries:
SWarp \citep{2002ASPC..281..228B},
CFITSIO \citep{1999ASPC..172..487P},
and WCSTools \citep{2002ASPC..281..169M}.

A basic overview of the routine is given here,
while more in-depth descriptions of the individual
subroutines appear in the following subsections.
Initially (monochromatic extraction),
\getso is run on each wavelength map separately:
the maps are decomposed into their separate spatial scales
and cleaned of background noise; sources are detected and measured.
This initial extraction is used to find source footprints
(the closest ellipse fitting to the source shape),
which will be used in a later extraction.
The spatial scale maps are now combined across all wavelengths
(combined extraction) and sources are detected and measured in the combined maps.
Using the source measurements from the previous steps,
extended background in the original maps is ``flattened''
to allow for better detection,
and the decomposition, cleaning, combination, detection,
and measurement subroutines are run again on these new maps
(final extraction).

\subsection*{Preparation}
Prior to the start of the routine, all input maps
must be prepared using the \textit{prepareobs} routine
(packaged with \getso, but not run as part of the main routine).
This resamples all images to the same pixel scale
(generally the finest pixel scale is used,
which in this paper was 1.4\as, as the run included 70\mum PACS data),
and converts the units to MJy/sr.
Two copies of each map are made:
the ``detection'' map, which is smoothed very slightly
(by about one third of the wavelength's observational beam),
and the ``measurement'' map, which is left unchanged.

Certain input parameters may also be chosen here;
for example, individual maps may be declared ``measurement-only'',
meaning that the maps are not used for source detection.
This can be used to include MIR and SMM maps, without filling the catalogue
with sources detected only at these wavelengths.

\subsection*{Decomposition}
The detection maps are decomposed into about
a hundred separate spatial scale maps.
The first scale map follows the equation
$\mathcal{I}_1 = \mathcal{I}_\mathrm{D} - \mathcal{G}*\mathcal{I}_\mathrm{D}$,
where $\mathcal{I}_\mathrm{D}$ is the detection map
and $\mathcal{G}$ is a smoothing Gaussian.
Larger-scale maps are formed using the smoothed map
($\mathcal{G}*\mathcal{I}_\mathrm{D}$) in place of the detection map,
and a slightly larger Gaussian.
The spatial scales used vary between
a few arcseconds and several thousand arcseconds, if the maps allow it.
The scale maps are not normalised, and so the original detection
map can be recreated by summing them together.

\subsection*{Cleaning}
To clean background noise from the spatial scale maps,
areas of high emission must first be masked.
Initially, all regions higher than 6$\sigma$ are masked,
and then a new value of $\sigma$ is found from the non-masked
area, which is used to mask out regions above the new 6$\sigma$
(we note that the exact factor of $\sigma$ used is generally 6,
but may differ depending on the scale map's skewness and kurtosis).
This process is iterated until there is no more change in either
the mask or the $\sigma$ value, at which point, 
every pixel below the new 6$\sigma$ value
(called the noise threshold) is set to zero.
Filaments are also extracted in this step, but the exact process
is covered in the following subsection.

\subsection*{Filament extraction}
During the decomposition phase, objects are split up according to their
spatial scales, and so extended objects will be seen only at larger scales,
while smaller objects will only be seen at smaller scales.
Filaments, however, will be assigned to scales close to their width,
rather than their length.
Consequently, large regions (the exact value depends on several variables,
but is generally several hundred times the beam size, at the smallest scales,
and much larger at larger scales) of connected pixels above the $\sigma$ value
in the smaller scale maps are likely to be due to filaments.
The filaments are extracted from the images, and sources within them
are detected by their small sizes and relatively high intensities,
and are removed from the filaments.
The filaments are then reconstructed across all spatial scales,
and are characterised by the creation of filament skeleton maps.

\subsection*{Combination}
The maps for each scale are summed across the wavelengths,
creating a set of wavelength-independent spatial scale maps.
We note that since the spatial scales have been separated,
the wavelengths' differing resolutions is not an issue.
Even so, there will not be much useful data from a scale smaller
than about one third of the wavelength's resolution,
and so such scales are not included in the summation.
Prior to summation, the maps are normalised,
so that weaker sources are not lost
(this will not be a problem, since these images are
only used for source detection, not for measurement).
This section of the routine is not performed in the first (monochromatic)
extraction, which is performed on all wavelengths separately.

\subsection*{Detection}
To detect sources in the cleaned scale maps, the routine
will select multiple values between the maximum intensity
and the noise threshold.
Pixels above these values are either used to make new sources,
or are added to existing (adjacent) ones.
The sources are correlated across the scale maps,
meaning that each source will have a scale range over which it appears.
This is used to distinguish overlapping sources with different spatial scales.
At each scale, the source's ``detection significance'' is measured;
this is equal to the maximum intensity at this scale
divided by the noise threshold,
and is used to determine how well-detected any source is.
Elliptical source footprints are also determined during this stage.

\subsection*{Measurement}
Measurement of source properties is performed on the ``measurement maps'',
which are essentially identical to the input maps.
Filaments are removed from the maps, and non-filamentary
background is estimated, using the source footprints measured above,
and also removed.
Sources are deblended by assuming a near-Gaussian profile
(with stronger, and thus more realistic, wings;
note that the profile is only used to allocate flux by deblending).
If the two sources are within one third of a beam size of one another,
and their sizes are within a factor of two,
they are assumed to be the same source and are thus combined.
The properties measured include peak intensity (the value at the peak, in Jy/beam),
total flux density (the integral over the whole source, in Jy),
and source shape (major and minor axes, and axis angle);
the final output is a catalogue listing these values
at each wavelength for every source.

\subsection*{Flattening}
Before the final extraction, a more robust form of background-subtraction
(``flattening'') is employed, in order to account for the problems raised by
the variable Galactic backgrounds present in most \her images.
The first two extractions (monochromatic and combined) were used to find
and characterise as many sources as possible, which are then removed
from the maps to create clean background maps.
This is used to create a map of local (3\,$\times$\,3 pixels) standard deviation,
which is smoothed, first by median filtering, and then by convolution
with a Gaussian three times the maximum source size, forming the scaling map,
which is used to scale the detection map.
This means that highly variable parts of the map will be deemphasised,
when compared with the regions with flatter background.
We note that this only affects the detection maps;
the measurement maps are not changed in any way.

\subsection*{Advantages and disadvantages}
The \getso routine works on all wavelengths at once, and at multiple
spatial scales at once, which gives it a great advantage over other source finders.
It deblends overlapping sources and utilises background flattening
to find the best sources possible.
Its scale-decomposition allows it to detect sources at both
large and small scales, something many other source finders do not do.
It is not without disadvantages, most of which stem from the sheer
size of the program.
For example, many other source finders can run over the five \her
wavelengths in under an hour, giving one or two output files for each of the five extractions;
on the same maps, \getso can run for a week, giving several GB of output.
Despite this, the advantages in enhanced detection will generally
outweigh the disadvantages in ease (and speed) of use.
Discussions of various source finders are given in \citet{2014prpl.conf...27A}
and \citet{2015rayner.....phdR}.
A look at the completeness of \getso in Mon~R2 is given in Appendix~\ref{app:comp}.

\section{The completeness of the \getso extraction}
  \label{app:comp}

    In order to test the completeness of the catalogue,
    the \getso program was run on a field of simulated data
    based on the original Mon~R2 maps,
    with additional sources superimposed on them.
    Though there are disadvantages to adding the sources
    to the original map, as the real sources on the map will
    add confusion to the output, the alternative
    (to add simulated sources to a source-free background map)
    may also have problems caused by artefacts from the source-removal process.
    The main advantage of using the original map is
    that the \getso output for the original map alone
    is already available, and any changes to this output
    will thus be due to the inserted sources.
    
    The basic technique was based on those of 
    \citet{2016MNRAS.459..342M} and \citet{2015A&A...584A..91K},
    albeit with modifications made due to the
    differences between the regions in question
    (Mon~R2 is at 830\,pc, with a very dense central hub,
    while Taurus and Aquila are at 140\,pc and 260\,pc respectively,
    without any regions comparable to the Mon~R2 hub).
    A total of 440 model sources were input into the maps.
    These included sources modelled as protostars (116),
    bound (186) and unbound (138) cores,
    with shapes approximating to 2D Gaussians.
    The mass range was 0.03--10\,M\msun and
    temperature and size ranges were chosen to fit with the
    observed parameter ranges within the region.
    The fluxes at each wavelength were calculated using the
    SED in Equation~\ref{equ:sed},
    and their positions on the map were chosen randomly,
    with protostars and bound sources placed
    in regions with column densities over
    $3 \times 10^{21}$cm$^{-2}\!$, and unbound sources
    (which do not necessarily trace the highest column densities)
    placed in regions with a 250\mum intensity over 57\,MJy/sr.
    
    The maps used for this procedure were all \her (PACS and SPIRE)
    maps, along with the 850\mum SCUBA-2 map, and the column density map.
    The 24\mum MIPS map was not included, as this was only
    used for luminosity calculations.
    For simplicity, the column density profiles of the sources
    were added directly into the column density map,
    rather than having the column density map recreated from
    the new flux maps.
    
\begin{figure}[htb]
      \mbox{\resizebox{\hsize}{!}{\includegraphics[trim=0 0 0 0,
        clip=true]{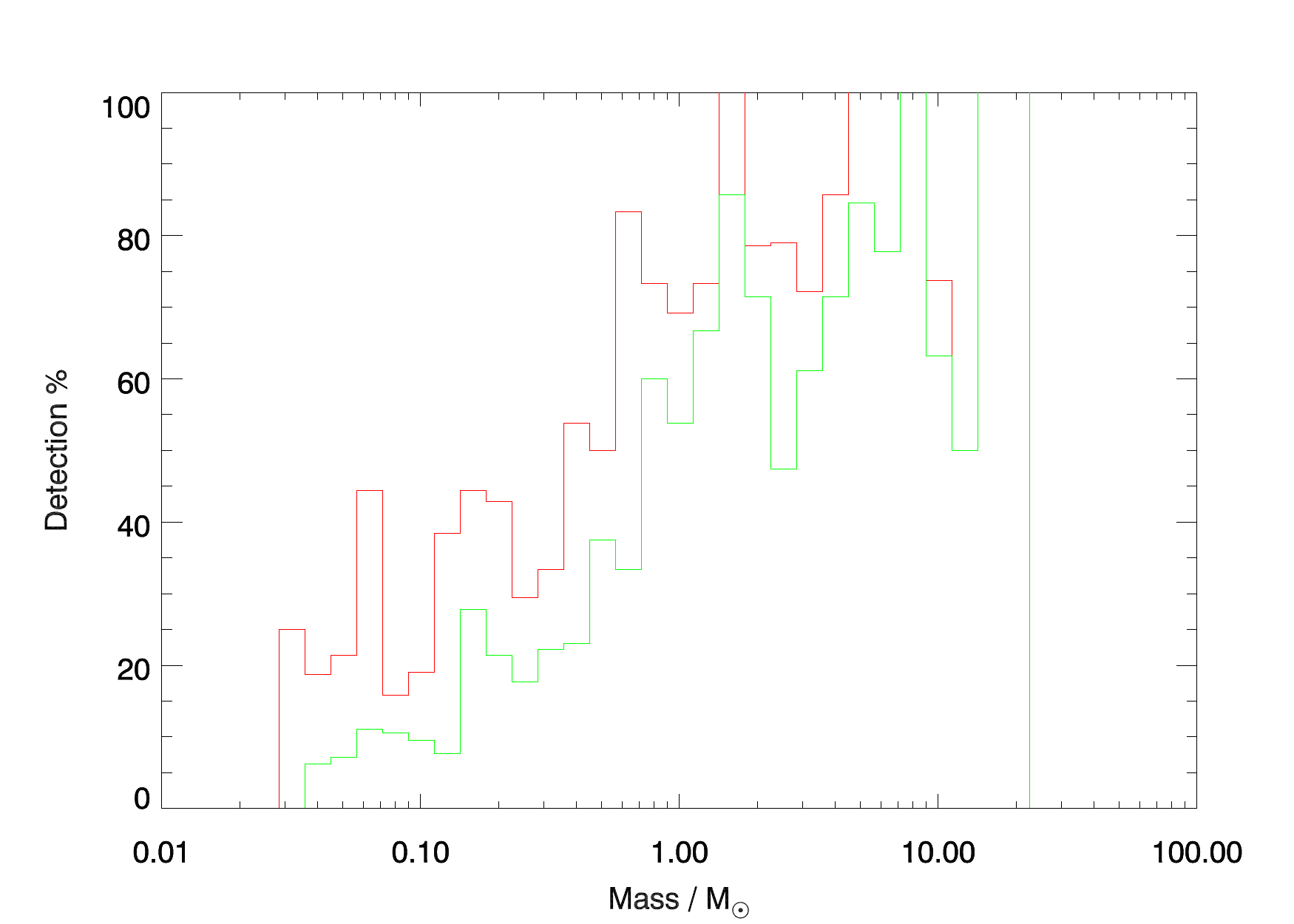}}}
      \caption{Completeness of sources.
      The red line shows the percentage of sources detected by \getso;
      the green line shows the percentage of sources added to the catalogue.
      }
      \label{fig:completeness}
    \end{figure}

    The \getso routine was run (see Section~\ref{app:getso}
    for details), and the source selection and cataloguing
    process was run on the output.
    Model sources were counted as detected when their centres
    lay within \rbeam\asec (the 160\mum beam size) or
    30\% of the model source size  (whichever was larger)
    of the centre of a detected source.
    In total, 40\% (175 of 440) model sources were detected,
    about 60\% of those above 1\,M\msun,
    but only 20\% of those below this value.
    Consequently, setting a completeness limit of 1\,M\msun
    would mean that the majority of sources had been detected.
    The percentage of detected sources against mass is given in
    Figure~\ref{fig:completeness}.
    A detected source also had a  $\gtrsim$75\% chance
    to be assigned the correct source type.

    The most likely reason for the low completeness,
    even at high masses, is the large sizes given to some of these sources
    by the model core creation routine.
    Although these were not much larger than the detected sources
    (the largest model source was 0.35\,pc across, compared to 0.30\,pc),
    this did have the consequence that even sources with masses over 20\,M\msun
    could have densities similar to those of much lower mass,
    and would thus be much harder to detect.
    If only sources with the smallest sizes (under 0.03\,pc) are considered,
    the completeness goes up to 80\% detection above 1\,M\msun,
    and 100\% detection above 4\,M\msun, but this is not
    a full representative sample of all sources in the region.
    
\begin{figure}[htb]
      \mbox{\resizebox{\hsize}{!}{\includegraphics[trim=0 0 0 0,
        clip=true]{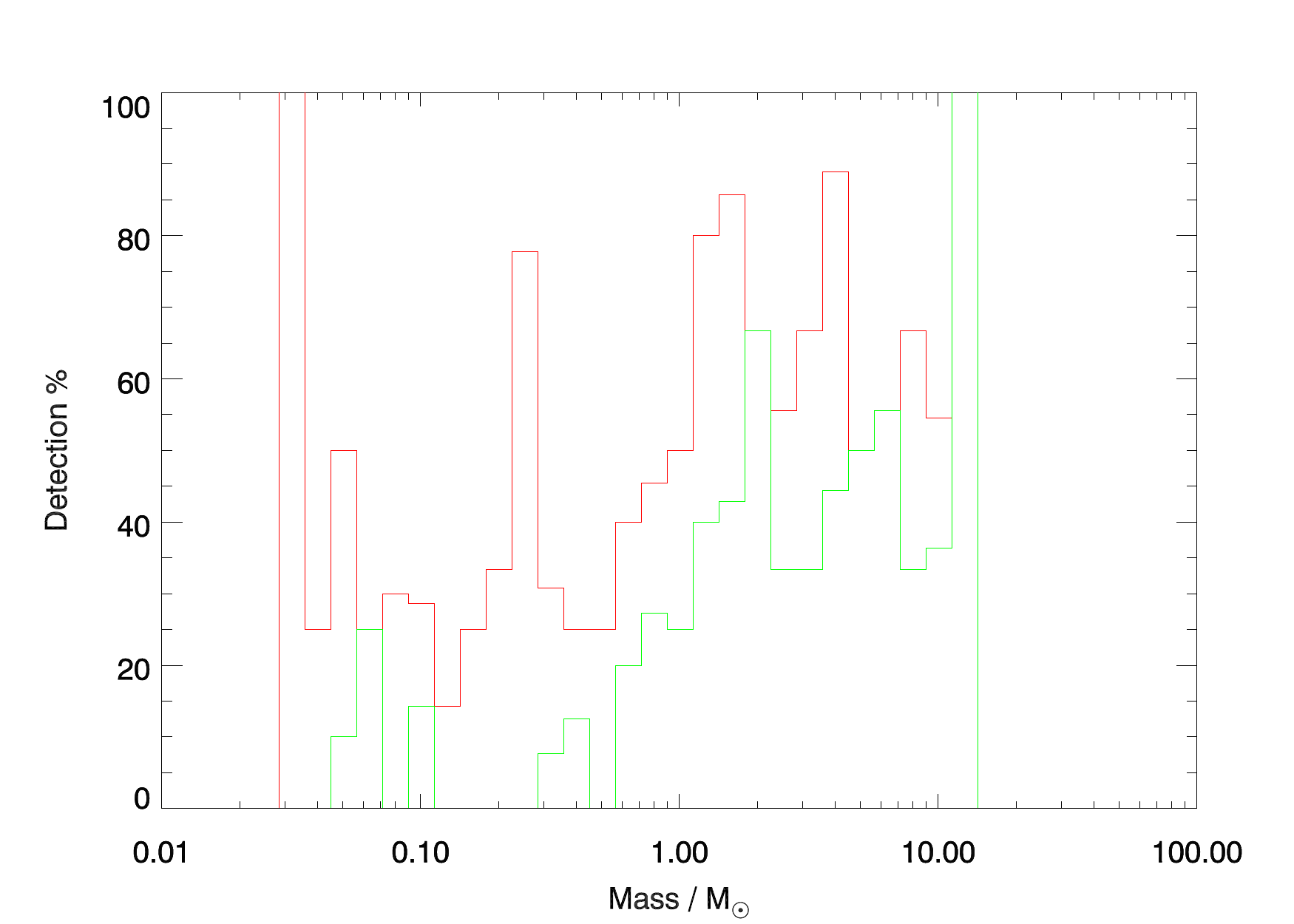}}}
      \caption{Completeness of sources in the hub region;
			otherwise, as for Figure~\ref{fig:completeness}.
      }
      \label{fig:compcore}
    \end{figure}

    A secondary completeness test was applied solely to the
    central hub, due to its higher confusion and average
    column density.
    A further five separate runs placed a total of 212 model sources
    into the hub (89 protostellar, 70 bound, and 53 unbound),
    with mass and size ranges equal to those of the original
    run, and a slight increase in the lower limit of the
    temperature range, to account for the higher temperatures
    Other than the area covered, and the number of sources,
    each of these runs was identical to the original run over the full region.
    In total, only 33\% of model sources with masses above
    1\,M\msun (or 20\% of all model sources) were detected,
    half as many proportionally as over the full maps.
    In addition, a dependence on size was also present,
    as 79\% of detected model sources were under 20\asec.
		The detections against mass for the hub region are shown in Figure~\ref{fig:compcore}.
    
    Overall, the completeness analysis shows that in the
    more quiescent parts of Mon~R2, more than 60\% of sources above 1\,M\msun
    will be detected robustly.
    This percentage depends strongly on the size of the object in question,
    and increases to 90\% when dealing strictly with objects under 0.035\,pc in size
    (and 100\% of for a mass limit of 4\,M\msun),
    while only 50\% of more extended sources are detected (above 1\,M\msun).
    This is seen to an even greater extent within the central hub,
    where \getso only robustly detects $\sim$30\% of sources.
    As such while it is not feasible to define a mass completeness limit
    for the larger, less-dense sources,
    a mass limit of 1\,M\msun for protostars and other compact sources
    under $\sim$0.035\,pc in size is suggested.

  \section{Catalogue}
    \label{app:Table}

\begin{table*}[h!]
\centering
\begin{tabu}{| r || c | c | c |}
\hline
\# & Name & RA$_{2000}$ & Dec$_{2000}$     \\
\hline
\hline
\rule{0pt}{3ex}$\!\!$
 1 & HOBYS J060740.3 $-$062447 & \hms{06}{07}{40.28} & $-$\dms{06}{24}{46.9} \\ 
 2 & HOBYS J060738.4 $-$062244 & \hms{06}{07}{38.44} & $-$\dms{06}{22}{44.3} \\ 
 3 & HOBYS J060746.1 $-$062312 & \hms{06}{07}{46.12} & $-$\dms{06}{23}{12.0} \\ 
 4 & HOBYS J060751.2 $-$062206 & \hms{06}{07}{51.20} & $-$\dms{06}{22}{06.4} \\ 
 5 & HOBYS J060752.2 $-$062327 & \hms{06}{07}{52.19} & $-$\dms{06}{23}{27.4} \\ 
 6 & HOBYS J060747.5 $-$062203 & \hms{06}{07}{47.51} & $-$\dms{06}{22}{02.7} \\ 
 7 & HOBYS J060743.6 $-$061028 & \hms{06}{07}{43.64} & $-$\dms{06}{10}{28.2} \\ 
 8 & HOBYS J060757.2 $-$062158 & \hms{06}{07}{57.23} & $-$\dms{06}{21}{57.6} \\ 
 9 & HOBYS J060706.3 $-$060904 & \hms{06}{07}{06.32} & $-$\dms{06}{09}{04.3} \\ 
10 & HOBYS J060747.9 $-$062501 & \hms{06}{07}{47.92} & $-$\dms{06}{25}{00.6} \\ 
11 & HOBYS J060621.9 $-$060519 & \hms{06}{06}{21.88} & $-$\dms{06}{05}{19.0} \\ 
\hline
\end{tabu}
\caption{
  Global parameters.
  This contains:
  source name; and
  the source Right Ascension (RA) and Declination (Dec) in J2000 coordinates.
}
\label{tab:base}
\end{table*}

\begin{table*}[h!]
\centering
\begin{tabu}{| r || c | c | c | c | c |}
\hline
\# & $F_{70\mathrm{P}}\pm\sigma_{70\mathrm{P}}$&
  $F_{70\mathrm{T}}\pm\sigma_{70\mathrm{T}}$&
  $A_{70}$ & $B_{70}$ & $\Theta_{70}$ \\
 & (Jy\,bm$^{-1}$) & (Jy) & (arcsec) & (arcsec) & (degrees) \\
\hline
\hline
\rule{0pt}{3ex}$\!\!$
 1 &    0.1\,$\pm$\,0.2    &   1.1\,$\pm$\,0.5   & 31.2 & 14.1 & 106.9 \\ 
 2 &    1.4\,$\pm$\,0.6    &     9\,$\pm$\,1     & 19.6 & 10.5 & 119.5 \\ 
 3 &    600\,$\pm$\,50     &  1730\,$\pm$\,80    & 14.3 &  5.9 &  69.6 \\ 
 4 &    0.9\,$\pm$\,0.6    &     6\,$\pm$\,1     & 20.6 & 11.3 &  42.2 \\ 
 5 &    0.5\,$\pm$\,0.5    &     3\,$\pm$\,1     & 22.2 &  9.4 & 116.1 \\ 
 6 &      4\,$\pm$\,4      &    14\,$\pm$\,5     & 11.4 &  6.7 & 169.6 \\ 
 7 &          ...          &  0.02\,$\pm$\,0.05  & 69.8 & 34.3 &  72.8 \\ 
 8 &          ...          &         ...         & 45.7 & 16.5 & 152.7 \\ 
 9 &  0.004\,$\pm$\,0.002  &  0.28\,$\pm$\,0.02  & 94.3 & 86.1 & 132.8 \\ 
10 &    0.2\,$\pm$\,0.2    &   2.2\,$\pm$\,0.5   & 23.4 & 12.1 & 152.9 \\ 
11 & 0.0116\,$\pm$\,0.0005 & 0.132\,$\pm$\,0.004 & 62.3 & 59.4 &  98.6 \\ 
\hline
\end{tabu}
\caption{
  70\mum parameters.
  This contains:
  $F_{70\mathrm{P}}$ and $\sigma_{70\mathrm{P}}$, the peak intensity and its uncertainty
  (left blank if the source was not detected at this wavelength);
  $F_{70\mathrm{T}}$ and $\sigma_{70\mathrm{T}}$, the total flux density and its uncertainty
  (we note that these are the direct output from \getso, and thus have not been modified
  by flux scaling);
  $A_{70}$, $B_{70}$ and $\Theta_{70}$, the major and minor FWHM axes
  and the axis orientation (degrees east of vertical).
}
\label{tab:f70}
\end{table*}

\begin{table*}[h!]
\centering
\begin{tabu}{| r || c | c | c | c | c |}
\hline
\# & $F_{160\mathrm{P}}\pm\sigma_{160\mathrm{P}}$ &
  $F_{160\mathrm{T}}\pm\sigma_{160\mathrm{T}}$ &
  $A_{160}$ & $B_{160}$ & $\Theta_{160}$ \\
 & (Jy\,bm$^{-1}$) & (Jy) & (arcsec) & (arcsec) & (degrees) \\
\hline
\hline
\rule{0pt}{3ex}$\!\!$
 1 &    5\,$\pm$\,2    &   15\,$\pm$\,3   & 21.5 & 14.7 &  27.2 \\ 
 2 &   11\,$\pm$\,4    &   26\,$\pm$\,5   & 24.3 & 11.7 & 135.8 \\ 
 3 &  770\,$\pm$\,40   & 1020\,$\pm$\,40  & 15.7 & 11.7 &  65.1 \\ 
 4 &   16\,$\pm$\,7    &   31\,$\pm$\,7   & 20.5 & 11.7 &  35.2 \\ 
 5 &   14\,$\pm$\,6    &   32\,$\pm$\,8   & 21.3 & 11.7 &  51.2 \\ 
 6 &   70\,$\pm$\,30   &  130\,$\pm$\,30  & 15.9 & 12.9 & 136.2 \\ 
 7 & 0.72\,$\pm$\,0.07 &  5.3\,$\pm$\,0.2 & 47.8 & 20.5 & 178.6 \\ 
 8 &  1.1\,$\pm$\,0.8  &    4\,$\pm$\,1   & 41.9 & 13.4 &  71.3 \\ 
 9 & 0.13\,$\pm$\,0.09 &  2.7\,$\pm$\,0.5 & 86.3 & 79.1 & 115.5 \\ 
10 &   10\,$\pm$\,3    &   26\,$\pm$\,5   & 21.2 & 12.6 & 141.7 \\ 
11 & 0.14\,$\pm$\,0.06 &  2.1\,$\pm$\,0.2 & 59.5 & 41.7 &   2.6 \\ 
\hline
\end{tabu}
\caption{
  160\mum parameters.
  As above.
}
\label{tab:f160}
\end{table*}

\begin{table*}[h!]
\centering
\begin{tabu}{| r || c | c | c | c | c |}
\hline
\# & $F_{250\mathrm{P}}\pm\sigma_{250\mathrm{P}}$ &
  $F_{250\mathrm{T}}\pm\sigma_{250\mathrm{T}}$ &
  $A_{250}$ & $B_{250}$ & $\Theta_{250}$ \\
 & (Jy\,bm$^{-1}$) & (Jy) & (arcsec) & (arcsec) & (degrees) \\
\hline
\hline
\rule{0pt}{3ex}$\!\!$
 1 &  16\,$\pm$\,5   &  25\,$\pm$\,5   & 23.9 & 19.0 &  23.7 \\ 
 2 &  13\,$\pm$\,4   &  18\,$\pm$\,4   & 24.9 & 18.2 & 122.3 \\ 
 3 & 486\,$\pm$\,9   & 654\,$\pm$\,8   & 21.5 & 18.2 &  95.1 \\ 
 4 &  18\,$\pm$\,8   &  17\,$\pm$\,7   & 19.0 & 18.2 &  41.9 \\ 
 5 &  19\,$\pm$\,7   &  24\,$\pm$\,7   & 22.6 & 18.2 &  50.8 \\ 
 6 & 100\,$\pm$\,10  & 103\,$\pm$\,9   & 18.5 & 18.2 & 107.8 \\ 
 7 & 3.1\,$\pm$\,0.5 &  12\,$\pm$\,1   & 30.5 & 27.3 &  80.0 \\ 
 8 &   4\,$\pm$\,1   &  12\,$\pm$\,2   & 38.1 & 24.6 &  45.9 \\ 
 9 & 0.5\,$\pm$\,0.2 & 5.1\,$\pm$\,0.6 & 81.9 & 56.8 &  91.0 \\ 
10 &  18\,$\pm$\,6   &  24\,$\pm$\,6   & 26.2 & 18.2 & 139.3 \\ 
11 & 0.8\,$\pm$\,0.1 & 3.2\,$\pm$\,0.3 & 42.7 & 26.8 & 169.2 \\ 
\hline
\end{tabu}
\caption{
  250\mum parameters.
  As above.
}
\label{tab:f250}
\end{table*}

\begin{table*}[h!]
\centering
\begin{tabu}{| r || c | c | c | c | c |}
\hline
\# &  $F_{350\mathrm{P}}\pm\sigma_{350\mathrm{P}}$ &
  $F_{350\mathrm{T}}\pm\sigma_{350\mathrm{T}}$ &
  $A_{350}$ & $B_{350}$ & $\Theta_{350}$ \\
 & (Jy\,bm$^{-1}$) & (Jy) & (arcsec) & (arcsec) & (degrees) \\
\hline
\hline
\rule{0pt}{3ex}$\!\!$
 1 &  12\,$\pm$\,2   &  11\,$\pm$\,5   & 30.5 & 26.1 &  14.0 \\ 
 2 &  17\,$\pm$\,2   &  27\,$\pm$\,2   & 33.2 & 26.2 &   2.2 \\ 
 3 & 221\,$\pm$\,4   & 246\,$\pm$\,3   & 24.9 & 24.9 & 128.0 \\ 
 4 &  13\,$\pm$\,4   &  18\,$\pm$\,3   & 30.8 & 28.9 & 144.5 \\ 
 5 &  14\,$\pm$\,4   &  15\,$\pm$\,4   & 24.9 & 24.9 & 130.0 \\ 
 6 &  67\,$\pm$\,4   &  72\,$\pm$\,4   & 24.9 & 24.9 &  76.3 \\ 
 7 & 3.5\,$\pm$\,0.4 & 8.2\,$\pm$\,0.6 & 35.9 & 28.1 &  79.7 \\ 
 8 &   4\,$\pm$\,2   &   6\,$\pm$\,2   & 42.4 & 24.9 &  54.5 \\ 
 9 & 0.6\,$\pm$\,0.2 & 3.3\,$\pm$\,0.4 & 70.8 & 45.7 &  79.8 \\ 
10 &  12\,$\pm$\,4   &  17\,$\pm$\,4   & 38.0 & 24.9 & 136.2 \\ 
11 & 1.1\,$\pm$\,0.2 & 2.8\,$\pm$\,0.2 & 41.9 & 30.2 & 172.2 \\ 
\hline
\end{tabu}
\caption{
  350\mum parameters.
  As above.
}
\label{tab:f350}
\end{table*}

\begin{table*}[h!]
\centering
\begin{tabu}{| r || c | c | c | c | c |}
\hline
\# &  $F_{500\mathrm{P}}\pm\sigma_{500\mathrm{P}}$ &
  $F_{500\mathrm{T}}\pm\sigma_{500\mathrm{T}}$ &
  $A_{500}$ & $B_{500}$ & $\Theta_{500}$ \\
 & (Jy\,bm$^{-1}$) & (Jy) & (arcsec) & (arcsec) & (degrees) \\
\hline
\hline
\rule{0pt}{3ex}$\!\!$
 1 &   11\,$\pm$\,2   &   52\,$\pm$\,4   & 89.1 & 49.2 & 109.6 \\ 
 2 &   10\,$\pm$\,2   &   12\,$\pm$\,2   & 39.0 & 36.3 &  46.0 \\ 
 3 &   75\,$\pm$\,2   &   73\,$\pm$\,2   & 36.3 & 36.3 & 137.4 \\ 
 4 & 15.0\,$\pm$\,0.4 & 20.8\,$\pm$\,0.4 & 38.0 & 36.3 & 177.2 \\ 
 5 &   16\,$\pm$\,2   &   35\,$\pm$\,3   & 67.8 & 36.3 &  19.0 \\ 
 6 &   25\,$\pm$\,2   &   24\,$\pm$\,2   & 36.3 & 36.3 & 175.0 \\ 
 7 &  3.2\,$\pm$\,0.3 &  4.4\,$\pm$\,0.4 & 38.4 & 36.3 &  82.6 \\ 
 8 &  2.5\,$\pm$\,0.7 &    4\,$\pm$\,1   & 41.7 & 36.3 &  94.4 \\ 
 9 &  0.5\,$\pm$\,0.2 &  1.5\,$\pm$\,0.3 & 59.6 & 46.5 &  83.2 \\ 
10 &   16\,$\pm$\,1   &   25\,$\pm$\,1   & 56.5 & 36.3 &  74.5 \\ 
11 &  1.0\,$\pm$\,0.2 &  1.5\,$\pm$\,0.2 & 44.8 & 36.3 & 172.8 \\ 
\hline
\end{tabu}
\caption{
  500\mum parameters. As above.
}
\label{tab:f500}
\end{table*}

\begin{table*}[h!]
\centering
\begin{tabu}{| r || c | c | c | c | c |}
\hline
\# &  $F_{24\mathrm{P}}\pm\sigma_{24\mathrm{P}}$ &
  $F_{24\mathrm{T}}\pm\sigma_{24\mathrm{T}}$ &
  $A_{24}$ & $B_{24}$ & $\Theta_{24}$ \\
 & (Jy\,bm$^{-1}$) & (Jy) & (arcsec) & (arcsec) & (degrees) \\
\hline
\hline
\rule{0pt}{3ex}$\!\!$
 1 &   0.01\,$\pm$\,0.01   &  0.05\,$\pm$\,0.03  &  33.2 & 22.4 & 156.0 \\ 
 2 &   0.00\,$\pm$\,0.02   &  0.04\,$\pm$\,0.04  &  10.4 &  6.0 &  81.8 \\ 
 3 &    0.0\,$\pm$\,0.1    &   0.0\,$\pm$\,0.1   &  19.8 &  6.5 &  68.2 \\ 
 4 &          ...          &         ...         &  17.7 & 13.6 & 108.4 \\ 
 5 &          ...          &         ...         &  29.7 & 17.0 &   6.4 \\ 
 6 &          ...          &         ...         &  19.0 &  6.0 & 168.9 \\ 
 7 & 0.0001\,$\pm$\,0.0002 & 0.031\,$\pm$\,0.001 &  69.4 & 36.9 & 151.0 \\ 
 8 &  0.005\,$\pm$\,0.008  &  0.15\,$\pm$\,0.04  &  48.3 & 41.9 &  81.6 \\ 
 9 & 0.0000\,$\pm$\,0.0003 &         ...         & 100.0 & 70.8 & 148.8 \\ 
10 &          ...          &  0.00\,$\pm$\,0.01  &  25.2 &  8.1 & 172.9 \\ 
11 &          ...          &         ...         &  63.0 & 25.2 & 163.6 \\ 
\hline
\end{tabu}
\caption{
  MIPS 24\mum parameters.
  As above.
}
\label{tab:f24}
\end{table*}

\begin{table*}[h!]
\centering
\begin{tabu}{| r || c | c | c | c | c |}
\hline
\# &  $F_{850\mathrm{P}}\pm\sigma_{850\mathrm{P}}$ &
  $F_{850\mathrm{T}}\pm\sigma_{850\mathrm{T}}$ &
  $A_{850}$ & $B_{850}$ & $\Theta_{850}$ \\
 & (Jy\,bm$^{-1}$) & (Jy) & (arcsec) & (arcsec) & (degrees) \\
\hline
\hline
\rule{0pt}{3ex}$\!\!$
 1 &   0.3\,$\pm$\,0.1   &  0.6\,$\pm$\,0.2  &15.9 & 13.0 &  50.0 \\ 
 2 &  0.13\,$\pm$\,0.1   &  0.3\,$\pm$\,0.1  &26.3 & 13.9 & 127.0 \\ 
 3 &   3.4\,$\pm$\,0.2   &  9.4\,$\pm$\,0.3  &21.3 & 17.3 &  89.3 \\ 
 4 &   0.1\,$\pm$\,0.1   &  0.3\,$\pm$\,0.1  &22.5 & 15.1 &  36.4 \\ 
 5 &   0.2\,$\pm$\,0.1   &  0.4\,$\pm$\,0.2  &22.9 & 13.0 &  53.6 \\ 
 6 &   1.1\,$\pm$\,0.2   &  2.9\,$\pm$\,0.4  &22.8 & 13.4 & 102.7 \\ 
 7 &  0.12\,$\pm$\,0.01  & 1.25\,$\pm$\,0.04 &40.3 & 29.9 &  59.6 \\ 
 8 &  0.05\,$\pm$\,0.03  & 0.22\,$\pm$\,0.06 &37.1 & 20.2 &  45.5 \\ 
 9 & 0.025\,$\pm$\,0.008 & 0.87\,$\pm$\,0.04 &83.9 & 78.3 &  90.2 \\ 
10 &  0.29\,$\pm$\,0.1   &  0.7\,$\pm$\,0.1  &24.0 & 14.3 & 105.8 \\ 
11 &  0.03\,$\pm$\,0.01  & 0.27\,$\pm$\,0.02 &52.8 & 25.5 & 161.8 \\ 
\hline
\end{tabu}
\caption{
  SCUBA-2 850\mum parameters.
  As above.
}
\label{tab:f850}
\end{table*}

\begin{table*}[h!]
\centering
\begin{tabu}{| r || c | c | c | c | c |}
\hline
\# & $F_{3.4}$ & $F_{4.6}$ & $F_{12}$ & $F_{22}$ & \multirow{2}{*}{WISE name} \\
 & (Jy) & (Jy) & (Jy) & (Jy) & \\
\hline
\hline
\rule{0pt}{3ex}$\!\!$
3  & 7.1  & 14   & 230  & 760 & J060746.08$-$062312.0 \\ % HOBYS J060746.1 -062312
\hline
\end{tabu}
\caption{
  WISE fluxes and name.
  Note that only one of the first eleven sources was on the WISE catalogue
  (the protostar HOBYS J060746.1 $-$062312/Mon~R2 IRS~1).
}
\label{tab:fwise}
\end{table*}

\begin{figure*}[htb]
\centering
\mbox{\resizebox{\hsize}{!}{\includegraphics{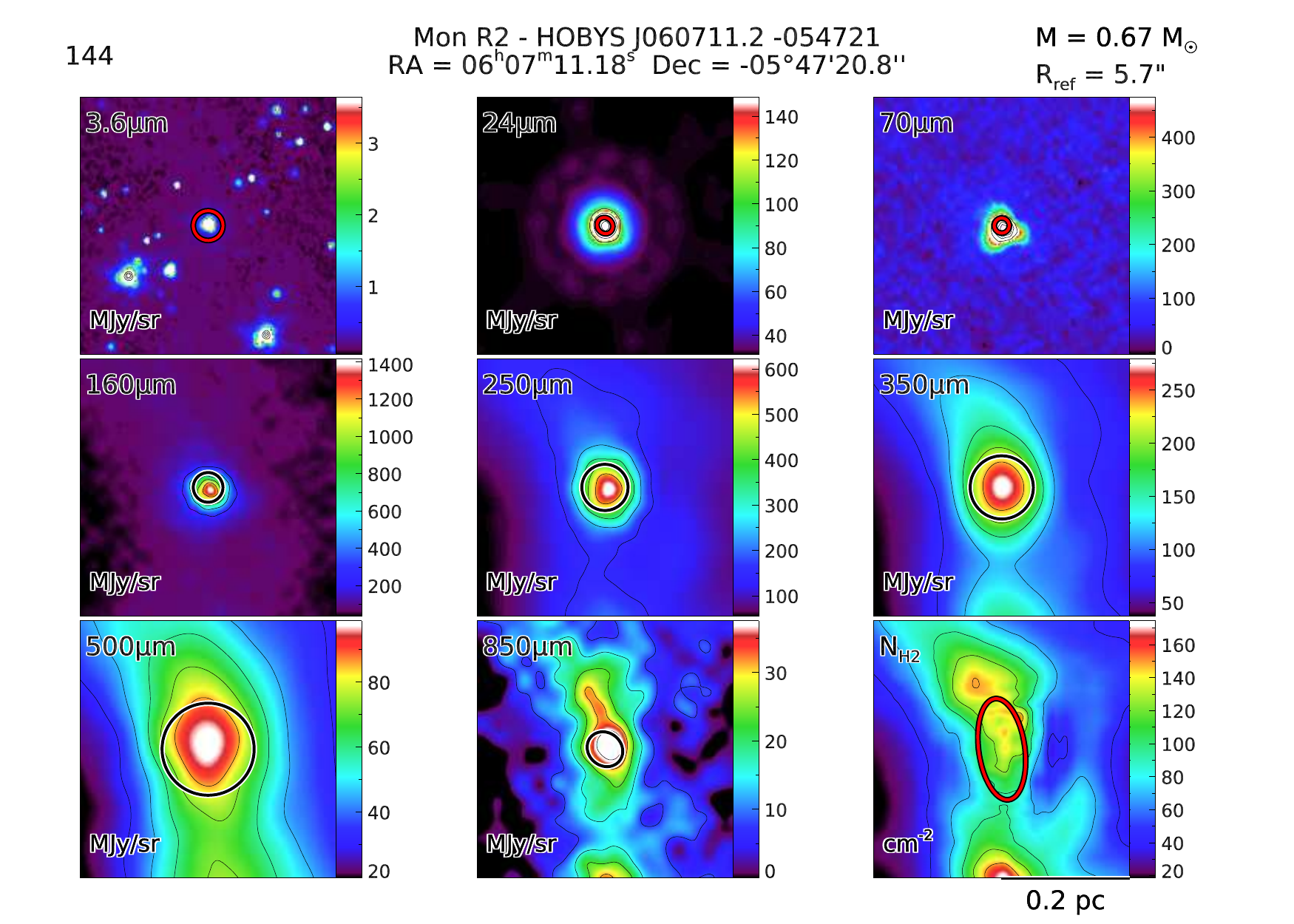}}}
\caption{The images of a probable protostellar core in the Monoceros R2 region
  at 3.6\mum (IRAC), 24\mum (MIPS), 70\mum, 160\mum (PACS), 250\mum, 350\mum, 500\mum (SPIRE), 850\mum (SCUBA-2) and in the
  high-resolution column density map (source HOBYS J060711.2 $-$054721; number 144 in the full catalogue).
  All images cover the same spatial scale (approximately $100''\,\times\,100''$, or $0.4\,$pc$\,\times\,0.4$\,pc);
  this is for consistency across all sources, some of which are much larger than HOBYS J060711.2 $-$054721.
  The ellipses show the extent of the source FWHM (assuming a 2d Gaussian shape)
  as determined by \getso for each individual wavelength.
  A black ellipse indicates that the detection at this wavelength was either
  considered reliable (160--850\mum), or was a 70\mum detection used for the SED fit.
  (The ellipse shown at 3.6\mum is the ellipse measured at the reference wavelength; here 160\mum.)
  The source mass and angular size at the reference wavelength are given at the upper right of the snapshots.
  We note that only one set of snapshots is included here;
  the full catalogue will contain snapshots for all robust sources.
  The SED of this source is given in Figure~\ref{fig:app_sed}.
}
\label{fig:app_cor}
\end{figure*}

\begin{figure*}[htb]
\centering
\mbox{\resizebox{\hsize}{!}{\includegraphics{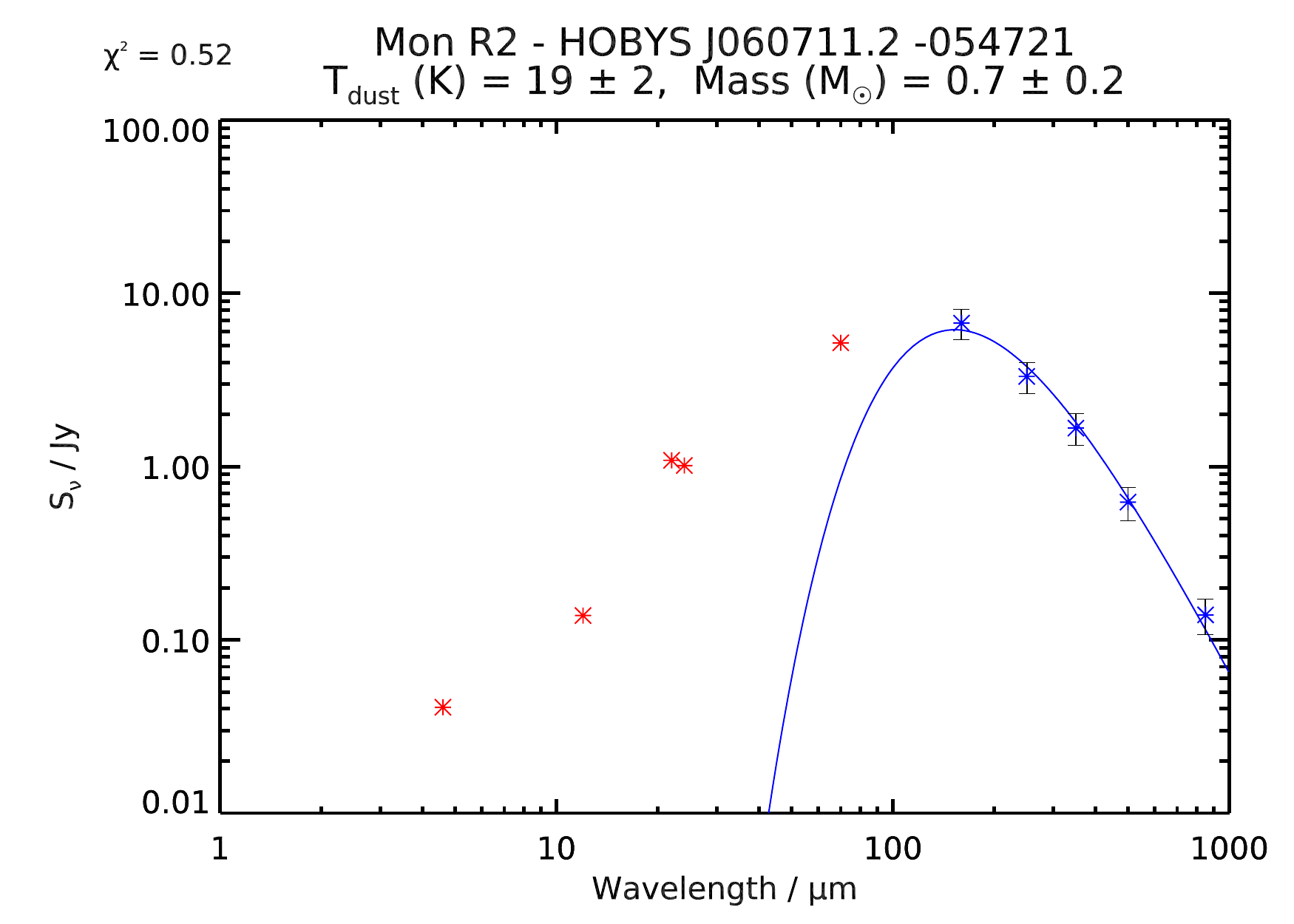}}}
\caption{The SED of a probable protostellar core in the Monoceros R2 region.
  Blue crosses represent reliable data points; red crosses
  represent short-wavelength data points used only for luminosity calculations
  (while not present here, some SEDs within the catalogue also show red crosses,
  albeit with long error-bars,
  which represent unreliable data points, which are used for SED fitting,
  but with greatly increased errors).
  The $\chi^2$ value of the fit is given at the upper left of the plot.
  The snapshots of this source are given in Figure~\ref{fig:app_cor}.
  We note that the fluxes are those used for the SED calculation,
  and have thus been modified using flux scaling.
}
\label{fig:app_sed}
\end{figure*}

The Mon~R2 HOBYS catalogue tables for the first eleven sources are given here
(ordered by mass; these are the sources over 10\,M\msun).
There are ten tables:
one containing the source name and position (Table~\ref{tab:base});
five containing \her \getso output (Tables~\ref{tab:f70}--\ref{tab:f850});
one for each wavelengths of non-\her \getso output
(in this case, MIPS 24\mum and SCUBA-2 850\mum);
one for WISE catalogue fluxes;
and one for derived parameters (this table is given in the main body of the
paper as Table~\ref{tab:deriv}).
We note that not all sources have associated WISE sources;
only those that do will be included here.
A full version of this table, including observed fluxes, and starless cores,
will be available online.
The full table will contain all robustly detected sources,
although with the caveat that the sample is likely incomplete,
especially below 1\,M\msun.
In addition to these tables, the catalogues will contain
% SEDs (such as Figure~\ref{fig:app_sed}), and 
a set of images for each source
(such as Figure~\ref{fig:app_cor}) showing their appearances at different wavelengths.
We note that all fluxes are direct outputs from \getso,
without flux-scaling or colour corrections applied.

\end{document}